\documentclass[%
twocolumn, onecolumn,
 aip,
 amsmath,amssymb,
 reprint,%
]{revtex4-1}

\usepackage{graphicx}
\usepackage{dcolumn}
\usepackage{bm}
\usepackage{float}
\usepackage{amsmath}
\usepackage{algorithm}
\usepackage{algorithmic}
\usepackage{xcolor}
\usepackage[utf8]{inputenc}
\usepackage[T1]{fontenc}
\usepackage{mathptmx}
\usepackage{multirow}
\begin{document}
\preprint{AIP/123-QED}
\newcommand{\sy}[1]{{\color{purple} Shuyang: #1}}
\newcommand{\dd}[1]{{\color{green!50!black} #1}}
 \title{Structure Prediction of Epitaxial Inorganic Interfaces by Lattice and Surface Matching with Ogre}



\author{Saeed Moayedpour}
\affiliation{Department of Chemistry, Carnegie Mellon University, Pittsburgh, PA 15213, USA}
\author{Derek Dardzinski}
 \affiliation{Department of Materials Science and Engineering, Carnegie Mellon University, Pittsburgh, PA 15213, USA}
 \author{Shuyang Yang}
 \affiliation{Department of Materials Science and Engineering, Carnegie Mellon University, Pittsburgh, PA 15213, USA}
 \author{Andrea Hwang}
 \affiliation{Department of Materials Science and Engineering, Carnegie Mellon University, Pittsburgh, PA 15213, USA}
\author{Noa Marom}%
 \email{nmarom@andrew.cmu.edu}
\affiliation{Department of Materials Science and Engineering, Carnegie Mellon University, Pittsburgh, PA 15213, USA}
\affiliation{Department of Chemistry, Carnegie Mellon University, Pittsburgh, PA 15213, USA}
\affiliation{Department of Physics, Carnegie Mellon University, Pittsburgh, PA 15213, USA}

\date{\today}

\begin{abstract}

We present a new version of the Ogre open source Python package with the capability to perform structure prediction of epitaxial inorganic interfaces by lattice and surface matching. In the lattice matching step a scan over combinations of substrate and film Miller indices is performed to identify the domain-matched interfaces with the lowest mismatch. Subsequently, surface matching is conducted by Bayesian optimization to find the optimal interfacial distance and in-plane registry between the substrate and film. For the objective function, a geometric score function is proposed, based on the overlap and empty space between atomic spheres at the interface. The score function reproduces the results of density functional theory (DFT) at a fraction of the computational cost. The optimized interfaces are pre-ranked using a score function based on the similarity of the atomic environment at the interface to the bulk environment. Final ranking of the top candidate structures is performed with DFT. Ogre streamlines DFT calculations of interface energies and electronic properties by automating the construction of interface models. The application of Ogre is demonstrated for two interfaces of interest for quantum computing and spintronics, Al/InAs and Fe/InSb.

\end{abstract}

\maketitle

\section{\label{sec:level1}Introduction}

Epitaxial inorganic interfaces play a crucial role in a wide range of modern day electronics, including semiconductor, spintronic, and quantum devices. For example, heterostructures of superconductors and semiconductors are considered as promising materials for the realization of topological quantum computing because they have shown evidence of Majorana fermions, which may help pave the way towards fault tolerant quantum computing. \cite{antipov2018effects,chang2015hard,das2012zero,Mourik2012,Zhang2016,Guel2017, su2020erasing,gazibegovic2017epitaxy,anselmetti2019end,de2018electric}
Another system of interest are interfaces between ferromagnetic and semiconducting materials which have been used in spintronic devices such as spin-filters and spin-valves. \cite{yang2020spin,sands1990stable,zhu2001room,lou2007electrical,schultz2009spin,crooker2005imaging} The functionality of such devices derives from the electronic and magnetic properties of the interface, which depend strongly on its structure at the atomistic scale. Different interface configurations may produce different properties.\cite{schultz2009spin, rath2018reduced, tung2014the, yang2021principles}
Moreover, defects and disorder may be detrimental to the functionality of a device. In particular, quantum devices may be extremely sensitive. Therefore, high-quality interfaces with precisely controlled structure and properties are required.   

 To grow high-quality interfaces, molecular beam epitaxy is often employed.\cite{Cho1983} If the lattice parameters and symmetries of the materials are closely matched, the film will strain to match the substrate and grow pseudomorphically. Three growth modes are possible:\cite{Sands1990,Narayan2002,Zheleva1994} two-dimensional, mono-layer by mono-layer growth via the Frank-Van der Merwe mode (FM); three-dimensional island growth via the Volmer-Weber mode (VW); and two-dimensional growth followed by three-dimensional growth via the Stranski-Krastanov mode (SK).\cite{Bauer1986} The balance between the surface energy of the film, $\gamma_{film}$, the surface energy of the substrate, $\gamma_{sub}$, and the interface energy, $\sigma$, determines which growth mode is favored.\cite{Bauer1986,Sands1990,Narayan2002} For the preferred FM growth to occur, the sum of the surface energy of the film and the interface energy must be smaller than the substrate surface energy: 
\begin{equation}
    \gamma_{film} + \sigma \leq \gamma_{sub}
    \label{eq: FM-cond}
\end{equation}
 This will always be true for homoepitaxy because $\gamma_{sub} = \gamma_{film}$ and $\sigma = 0$. In the case of heteroepitaxy, where $\gamma_{sub} \neq \gamma_{film}$ and $\sigma \neq 0$, the film is strained and the excess strain energy accumulates in the $\sigma$ term as the film grows. If the FM condition is not met initially, VW growth will take place, and if the FM condition is met initially, but is broken after a certain number of layers, SK growth will occur.\cite{Bauer1986}

An example of psuedomorphic growth can be found in the closely lattice matched heteroepitaxial interface of CdTe and HgTe.\cite{Chow1985, Faurie1982} Both materials assume the cubic zinc-blende structure and the $F \bar{4}3m$ space group, with lattice paramters of 6.45 {\AA} and 6.48 {\AA}, respectively, amounting to a lattice mismatch of 0.465\%.\cite{west1999basic} For such a heterostructre, determining the orientations of the film is trivial because the film will grow in the same direction as the substrate to minimize the lattice strain.
If the lattice parameters of the film and substrate do not closely match, the film may assume different orientations. Domain matching epitaxy may occur, where the lattices are matched using domains that contain integer multiples of film and substrate lattice plane spacings.\cite{Narayan2002, Xie2016, Zheleva1994, Trampert2000, Erwin2011,Krogstrup2015,Zur1984} For example, if the film has a lattice parameter that is 75\% of the substrate, then four lattice planes of the film will match with three unit lattice planes of the substrate for so called  cube-on-cube growth in the $[0 0 1]$ direction. The effective strain, $\epsilon_{\text{\textit{eff}}}$, produced by domain matching is given by:\cite{Narayan2002, Xie2016, Zheleva1994, Trampert2000, Erwin2011,Krogstrup2015,Zur1984}
\begin{equation}
    \epsilon_{\text{\textit{eff}}} = \frac{m d_{f} - n d_{s}}{n d_{s}}
\end{equation}
where $d_{f}$ and $d_{s}$ are the film and substrate lattice plane spacings, and $m$ and $n$ are the integer multiples required to match the domains. For many heterostructures, $\epsilon_{\text{\textit{eff}}}$ may be small for several orientations of the film, giving rise to the coexistence of multiple domains.\cite{Krogstrup2015, Trampert2000, rath2018reduced}

For example, one of the most studied superconductor/ semiconductor interfaces for Majorana-based quantum computing is Al deposited on InAs nanowires.\cite{antipov2018effects,chang2015hard,das2012zero,Mourik2012, gazibegovic2017epitaxy,de2018electric} 
Al and InAs are not lattice matched. When wurtzite InAs nanowires grown in the $\langle 0 0 0 1 \rangle$ direction serve as the substrate, face-centered cubic Al has been observed to grow on the $\{ 1 \bar{1} 0 0 \}$ InAs facets with both $\langle 1 1 1 \rangle$ and  $\langle 1 1 \bar{2} \rangle$ out of plane directions. 
For the $\langle 1 1 1 \rangle$ interface, there is a 7:5 match between the lattice planes in the Al $[1 1 \bar{2}]$ direction and the InAs $[0 0 0 1]$ direction, as well as a 3:2 match in the lattice planes between the Al $[1 \bar{1} 0]$ and the InAs $[1 1 \bar{2} 0]$ directions, resulting in effective strains of -0.5\% and 0.3\% respectively. The lattice matched domains for the $\langle 1 1 \bar{2} \rangle$ interface are a 1:1 match between the lattice planes in the Al $[1 1 1]$ direction and the InAs $[0 0 0 1]$ direction, as well as a 3:2 match in the lattice planes between the Al $[1 \bar{1} 0]$ direction and the InAs $[1 1 \bar{2} 0]$ direction, both resulting in effective strains of 0.3\%.\cite{Krogstrup2015} Because both growth directions of Al result in low effective lattice strain via domain matching epitaxy, they are both likely candidates for the interface structure. The $\langle 1 1 \bar{2} \rangle$ orientation is preferred at larger thicknesses of Al and forms a faceted surface, whereas the $\langle 1 1 1 \rangle$ orientation is preferred at smaller thicknesses and forms a planar surface. At an intermediate thickness range domains of the two orientations coexist.

The structure of interfaces may be characterized experimentally by observing a cross section with high resolution electron microscopy. \cite{Krogstrup2015,kanne2020epitaxial,Zheleva1994, rath2018reduced, zega2006determination} However, the exact alignment of the substrate and film may be difficult to determine precisely, especially if there are multiple domains present at the interface. Computer simulations may help interpret experiments and assist in the structural characterization of experimentally grown interfaces.\cite{rath2018reduced, liu2019coherent, zega2006determination}
Beyond interpreting experiments, computer simulations may aid in the search for new materials systems for semiconductor, spintronic, and quantum devices. The structure and properties of interfaces comprising various materials combinations may be predicted theoretically to guide synthesis efforts in the most promising directions. For example, simulations can be useful for epitaxial stabilization of metastable crystal structures with desirable properties.\cite{Wittkamper2017, Xu2017, Mehta2014, Ding2016}
Therefore, it is imperative to develop accurate and efficient methods for structure prediction of epitaxial inorganic interfaces. 

While structure prediction of inorganic crystals is fairly well-established, \cite{Oganov2019, yang2020spin,glass2006uspex,lonie2011xtalopt,trimarchi2007global} 
little work has been done on structure prediction of interfaces.\cite{Chua2010,Zhu2018, Raclariu2015, Mathew2016,Gao2019} Several of the codes developed for interface structure prediction,\cite{Mathew2016, Raclariu2015, ong2013python, larsen2017atomic} including the one introduced here, rely on implementations of the lattice matching algorithm by Zur and McGill,\cite{Zur1984} which works by generating domain matched superlattices between the film and substrate within user-defined mismatch and area tolerances. 
Once candidate interface structures are generated, their stability must be evaluated. Classical force fields are often used for this,\cite{Chua2010,Zhu2018,Gao2019} 
but general-purpose force fields may lack the accuracy to resolve small energy differences between interface configurations with similar stability.\cite{frederiksen2004bayesian, chen2017accurate} 
To achieve the required accuracy, \textit{ab initio} density functional theory (DFT) can be utilized to calculate the interface energies.\cite{li2016first,liu2004first,zhuo2018density,wang2020first} However, the increase in  accuracy entails an increase in computational cost. Owing to quantum size effects, a large number of layers of each material must be included in the interface model to converge its properties.\cite{yang2020electronic, yang2020topological, yang2021principles} 
Moreover, domain matched interfaces may require large supercells. This often amounts to models containing hundreds of atoms. Depending on the mismatch and area tolerances, many candidate interface structures may be produced by the lattice matching algorithm. Additionally, once a commensurate interface is identified, the separation between the substrate and film at the interface, as well as the registry in the plane of the interface may still be varied. Therefore, surface matching should be performed to find the optimal configuration(s). This may require sampling hundreds of points in the 3D search space. Therefore, a fast pre-screening method is useful to reduce the number of candidate interface structures to be considered by DFT.\cite{Raclariu2015, Mathew2016} 
 A solution proposed by Raclariu \textit{et al.}\cite{Raclariu2015} is a score function that ranks structures based on nearest neighbor distances and electronegativty differences at the interface, where bond lengths closer to the ideal bond length of a material and larger electronegativity differences between neighboring atoms lead to better scores. 
 
Here, we introduce a new version of the open source python package, Ogre, which was previously developed to generate surface models of molecular crystals and streamline the calculation of surface energies and Wulff shapes.\cite{yang2020ogre} We have implemented in Ogre a new functionality of predicting the structure of epitaxial inorganic interfaces by lattice and surface matching. Similar to Refs. \cite{Raclariu2015, Mathew2016}, the workflow of Ogre begins by using Zur and McGill's lattice matching algorithm and proceeds to perform surface matching. A new score function is proposed to optimize and rank candidate interface structures from purely geometric considerations without performing any energy evaluations. The score function is based on the overlap and empty space between atomic spheres at the interface. It is efficiently implemented using tensor algebra and demonstrated to correctly reproduce the extrema of the DFT potential energy surface. Bayesian optimization is then performed to explore the 3D space of interfacial distance and registry and find the most stable interface configurations. The optimal configurations of all domain matched interfaces are ranked using a geometric score function. 
In the final stage, a small subset of the most promising candidate structures are evaluated with DFT. Ogre streamlines the convergence of the interface thickness and the evaluation of interface energies with DFT. The application of Ogre is demonstrated for interfaces of interest for quantum computing and spintronics, including Al/InAs and Fe/InSb.

\section{\label{sec:level1}Code Description}

\begin{figure}[htbp]
\centering
\includegraphics[scale = 0.43]{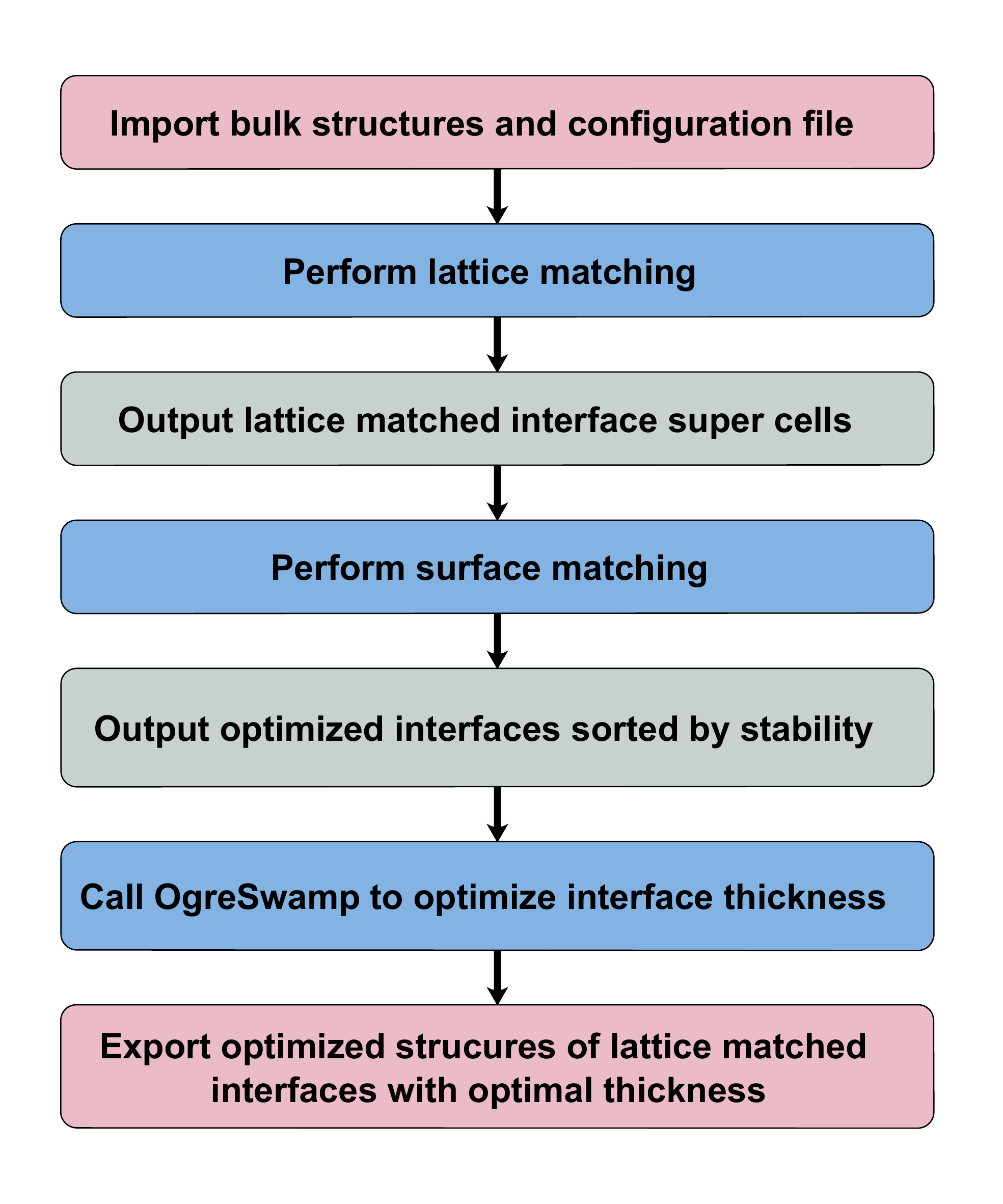}
\caption{Workflow of interface structure prediction with Ogre. The purple boxes represent code inputs and outputs. The blue boxes represent different code modules. The gray boxes show module outputs that serve as inputs of the subsequent module. 
}
 \label{fig:workflow}
\end{figure}

Ogre is written in Python 3 and utilizes the Python Materials Genomics (pymatgen) \cite{ong2013python} and Atomic Simulation Environment (ASE) \cite{larsen2017atomic} libraries. The package is available for download from www.noamarom.com under a BSD-3 license. The inputs to the code are the bulk structures of the substrate and film materials, as well as a configuration file that contains user-specified settings. Ogre supports several common formats of input structure  files, including crystallographic information files (CIF), the geometry.in format of the FHI-aims code \cite{Blum2009}, and the POSCAR format of the Vienna \emph{ab initio} Simulations Package (VASP).\cite{Joubert1999,Kresse1996,Kresse1996a,Kresse1993,Kresse1994} 
An overview of the workflow of interface structure prediction with Ogre is shown in Figure \ref{fig:workflow}. The hierarchical screening workflow of Ogre comprises three main steps: lattice matching (Section II.A), surface matching (Section II.B), and final ranking with DFT (Section II.C).  

The lattice matching step identifies all domain-matched supercells of the substrate and film within the user-defined misfit and area tolerances. If there are several possible orientations of the substrate and film, a Miller index scan may be performed. The input parameters for lattice matching are the maximum interface area, the misfit tolerance, and the Miller indices to be considered for the substrate and film. The output of the lattice matching step is a list of structures sorted by their super cell area misfit values. We note that the area misfit is a more stringent criterion than the effective strain along one direction.
From an experimental perspective, robust epitaxial growth usually occurs with misfit values below one percent.\cite{chang2012molecular} Therefore, the default criteria for selecting the interfaces that proceed to the surface matching step are supercell area misfit, lattice vector length misfit, and angle misfit below 1\%. The selection criteria may be modified by user input.

The surface matching step uses a geometric score function based on the overlap and empty space between atomic spheres at the interface to find the optimal distance in the $z$ direction and registry in the $xy$ plane between the substrate and film. For surface matching and ranking with the score function, the Hirshfeld radii of each chemical species must be specified in the configuration file. To this end, the Ogre radii calculation module streamlines the calculation of the Hirshfeld\cite{Spackman2009} radii using the FHI-aims code. If no Hirshfeld radii are provided, Ogre will automatically use the van der Waals radii tabulated in the Pymatgen periodic table module.
The score function serves as the objective function for Bayesian optimization to explore the 3D space above the substrate by shifting the film in $x$,$y$, and $z$ directions to optimize the structure of generated lattice matched interfaces. 
Subsequently, the optimized structures are ranked based on the deviation of the overlap between species at the interface from the respective bulk structures. The optimized interfaces are sorted based on their predicted stability and users can select a certain percentage to output.

For the most promising structures the interface energy is calculated with DFT. 
Ogre includes a module for streamlining interface energy calculations with DFT. Interface slab models are constructed with a user-defined number of layers and a vacuum region. Automatic passivation with pseudo-hydrogen atoms can be applied to terminate dangling bonds at the surfaces. An option of generating periodic heterostructure models without a vacuum region is also available. 
The number of layers of each material is converged by adding layers of either material to the interface model. For surface energy calculations, the linear method has been found to exhibit good convergence behavior.\cite{Scholz2019,Sun2013} Here, we use a modified linear approach for the calculation of interface energies. Finally, for the most stable interface structures further analysis may be performed with DFT, including structural relaxation and calculation of electronic properties.

\subsection{\label{sec:level2}Lattice Matching}

\begin{figure}[htbp]
\centering
\includegraphics[scale = 0.43]{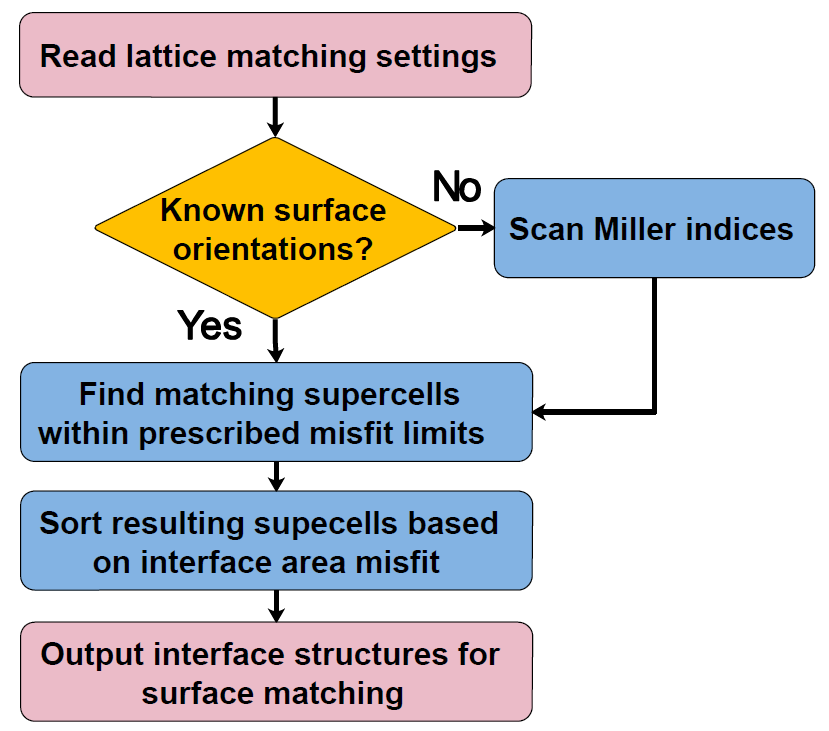}
\caption{The workflow of lattice matching in Ogre. The blue boxes represent different lattice matching modules. The yellow diamond represents a decision whether a Miller index scan should be performed
}
 \label{fig:lattice_workflow}
\end{figure}

The workflow of Ogre's lattice matching module is illustrated in Figure \ref{fig:lattice_workflow}. The algorithm proposed by Zur and McGill\cite{Zur1984} is utilized to identify matching supercells of the film and substrate.
First, bulk structures are cleaved along the specified Miller planes using ASE to determine the basis vectors of the substrate and film surface slab models. The resulting surface basis vectors are reduced to a pair of primitive basis vectors using Pymatgen to obtain a unique representation of two-dimensional lattices, which is necessary for the comparison of substrate and film lattice properties. Second, the Pymatgen substrate analyzer module (which follows the Zur and McGill algorithm) is utilized to generate all the transformation matrices that  
would produce lattice matched super-cells with a low misfit of lattice vector lengths and angles.  Third, a reduction scheme is used to find all unique commensurate interfaces of a given system. 
Finally, the interface structure is constructed. By default, the lattice parameter of the substrate is fixed and the film layer is strained to match the substrate to simulate an epitaxial growth experiment. Ogre creates the matching substrate and film super-cells and aligns the atomic coordinates to build a lattice matched interface with specified structural properties, including the interfacial distance in the $z$ direction, the shifts in the $xy$ plane, the number of layers, and the amount of vacuum.

Ogre's lattice matching module takes three sets of input parameters that determine the matching criteria and interface structural properties: the substrate and film Miller indices; maximum super-cell area; and misfit tolerances for the unit cell area, length, and angle. The default for all misfit tolerances is one percent and the default maximum super-cell area is 500 {\AA}$^2$. Specifying a larger maximum area or higher misfit tolerances would generate additional candidate interfaces, however they are likely to be less stable. The user may also specify parameters for constructing the interface model, such as the amount of vacuum, the interfacial distance, the shift in the $xy$ plane, and the number of layers. 
The default values of the vacuum region and interfacial distance are 40 {\AA}$ $ and 2 {\AA}$ $, respectively. By default, no $xy$ shifts are applied to the initial interface structure. The optimal registry is later found by Ogre's surface matching module. The substrate and film thickness can be modified by specifying the number of layers or a range of values to calculate.
The surface termination may be defined by the user. If a surface termination is not set by the user, Ogre can identify all possible terminations and automatically generate the corresponding interfaces. For example, for  InAs(111) the user may specify an As-terminated or In-terminated surface, otherwise Ogre will generate both.

\begin{figure}[htbp]
\centering
\includegraphics[scale = 0.30]{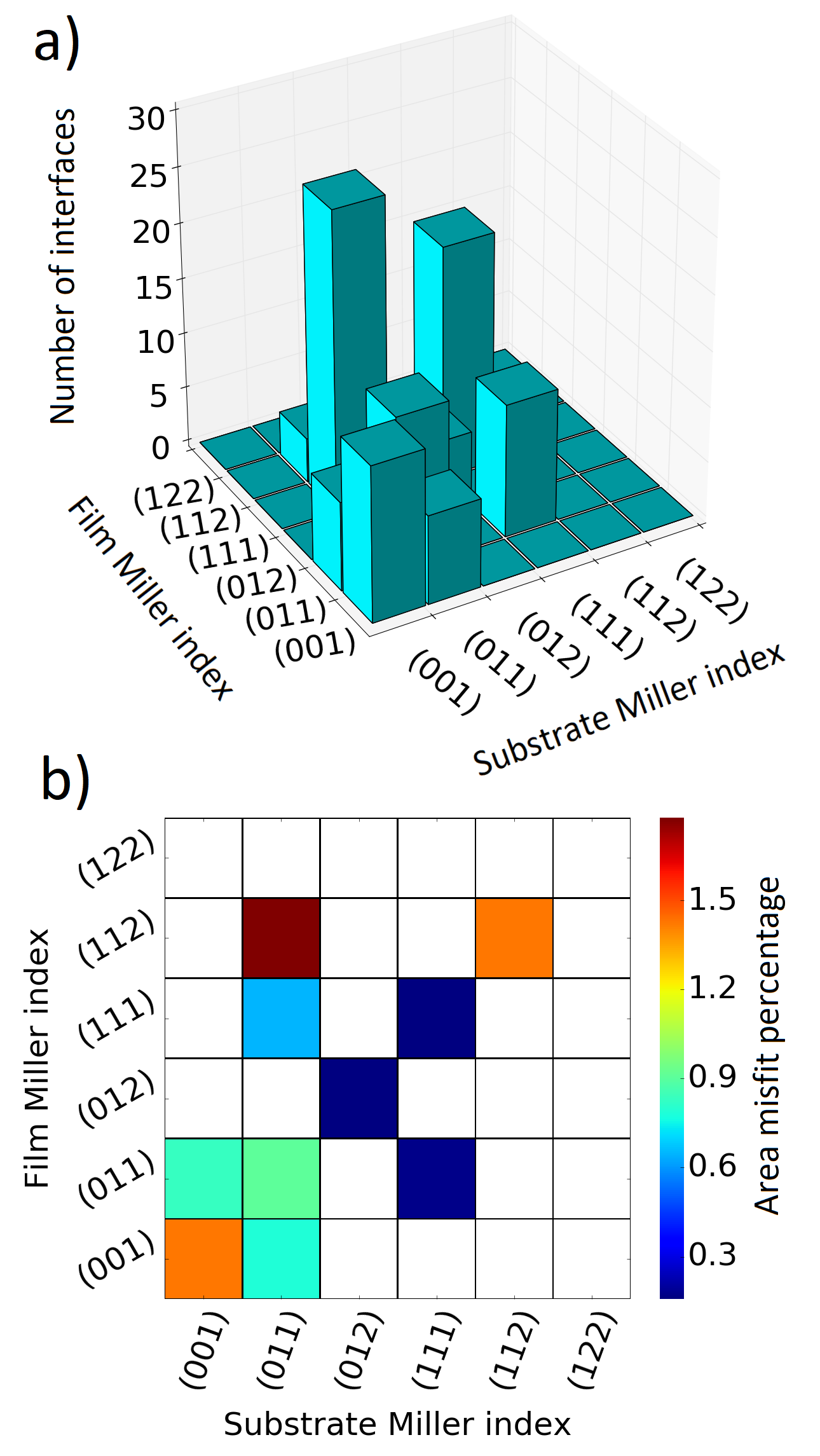}
\caption{Results of a Miller index scan for the Al/InAs interface with a maximal Miller index of 2, maximum interface area of 500 {\AA}$^2$,  and area misfit tolerance of 2\% a) Histogram of the number of domain-matched interfaces generated for each set of Miller indices. b) A heat map plot showing the lowest area misfit obtained for each set of Miller indices. White cells represent Miller index combinations for which no structures with an area misfit below 2\% were found.}
 \label{fig:Miller}
\end{figure}

If the orientation of the substrate and film is known, the user may specify the corresponding Miller indices in the configuration file. In some cases, several substrate orientations are possible and/or the film orientation is not known. For example, when a superconductor is grown on top of a semiconductor nano-wire, growth on different facets may result in different orientations.\cite{krogstrup2015epitaxy, gusken2017mbe}
In such cases, a Miller index scan may be conducted, where lattice matching is performed for each combination of substrate and film Miller indices. The Miller index search module takes the maximum single index as input and finds all possible symmetrically unique Miller indices as described in Ref. \cite{yang2020ogre}.  
For example, Figure \ref{fig:Miller} shows the results of a Miller index scan for an Al/InAs interface with a maximal Miller index of 2. The results were generated using a maximum interface area of 500 {\AA}$^2$ and a misfit tolerance of 2\%. Figure \ref{fig:Miller}(a) displays a histogram of the number of interfaces generated for each combination of Miller indices. In total, 116 candidate lattice matched interfaces are generated with these settings. 
Figure \ref{fig:Miller}(b) shows the minimum misfit percentage obtained for each interface orientation. 
In this case, seven of the ten possible orientations found in the Miller index scan have a misfit under 1\%. Further screening of these structures is performed in Section IV below.

\subsection{\label{sec:level2}Surface Matching}

\begin{figure}[htbp]
\centering
\includegraphics[scale = 0.43]{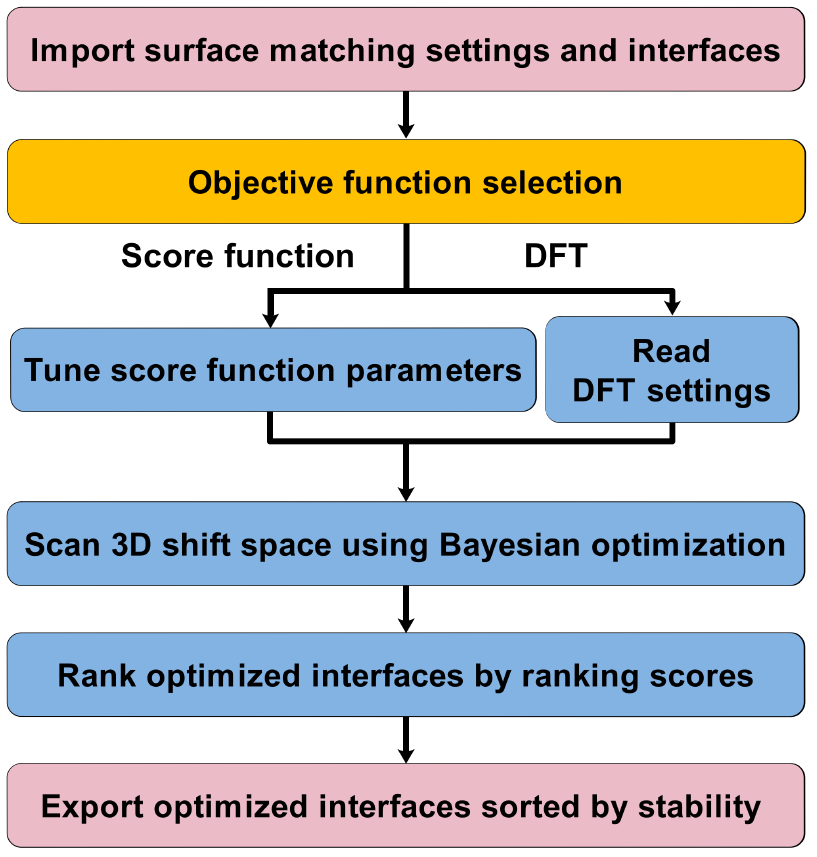}
\caption{The workflow of surface matching in Ogre. The purple boxes represent code inputs and outputs. The blue boxes represent different code modules. The yellow box represents a decision on which objective function to use
}
 \label{fig:surf_flow}
\end{figure}

Once commensurate interface supercells are identified via lattice matching, surface matching is performed to determine the optimal distance in the $z$ direction and registry in the $xy$ plane between the substrate and film.\cite{Raclariu2015, Mathew2016} Owing to the size of the configuration space to be searched (the number of lattice matched candidates multiplied by the number of displacements to be sampled in the $x$, $y$, and $z$ directions for each interface) it is desirable to avoid computationally expensive DFT calculations. Therefore, we have developed a geometric score function.  In Ref. \cite{Marom2010}
it has been shown that a simple geometric model based on the overlap of circles captures the main features of the DFT potential energy surface for the interlayer sliding of hexagonal boron nitride. Here, we define a score function based on the overlap and empty space between atomic spheres at the interface.

The workflow of Ogre's surface matching module is shown in Figure \ref{fig:surf_flow}. After importing the interface structures constructed in the lattice matching step, the user may choose to utilize the geometric score function or DFT to perform surface matching. Bayesian optimization is then used to efficiently scan the 3D parameter space to find the optimal position of the film above the substrate. Finally, if the geometric score function is selected, a geometry based metric is used to rank the optimized structures based on their predicted stabilities. If DFT is selected, the structures are ranked based on their DFT energy. 
The optimized interface structures are written in the appropriate geometry file format (e.g., POSCAR for VASP).

\subsubsection{\label{sec:level3}Geometric Score Function}

The score, $S$, is defined as:
\begin{equation}
S = (1 + \bar{O})^2 + c\bar{E}
\label{eq:score}
\end{equation}
where the scaled overlap is defined as the overlap volume divided by the total volume occupied by atoms at the interface:
\begin{equation}
\bar{O}  = \frac{V_O}{\sum_{atoms}V_{at}} \\
\label{eq:scaled_overlap}
\end{equation}
and the scaled empty space is defined as the volume of empty space divided by the unit cell volume:
\begin{equation}
\bar{E}  = \frac{V_E}{V_{cell}} 
\label{eq:scaled_empty}
\end{equation}

We note that the unit cell volume changes when the interfacial distance is changed along the $z$ direction. 
The effective atomic volume of each element, used for the calculation of the score function, is a system dependent parameter. Here, it is obtained using Hirshfeld partitioning of the bulk material's DFT charge density. Ogre's volume determination module streamlines the calculation of Hirshfeld volumes with the FHI-aims code and converts them into scaled Hirshfeld radii. The scaled Hirshfeld radius of species M, $R_M$, is given by:
\begin{equation}
R_M = \frac{\rho}{1 + \rho} \times D_{MX} \qquad \rho  = \sqrt[3]{\frac{H_{V,M}}{H_{V,X}}}
\label{eq:scaled_rad1}
\end{equation}
where $D_{MX}$ is the minimum distance between species M and X in the corresponding bulk crystal structure, and $H_{V,M/X}$ is the Hirshfeld volume of species M/X in the bulk crystal structure. The species radii for the substrate and film are calculated separately so if a certain species exists in both the substrate and film slabs (e.g., Te in the SnTe/CaTe interface) two different radii are considered. If there are more than two species in the bulk crystal, the resulting Hirshfeld radii for each element are averaged.

To calculate overlap and empty space at the interface Ogre applies an efficient vectorized method based on mesh voxelization, as shown in Figure S1 in the SI. Voxel representations of the substrate and film slabs are generated and stored as two 3D binary matrices, in which voxels occupied by atoms are assigned a value of one and empty voxels are assigned a value of zero. The default voxel volume is 0.001{\AA}$^3$. This produces a sufficiently fine grid to sample the atomic surface accurately. The user may set a different voxel size. For the calculation of empty space and overlap logical NOR and logical AND operators are performed on the two binary matrices, which results in overlap and empty space matrices. To calculate the overlap volume and empty space in {\AA}$^3$ the voxel volume is multiplied by the number of "one" entries in the overlap and empty space matrices respectively.

\begin{figure}
\centering
\includegraphics[scale = 0.52]{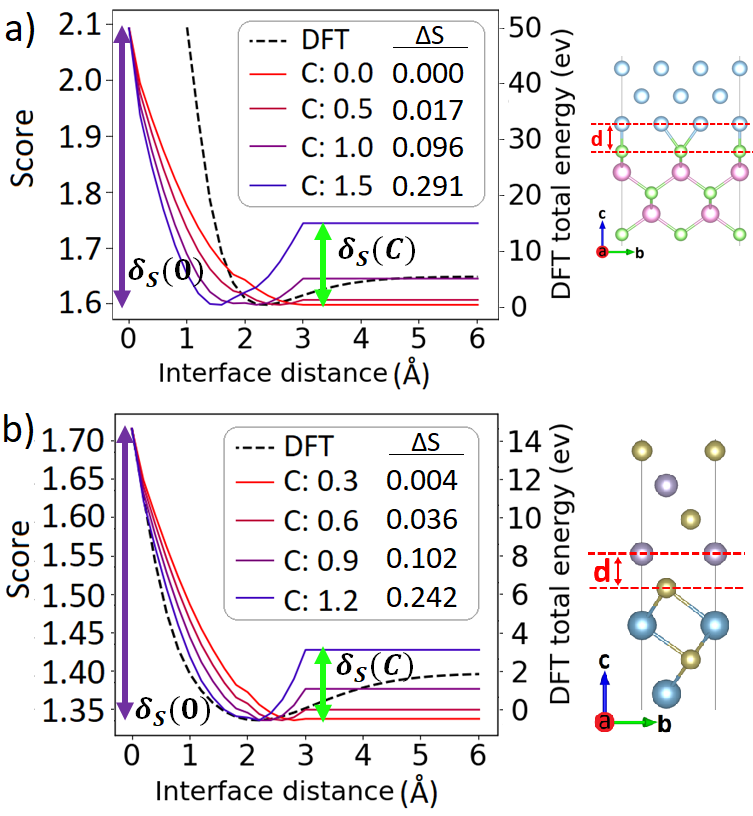}
\caption[Two numerical solutions]{Performance of the geometric score function for determining the interfacial distance: The score obtained with different values of $c$ as a function of the interfacial distance in the z direction compared to the DFT total energy curves for (a) the Al(100)/InAs(100) interface and (b) the SnTe(111)/CaTe(111) interface.}  
\label{fig:Deltas}
\end{figure}

 In Eq. \eqref{eq:score}, the scaled overlap and empty space terms effectively serve the function of repulsive and attractive terms, respectively. The effective repulsion term is squared to reflect the short-ranged nature of repulsive forces. The coefficient $c$ balances the weights of the overlap and empty space terms. The value of $c$ significantly affects the performance of the score function. For example, Figure \ref{fig:Deltas} shows the score as a function of the interfacial distance obtained with different values of $c$ for interfaces of Al(100)/InAs(100) and SnTe(111)/CaTe(111). DFT total energy curves are shown for comparison with the minimum energy of each curve referenced to zero (for DFT settings, see Section III). The interface distance is calculated by subtracting the height (z value in Cartesian coordinates) of the top atom of the substrate slab from the height of the bottom atom in the film slab.

 We have developed a procedure for finding the optimal value of $c$ without relying on DFT calculations. We define $\Delta_S(c)$ as the the difference between the asymptotic and minimal score values obtained with a given value of $c$, $\delta_S(c)$, divided by the difference between the maximal and minimal score values obtained with $c=0$, $\delta_S(0)$, as illustrated in Figure \ref{fig:Deltas}:   
 \begin{equation}
\Delta_S(c)  = \frac{\delta_S(c)}{\delta_S(0)}
\label{eq:score_ratio}
\end{equation}
 Increasing $c$ leads to an increase in $\Delta_S(c)$ and a decrease in the optimal interface distance produced by the score function. We have found empirically that the optimal distance produced by the score function is closest to the DFT result when $\Delta_S(c)$ is 0.1,  as shown in Figure \ref{fig:Deltas}.   
 Therefore, to find the optimal $c$ coefficient for a certain interface Ogre starts from $c$=0 and incrementally increases $c$ until $\Delta_S(c)$ reaches 0.1.    
 As the film is shifted in the $xy$ plane, the optimal $z$-distance and $c$ parameter may also vary. Examples are shown in Figure S2 in the SI. This requires Ogre to find an optimal $c$ parameter that performs well over the entire search space. To this end, Ogre first generates a preliminary 2D score contour at the initial user-defined interface distance to locate the positions of the minimum and maximum.  The $c$ coefficient optimization process is then performed for the structures with  the minimal and maximal score values. Finally, the value of the two $c$ coefficients obtained for extrema of the score contour is averaged and used for the 3D surface matching. The averaged value of $c$ is found to be 0.43 for the Al(100)/InAs(100) interface and 0.54 for the SnTe(111)/CaTe(111) interface.

\begin{figure}[htbp]
\centering
\includegraphics[scale = 0.155]{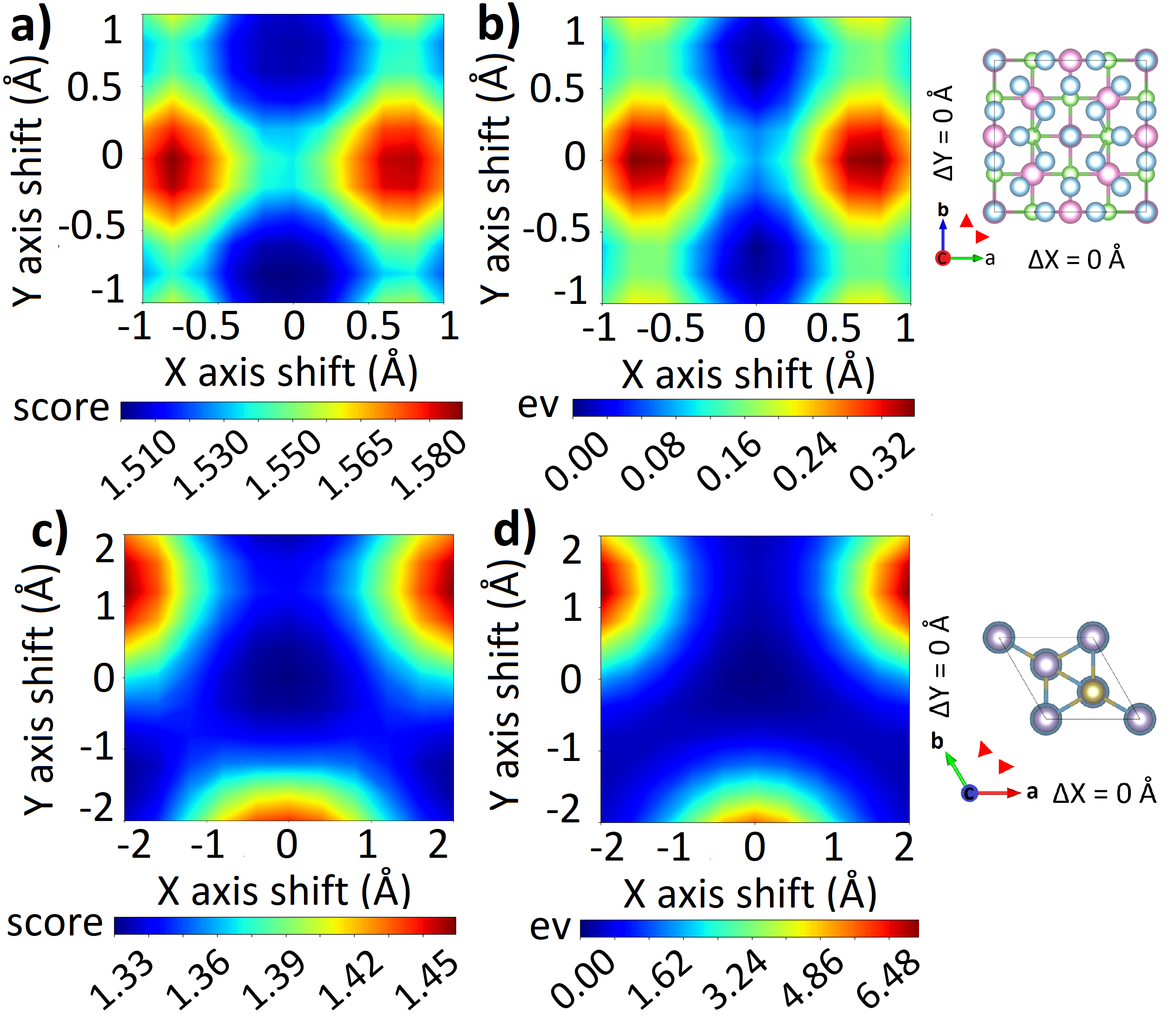}
\caption{ Performance of the geometric score function for the registry in the $xy$ plane: (a) Score function contour plot compared to (b) the DFT potential energy surface at a fixed interfacial distance of 2.2 {{\AA}} for the Al(011)/InAs(001) interface. (c) Score function contour plot compared to (d) the DFT potential energy surface at a fixed interfacial distance of 2.0 {{\AA}} for the SnTe(111)/CaTe(111) interface. } 
\label{fig:Contours}
\end{figure}

To validate the score function, we compare its results to DFT. Figure \ref{fig:Contours} shows  contour plots of the score obtained with the optimal value of $c$, compared with the DFT potential energy surface as a function of the displacement in the $xy$ plane at a fixed interfacial distance of 2.2 {{\AA}} for Al(100)/InAs(100) interface and 2.0 {{\AA}} for SnTe(111)/CaTe(111) interface. In both cases the score function reproduces well the features of the DFT potential energy surface and the positions of the extrema. An additional example for a ternary compound is provided in Figure S3 in the SI.

\subsubsection{\label{sec:level3}Bayesian Optimization}
 
 Once the atomic radii and the $c$ parameter of the score function are determined, a three-dimensional search is performed to find the $(x,y,z)$ coordinates of the global minimum structure. To this end, we use Bayesian optimization (BO).\cite{Frazier2018}
 BO is a machine learning algorithm that efficiently samples points from a black-box objective function to find the global optimum. The objective function is first estimated by a Bayesian statistical model using a Gaussian process prior.\cite{brochu2010tutorial} An acquisition function is then used to predict the optimal position for the next point to be sampled, and the prior is updated with the new information to create a new surrogate function called the posterior. Once the posterior is calculated, it becomes the new prior. The process is repeated for a specified number of steps or until convergence is achieved.  
 BO is a superior optimization technique compared to a grid search because it maximizes the information gained about the black-box function, while sampling a minimal number of points.   
 
 To perform Bayesian optimization, Ogre utilizes the  bayesian-optimization Python package.\cite{Nogueira2014} Here, the BO objective function for surface matching is defined as the negative of the score value or the negative of the DFT total energy, such that the score/energy value is minimized by maximizing the objective function. If the DFT total energy is used as the objective function, Ogre is compatible with FHI-aims and VASP. The DFT calculation settings are read from an input file provided by the user. Ogre automatically performs the necessary DFT calculations and analyzes the results. 
 
 The parameter space to be searched is defined by the default bounds for shifts in the $x$, $y$,and $z$ directions, which are (0, $a$), (0, $b$), and ($d-1$ {\AA}, $d+1$ {\AA}), respectively, where $a$ and $b$ are the interface unit cell lattice parameters in the $xy$ plane and $d$ is the initial interface distance. To determine the next point ($\vec{r}$) to be sampled in each iteration, the upper confidence bound acquisition function,\cite{Frazier2018, brochu2010tutorial, williams2006gaussian} is used:
 \begin{equation}
     \vec{r}_{n+1} = \text{argmax}(\mu_{n}(\vec{r}) + \kappa \sigma_{n}(\vec{r}))
     \label{eq:UCB}
 \end{equation}
 where $\mu$ and $\sigma$ are the mean and standard deviations at each point and $\kappa$ is a hyperparameter that
 controls the trade-off between exploration and exploitation. By default, $\kappa$, is set to 5 and the number of iterations, \emph{N}, is set to 100. Both parameters can be modified through the surface matching settings file. Once the maximal number of iterations is reached the code outputs the most stable structure, as well as any structures whose score or DFT energy is within a user defined tolerance of the minimum. Although BO is the default and recommended optimization method in Ogre, a grid search option is available, \textit{e.g.,} to generate potential energy surfaces or binding energy curves. After the structure optimization is completed the optimized interface structures may be exported to input geometry files for DFT calculations or proceed to Ogre's structure ranking module.

\subsubsection{\label{sec:level3}Preliminary Ranking}

\begin{figure}[htbp]
\centering
\includegraphics[scale = 0.75]{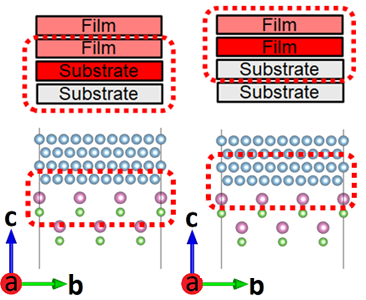}
\caption{Illustration of the selected layers for the calculation of the substrate and film relative overlaps at the interface. }
\label{fig:ranking}
\end{figure}

Surface matching is performed for every candidate structure passed from the lattice matching step, which may still amount to a large number of structures. Therefore, it is desirable to perform preliminary ranking in order to select a smaller number of the most promising candidate structures for the final evaluation with DFT. The geometric score function used for surface matching cannot be used for structure ranking because the score function parameter, $c$, is optimized for each interface, such that interfaces between the same materials with different orientations may have different $c$ parameters.
A  ranking score function, $R$,  is formulated based on the assumption that a stable interface is more likely to form when the chemical environment at the interface is similar to that of the bulk materials. The scaled overlap, defined in Eq. \eqref{eq:scaled_overlap}, is used as a measure of similarity between the bulk and the interface (int) environment for the film and substrate (sub): 
\begin{equation}
R=|\bar{O}_{int}^{film}-\bar{O}_{slab}^{film}|+|\bar{O}_{int}^{sub} - \bar{O}_{slab}^{sub}| 
\label{eq:ranking_metric}
\end{equation}
$\bar{O}_{slab}$ is calculated for unstrained  film and substrate slab structures with the same orientation as the interface, such that the interface strain is taken into account in the ranking.
$\bar{O}_{int}$ is calculated for the first few layers of the film and substrate at the interface, as shown in Figure \ref{fig:ranking}. Atoms that are not in the vicinity of the interface do not affect the surface contours (see SI for an example of the voxel representation of the surface contour). Ogre automatically finds the minimum number of layers required for full representation of the substrate and film surface contours at the interface. 
A smaller value of $R$ means that the chemical environment at the interface is more similar to that of the bulk materials. Hence, the structure with the lowest value of $R$ is expected to be the most stable. 

\begin{figure}[htbp]
\centering
\includegraphics[scale = 0.22]{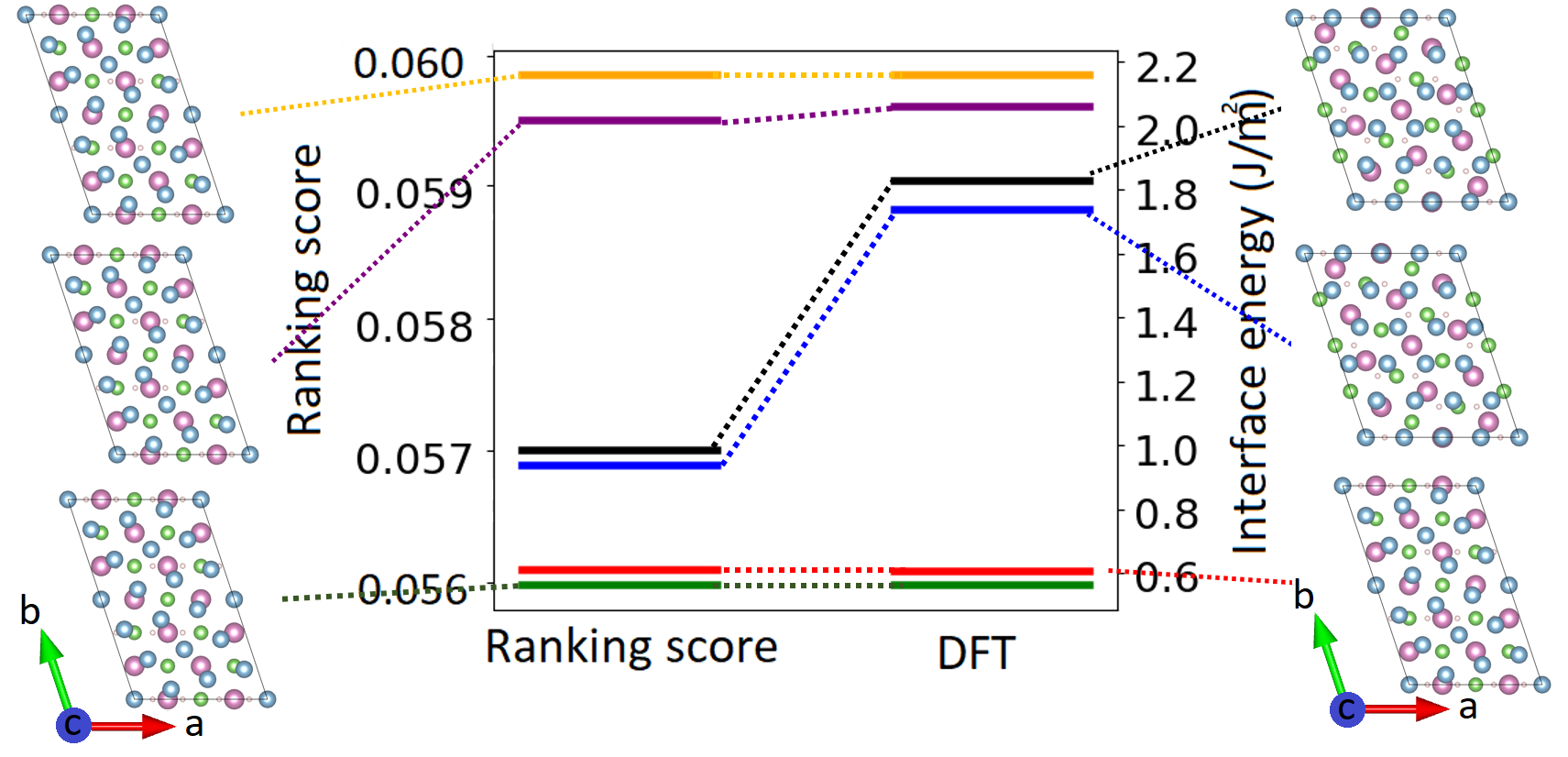}
\caption{The ranking score compared with DFT interface energies for the most stable structures of the Al(011)/InAs(100) interface. 
The ranking score correctly reproduces the order of stability obtained from DFT.}
\label{fig:ranking_result}
\end{figure}

In Figure \ref{fig:ranking_result}, the ranking score is compared to DFT interface energies (calculated as described in Section C) for six candidate structures of the Al(011)/InAs(001) interface. 
 The ranking score, which is based on purely geometric considerations, successfully reproduces the ranking order obtained with DFT and identifies the most stable candidate structures. The ranking score is found to perform similarly well for several additional interfaces, as shown in the Applications Section below.
 Based on the ranking score, the list of structures selected to proceed to the final DFT ranking stage may be narrowed down. 

\subsection{\label{sec:level2} DFT Ranking}

\subsubsection{\label{sec:level3}Automated Surface Passivation}

\begin{figure}[htbp]
\centering
\includegraphics[scale=0.19]{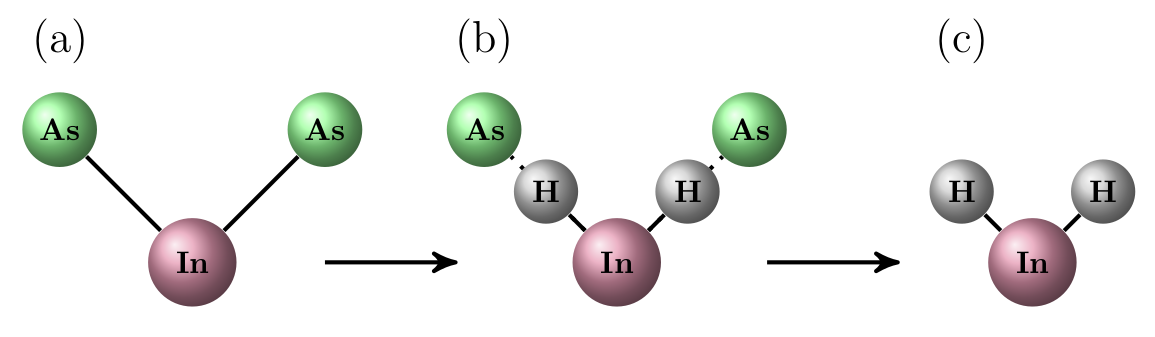}
\caption{A schematic representation of the pseudo-hydrogen passivation function: (a) The bonds are identified in the initial structure (b) Pseudo-hydrogen atoms are inserted along the bond at a distance proportional to the covalent radii of the hydrogen and terminating species. (c) The old surface atoms are removed to leave a passivated surface}
 \label{fig:passivation-schematic}
\end{figure}

In DFT calculations of surface and interface slab models dangling bonds on the surface are terminated by pseudo-hydrogen atoms with appropriate fractional charges to avoid surface states.\cite{huang2005surface, deng2012effect, zhang2016pseudo} 
Ogre automates the surface passivation during the construction of interface models, as illustrated in in Figure \ref{fig:passivation-schematic}. 
First, one additional layer is added to the top and bottom surfaces of the slab. Second, vectors in the directions of the dangling bonds are constructed by performing a nearest neighbor search for the atoms on the surface to obtain the proper coordination. Third, the relative positions of each neighboring atom are converted from Cartesian to spherical coordinates and the pseudo hydrogen is inserted between the covalent radii of the two atoms. The charge of the pseudo hydrogen ($q_{H}$) is calculated using the number of valence electrons of the atom being replaced by pseudo hydrogen ($N_{v}$) and the coordination of that atom ($X$):
\begin{equation}
    q_{H} = \frac{N_{v}}{X}
\end{equation}
The spherical coordinates are then converted back to Cartesian coordinates. Lastly, the top and bottom atomic layers are removed, leaving only the pseudo-hydrogen passivation layer on the surface. After the passivated structure is generated, the pseudo-hydrogen positions should be relaxed using DFT. For computational efficiency, this is performed using a structure with a small number of layers and the relaxed pseudo hydrogen positions are subsequently transferred to larger slab models.

\subsubsection{\label{sec:level3}Interface Energy Evaluation}

The interface energy, $\sigma$, is defined as the energy of eliminating two surfaces and creating an interface:\cite{xiong2017first, christensen2002first, Arya2003, Arya2004, Qi2004}
\begin{equation}
\sigma = \gamma_{sub} + \gamma_{film} - W_{ad} 
\label{eq:interface_energy}
\end{equation}
where $W_{ad}$ is the adhesive energy of the interface and $\gamma_{sub/film}$ are the surface energies of the substrate and film, calculated with the OgreSwamp module of Ogre,\cite{yang2020ogre} using the linear method:\cite{Scholz2019,Sun2013} 
\begin{equation}
    E_{slab} = N E_{bulk} + 2A\gamma_{slab}
    \label{eq:surface_energy}
\end{equation}
where $E_{slab}$ and $E_{bulk}$ are the DFT total energies of a surface slab with $N$ layers and a bulk unit cell, respectively and $A$ is the cross-section area of the interface. We note that $\gamma_{sub/film}$ must be recalculated if the respective lattice is strained (by default only the film is strained in Ogre). Examples are provided in Figure S4 in the SI.
The adhesive energy of an interface is given by:\cite{li2016first,liu2004first,zhuo2018density,wang2020first, Arya2003, Arya2004, Qi2004}
\begin{equation}
W_{ad} = \frac{1}{A}(E_{sub} + E_{film} - E_{int})
\label{eq:adhesive_energy}
\end{equation}
where $E_{sub}$, $E_{film}$, and $E_{int}$ are the DFT total energies of the substrate slab, the film slab, and the interface slab, respectively. 
Because $E_{sub}$,$E_{film}$, and $E_{int}$ are dependent on the number of layers in the system, $W_{ad}$ must be converged with respect to the number of layers used in the interface slab construction. Often,  the number of layers of the substrate and film are converged separately and then used to build the interface model.\cite{xiong2017first,wang2020first}
However, we find that owing to the electronic and magnetic interactions at the interface, the number of substrate/ film layers required to converge the interface energy may be different than the number of layers required to converge the surface energy of either material. Similar to the calculation of surface energy, we may define a linear method for the calculation of the interface energy by substituting Eq. \eqref{eq:adhesive_energy} and Eq. \eqref{eq:surface_energy} into Eq. \eqref{eq:interface_energy}:
\begin{equation}
E_{int}   = E_{bulk}^{sub}N_{sub} + E_{bulk}^{film}N_{film} + A(\sigma + \gamma_{sub} + \gamma_{film})  
\label{eq:interface_conv}
\end{equation}
Where $E_{int}$ is the total energy of the interface, $N_{sub}$ and $N_{film}$ are the number of layers of the substrate and film in the interface model, $E_{bulk}^{sub}$ and $E_{bulk}^{film}$ are total bulk energies of the substrate and film. Using Eq. \eqref{eq:interface_conv}, the interface energy, $\sigma$, can be obtained by performing two linear regressions: one where the number of slab layers is held constant and the number film layers is increased, and one where the number of film layers is held constant and the number of slab layers is increased. In each case, after a user-defined number of layers of the film/substrate are added to the interface, linear regression is performed, and $\sigma$ can be extracted from the intercept, $b$:

\begin{equation}
    \sigma_{film/sub}(N) = \frac{b - E_{bulk}^{film/sub} N_{film/sub}}{A} - \gamma_{sub} - \gamma_{film}\\
\end{equation}

In these regressions a straight line is fitted to all of the interface total-energy data versus the number of layers except for the structures with that contain only one atomic layer of the film/ substrate, following the method of Refs. 
\cite{Scholz2019,Sun2013}.
The interface energy is taken as the average of the $\sigma_{film/sub}$ values obtained from both regressions: 
\begin{equation}
    \sigma(N) = \frac{\sigma_{film}(N) + \sigma_{sub}(N)}{2}
\end{equation}
The default convergence criterion
is that the relative deviation of the averaged interface energy should be within 0.3\%  with the addition of one layer.

\begin{figure}[htbp]
\centering
\includegraphics[scale = 0.185]{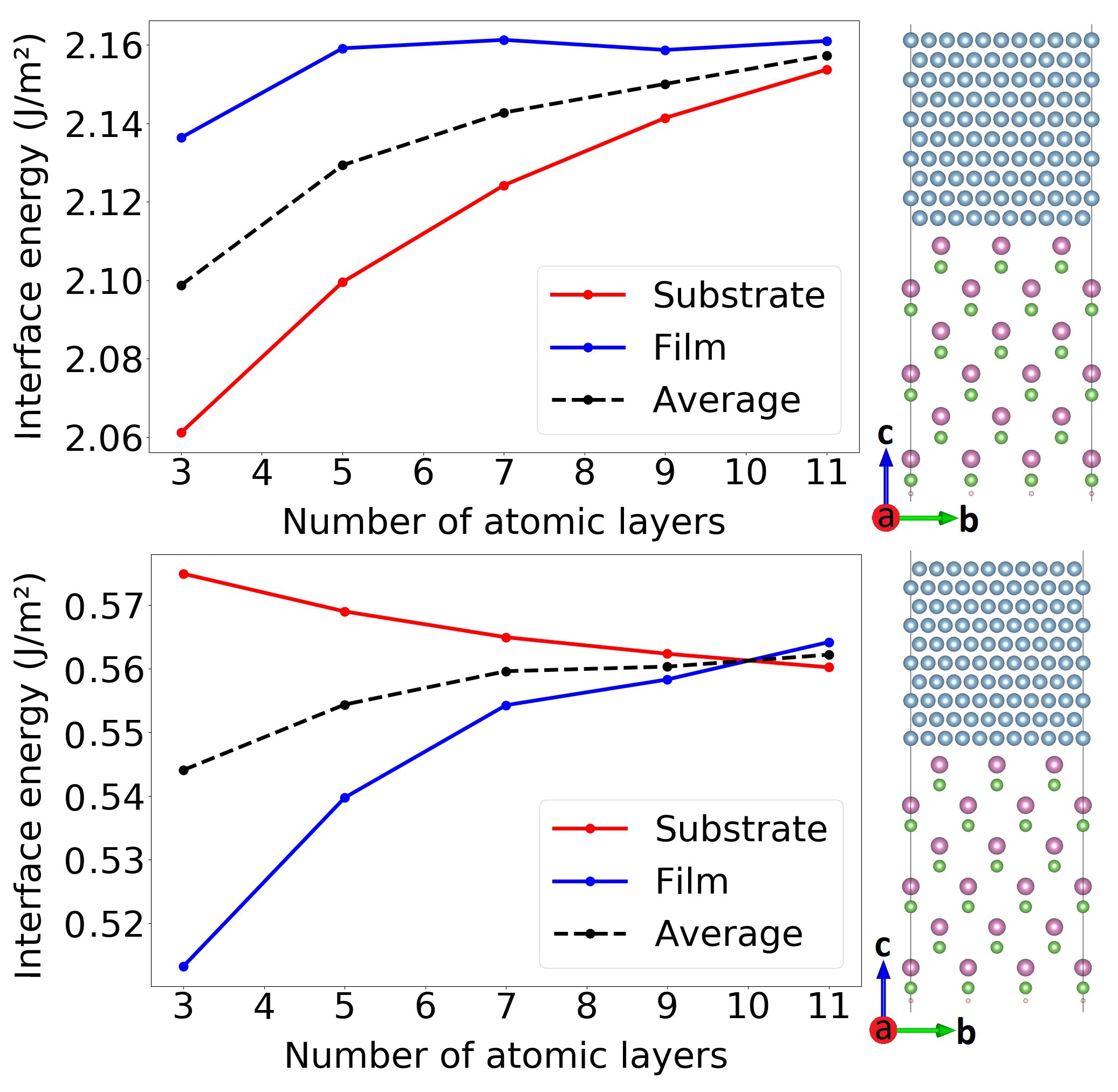}
\caption{Interface energy convergence plots for two  structures of the Al(111)/InAs(001) interface. The respective structures are shown on the right of each plot.}
\label{fig:Interface_energy_conv}
\end{figure}
 
 Figure \ref{fig:Interface_energy_conv} shows the interface energy as a function of the number of layers for two structures of the Al(111)/InAs(001) interface. The curves are obtained as described above, using 10 layers for the fixed substrate/film in each regression. As expected, both lines converge to a similar value. Additional interface energy convergence plots for Al(111)/InAs(111) are provided in Figure S5 in the SI. 
 The workflow of calculating converged interface energies is fully automated in Ogre, and Eq. (\ref{eq: FM-cond}) is used to estimate which growth mode will dominate based on the relative interface and surface energies. We note that Ogre does not take into account growth conditions and kinetics, which may lead to the formation of different structures.\cite{COSANDEY20011, BANSAL2011224} 
 In addition, Ogre does not take into account interdiffusion, which may lead to formation of substitutional impurities and interface phases.\cite{schultz2009spin, zega2006determination}

\section{\label{sec:level1}Computational Details}

DFT calculations were performed using the VASP code\cite{Joubert1999,Kresse1996,Kresse1996a,Kresse1993,Kresse1994} 
with the  projector-augmented wave (PAW) method.\cite{blochl1994projector, Kresse1999} The generalized gradient approximation (GGA) of Perdew, Burke, and Ernzerhof\cite{perdew1996generalized} (PBE) was employed for the description of the exchange-correlation interactions between electrons. For InAs and InSb, a Hubbard U correction within the Dudarev formalism \cite{dudarev1998electron} was applied to the $p$ orbitals of In, As, and Sb. The values of $U_{\text{eff}}$ were determined by Bayesian optimization\cite{Yu2020} to be: $U_{\text{eff,In}} = -0.5$ eV and  $U_{\text{eff,As}} = -7.5$ eV for InAs; $U_{\text{eff,In}} = -0.2$ eV and  $U_{\text{eff,Sb}} = -6.1$ eV for InSb. These U values have produced band structures in good agreement with angle-resolved photoemission spectroscopy (ARPES) for InAs and InSb surfaces.\cite{yang2021principles}
The Tkatchenko–Scheffler (TS)\cite{tkatchenko2009accurate} pairwise dispersion method was used to account for the van der Waals interactions at the interface. A plane-wave cutoff of 450 eV was adopted.
The convergence criterion used in the structural relaxation was  for the Hellman-Feynman
forces acting on ions to be below 0.001 eV/{\AA}.
A k-point mesh of $5\times5\times1$ was used for SCF calculations; and a k-point mesh of $7\times7\times1$ was used for DOS calculations. Key VASP INCAR file tags for convergence were ALGO=Fast, AMIN = 0.01, and BMIX = 3. In all calculations, dipole corrections were applied along the $z$-axis \cite{neugebauer1992adsorbate} and  spin-orbit coupling\cite{steiner2016calculation} was applied with the $z$ spin quantization axis. For the Fe/InSb interface, spin-polarized calculations were performed to study the local spin-polarization induced in the InSb. The lattice parameters of InAs, InSb, Al, Fe, CaTe, and SnTe were 6.0584 {\AA}, 6.4794 {\AA}, 4.03893 {\AA}, 2.866 {\AA}, 6.4010 {\AA}, and 6.4002 {\AA} respectively. For slab models, a vacuum region of 40 {\AA} was added. 
DFT calculations for Hirshfeld analysis were performed with the all-electron electronic structure code FHI-aims\cite{Blum2009} using the PBE exchange-correlation functional with the Tkatchenko–Scheffler (TS) pairwise dispersion correction \cite{Tkatchenko2009}, the light numerical settings, tier 1 basis sets, and a k-point mesh of $8\times8\times1$. The tag \texttt{vdw\_correction\_hirshfeld} was used for exporting Hirshfeld analysis results.   

\section{\label{sec:level1}Applications}

\subsection{\label{sec:level2}Al/InAs}
vbvvv
Based on Figure \ref{fig:Miller}, the Al(111)/InAs(111), Al(011)/InAs(111), and Al(012)/InAs(012) interfaces have the lowest area misfit values of 0.158\%, 0.172\%, and 0.158\%, respectively. The (111) orientation is more commonly used for InAs substrates in experiments than the (012) orientation. With this substrate orientation, Al(111) results in a minimal interface area of 47.68 {\AA}$^2$ whereas Al(011) results in an interface area of 197.30 {\AA}$^2$. Therefore, we proceed with the Al(111)/InAs(111) interface orientation. As-terminated InAs is chosen, based on the experiment reported in Ref. \cite{krogstrup2015epitaxy}. 

\begin{figure}[htbp]
\centering
\includegraphics[scale=0.145]{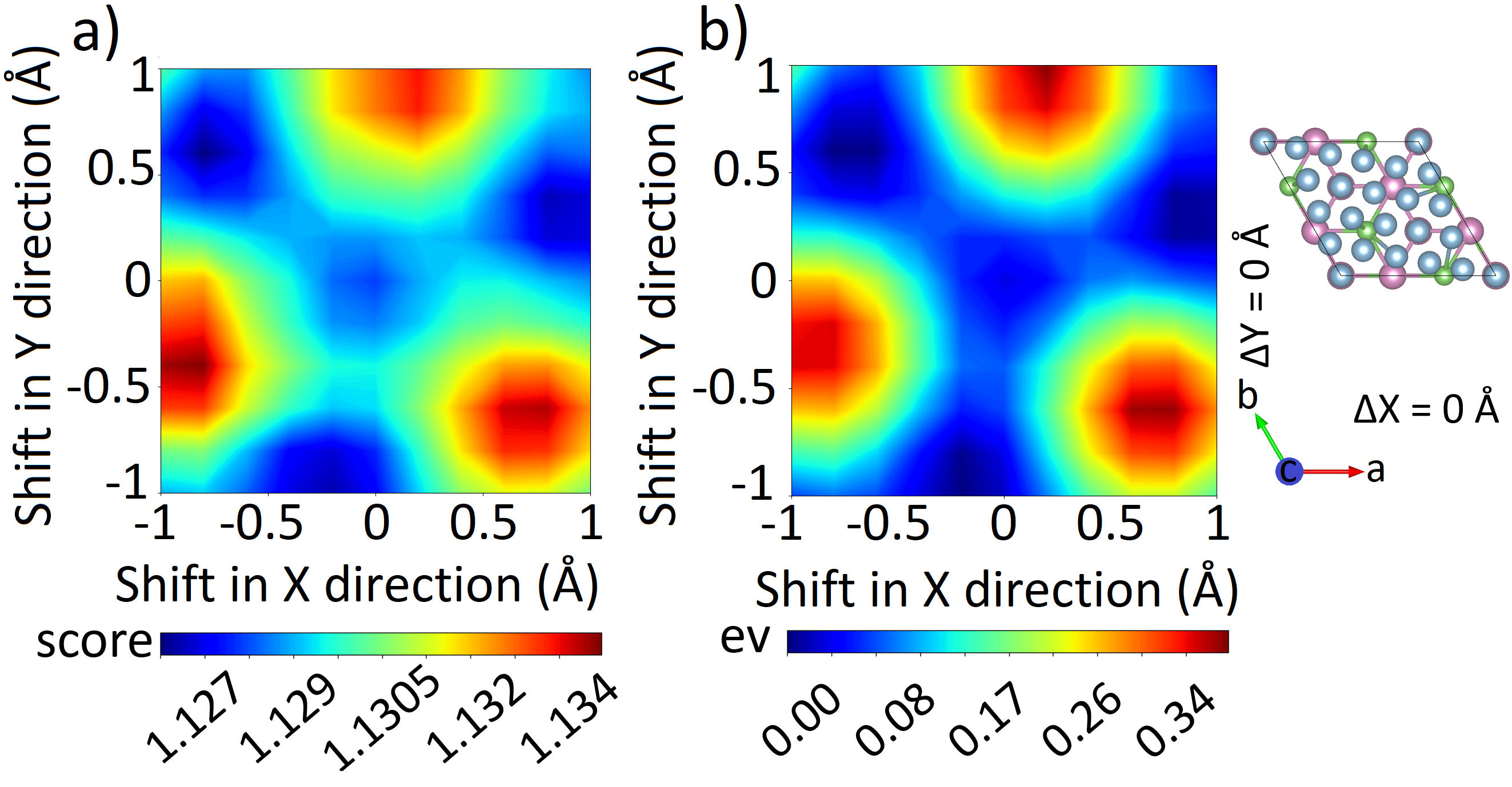}
\caption{Performance of the geometric score function for the registry in the $xy$ plane: (a) Score function contour plot compared to (b) the DFT potential energy surface at a fixed interfacial distance of 2.2 {{\AA}} for the Al(111)/InAs(111) interface.
} 
 \label{fig:PES_111_111}
\end{figure}

We obtain three Al(111)/InAs(111) interface structures with interface areas of 47.68 {\AA}$^2$, 111.25 {\AA}$^2$, and 63.57 {\AA}$^2$, respectively and an  area misfit value of 0.158\%. 
The structures also differ in terms of the film rotation with respect to the the substrate. Surface matching was performed for these three  structures to find their optimal interface distance and $xy$ registry. The score function was validated by comparing to DFT results for a representative structure. As shown in Figure \ref{fig:PES_111_111}, good agreement is obtained for the optimal registry in the $xy$ plane.

\begin{figure}[htbp]
\centering
\includegraphics[scale=0.36]{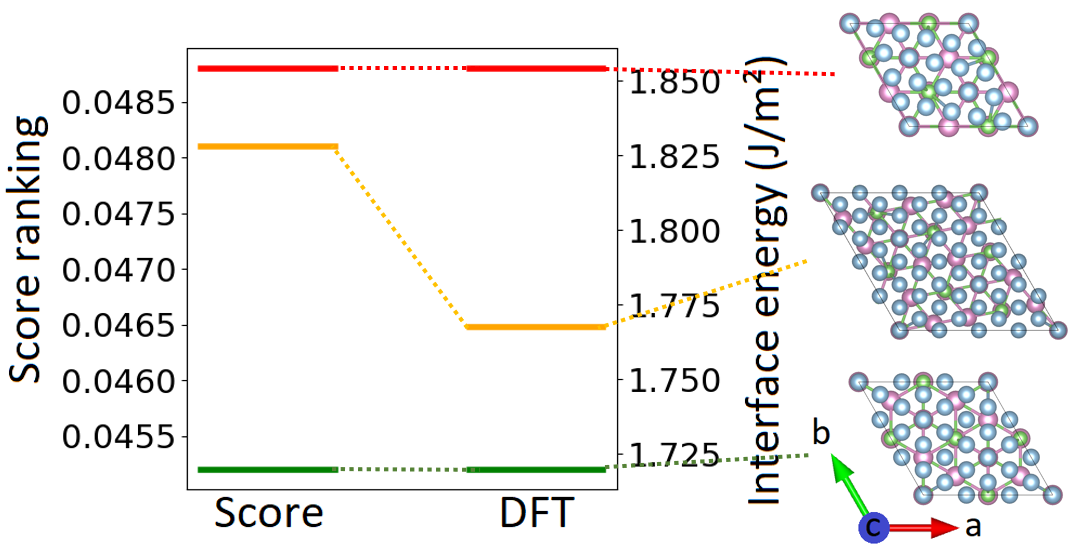}
\caption{Ranking score compared to interface energies obtained with DFT for Al(111)/InAs(111) interface structures.
} 
 \label{fig:levels_111_111}
\end{figure}

The optimized interface structures were ranked using Ogre's ranking score and subsequently their interface energies were calculated with DFT. The interface energies converged with 12 layers of InAs and 8 layers of Al. convergence plots are provided in Figure S5 in the SI. Figure \ref{fig:levels_111_111} shows that the Ogre ranking score successfully predicts the DFT ranking. The lowest energy interface structure is similar to the one observed by tunneling electron microscopy (TEM) in Ref. \cite{krogstrup2015epitaxy}.

\begin{figure}[htbp]
\centering
\includegraphics[scale=0.8]{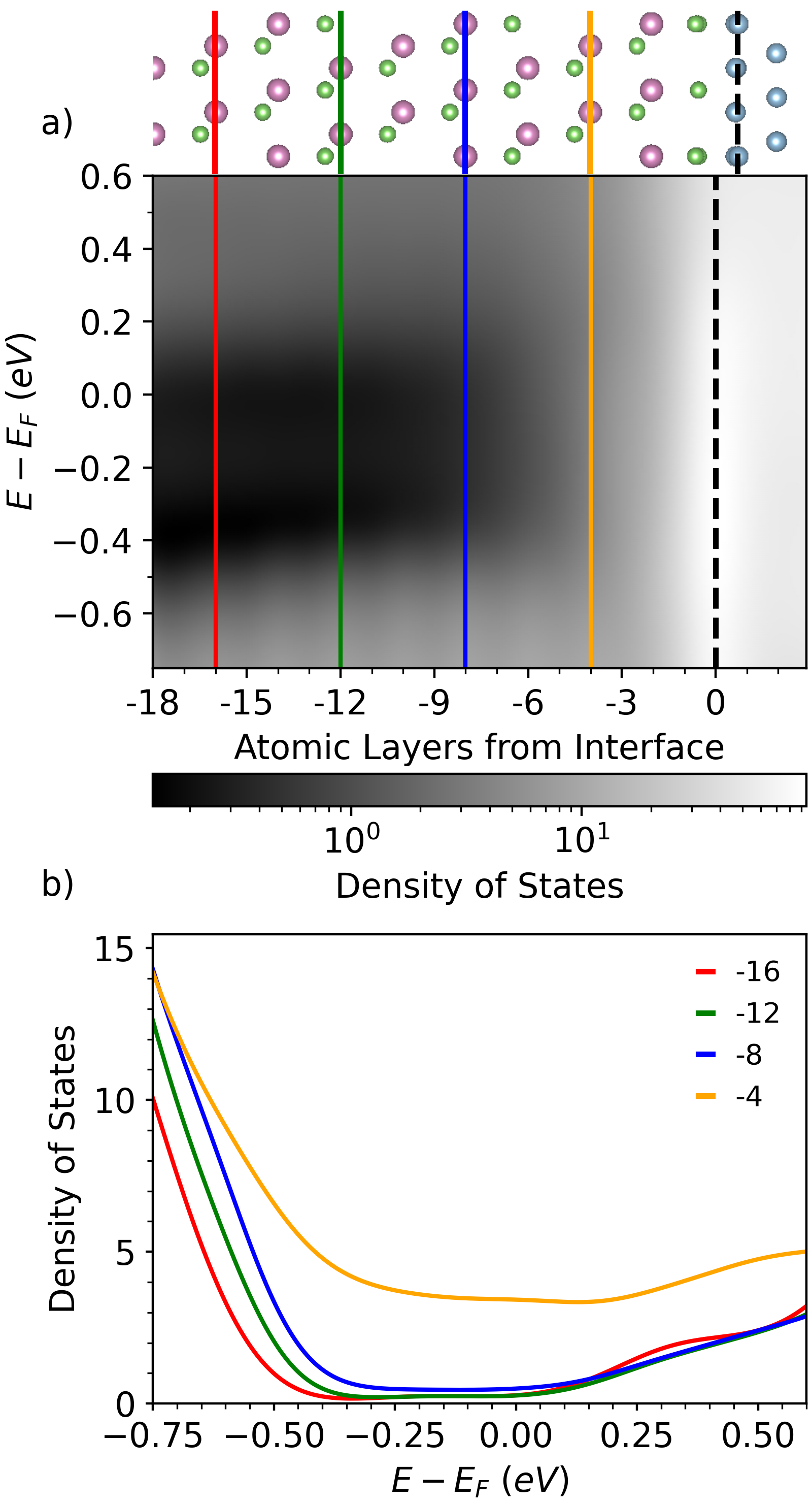}
\caption{Electronic structure of the most stable Al(111)/InAs(111) interface structure: (a) the density of states as a function of distance from the interface with the interface structure illustrated on top; (b) the local density of states of the InAs at 4, 8, 12, and 16 layers from the interface, indicated in panel (a) by vertical lines in the same colors. }
\label{fig:dos_AlInAs}
\end{figure}

Finally, electronic structure calculations were performed for the most stable interface configuration. To eliminate the effect of the surfaces, these calculations were performed for a periodic heterostructure with 7 atomic layers of Al and 31 atomic layers of InAs. Geometry relaxation was performed for the four atomic layers of InAs and the three atomic layers of Al closest to the interface. The atoms were constrained in the $x$ and $y$ directions and allowed to relax along the $z$-direction. Figure \ref{fig:dos_AlInAs} shows the variation of the InAs local DOS as a function of the distance from the interface. The Fermi level position is at the edge of the InAs conduction band. Close to the interface, (\textit{e.g.,} 4 atomic layers from the interface) a significant density of metal-induced gap states (MIGS)\cite{volker1965theory, monch1999winfried, nishimura2007evidence} is found in the gap of the InAs. The MIGS decay gradually with the distance from the interface. 16 layers away from the interface the bulk DOS of InAs is recovered. We note that due to the quantum size effect the band gap of InAs is 0.45 eV, which is 0.14 eV larger than the bulk PBE+U(BO) value of 0.31 eV.\cite{yang2021principles}

\subsection{\label{sec:level2}Fe/InSb}
\begin{figure}[htbp]
\centering
\includegraphics[scale=0.30]{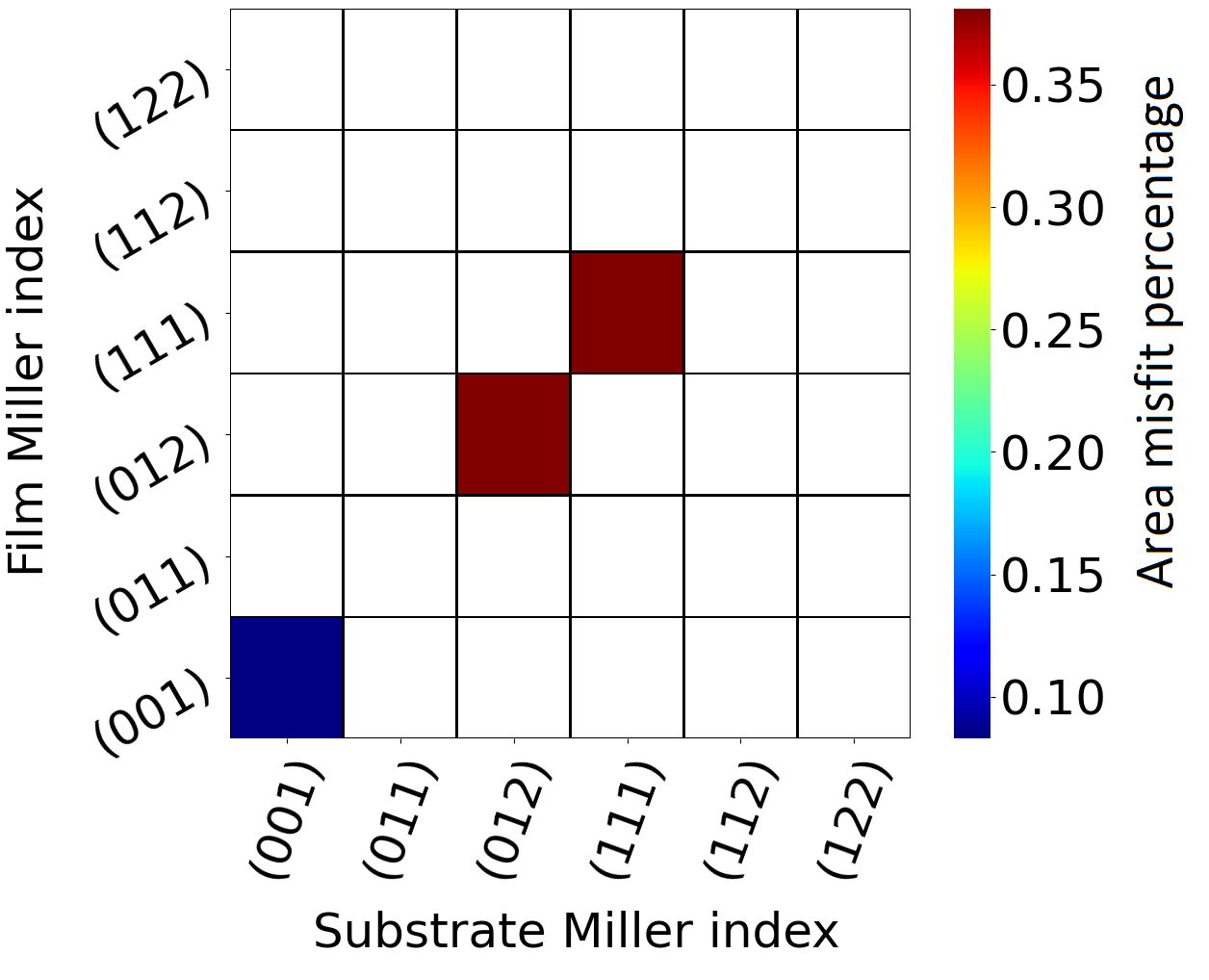}
\caption{Results of a Miller index scan for the Fe/InSb interface with a maximal Miller index of 2, maximum interface area of 500 {\AA}$^2$,  and area misfit tolerance of 1\%. 
}
\label{fig:MI_InSbFe}
\end{figure}

The Fe/InSb  ferromagnet/ semiconductor interface has been shown to achieve spin-filtering and spin-valve effects.\cite{yang2020spin} The Fe/InSb interface is similar to the Fe/GaAs interface, in which spin injection has also been observed.\cite{schultz2009spin} The Fe and GaAs lattice parameters are matched nearly 2:1, leading to coherent "cube on cube" growth. In contrast, the lattice parameters of Fe, 2.866 {\AA}, and InSb, 6.4794 {\AA}, result in a lattice mismatch of 56\%, making domain-matched epitaxy more likely. To find possible domain matched interfaces, a Miller index scan was performed with a maximal Miller index of 2, a maximum interface area of 500 {\AA}$^2$, and an area misfit tolerance of 1\%. Only Sb-terminated InSb was considered, based on the experiment reported in Ref. \cite{yang2020spin}. The results are shown in Figure \ref{fig:MI_InSbFe}. Three Miller index pairs are identified to have an area misfit below 1\%. The Fe(012)/InSb(012) interface  is not an obvious choice from a practical perspective because (012) is not a common orientation for the InSb substrate. The Fe(111)/InSb(111) interface generates only one possible interface structure with the given inputs, making it a trivial example and thus not interesting for demonstration purposes. The Fe(001)/InSb(001) interface produces four possible interface structures with the same cross-section area of 104.96 {\AA$^2$} and effective strain of 0.86\%. The four configurations differ only by relative rotations of the Fe film on top of the InSb substrate.

\begin{figure}[htbp]
\centering
\includegraphics[scale=0.14]{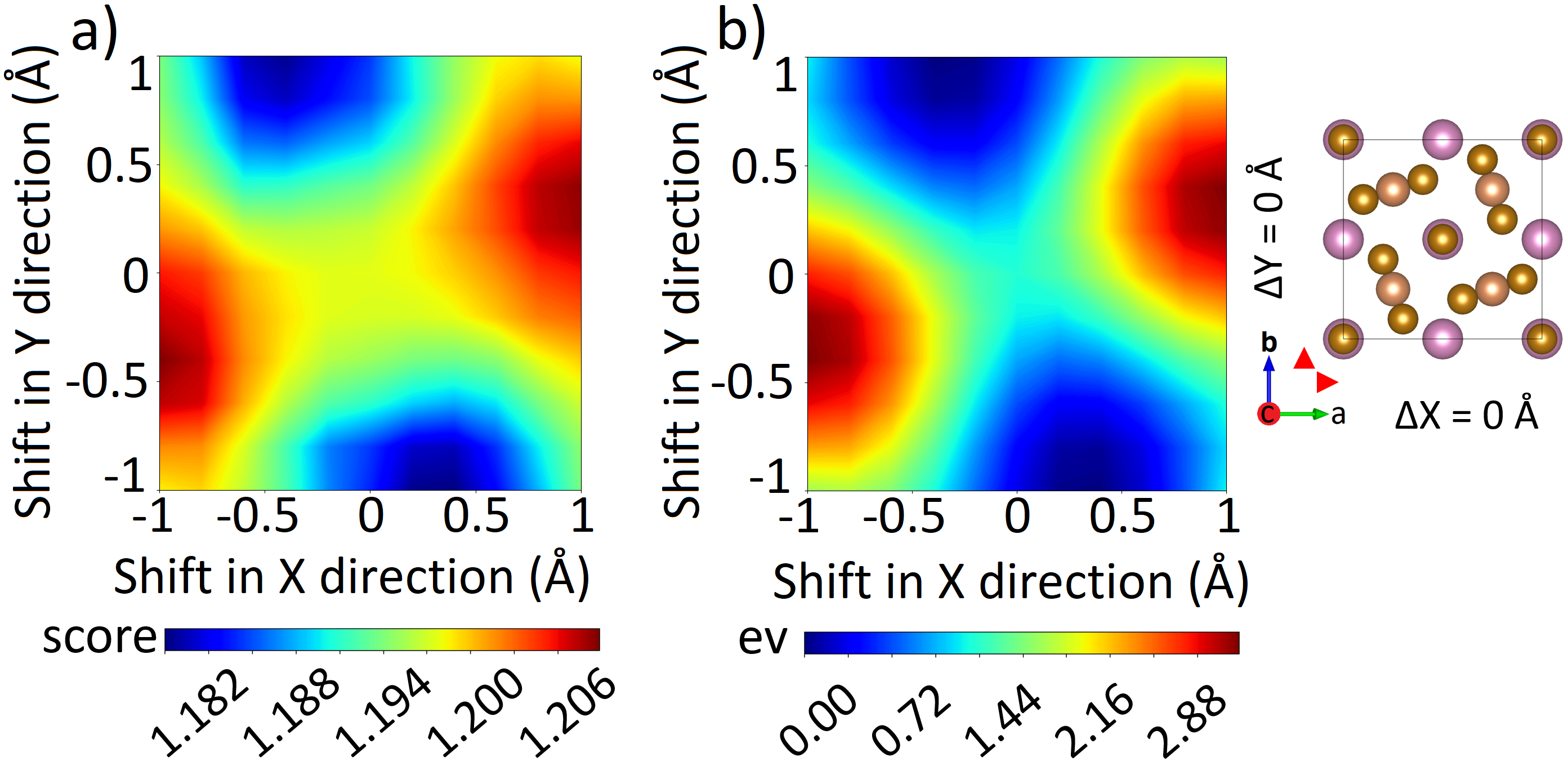}
\caption{Performance of the geometric score function for the registry in the $xy$ plane: (a) Score function contour plot compared to (b) the DFT potential energy surface at a fixed interfacial distance of 1.9 {{\AA}} for the Fe(001)/InSb(001) interface.
}
 \label{fig:PES_InSbFe_100_1000}
\end{figure}

\begin{figure}[htbp]
\centering
\includegraphics[scale=0.29]{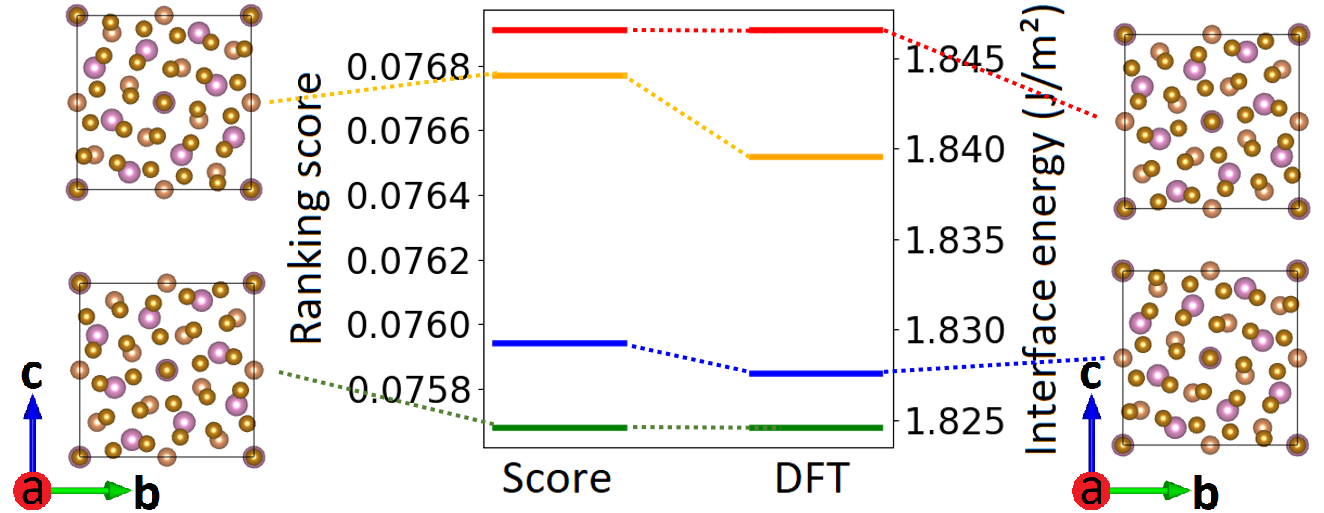}
\caption{Ranking score compared to interface energies obtained with DFT for Fe(001)/InAs(001) interface structures.
}
 \label{fig:levels_InSbFe}
\end{figure}

For the four Fe(001)/InSb(001) candidate structures, surface matching was performed using the Ogre score function to find their optimal interface distance and $xy$ registry. The geometric score function was validated by comparison to DFT for a Fe(001)/InSb(001) interface structure with a smaller area of 45.55 {\AA}$^2$ and a mismatch of 1.49\%. Figure \ref{fig:PES_InSbFe_100_1000} shows that the score function is in good agreement with the DFT potential energy surface with respect the positions of the extrema and the optimal registry in the $xy$ plane.  
Figure \ref{fig:levels_InSbFe} shows that the Ogre ranking score successfully predicts the DFT ranking. The four configurations are very close in energy and may co-exist.

\begin{figure}[htbp]
\centering
\includegraphics[scale=0.75]{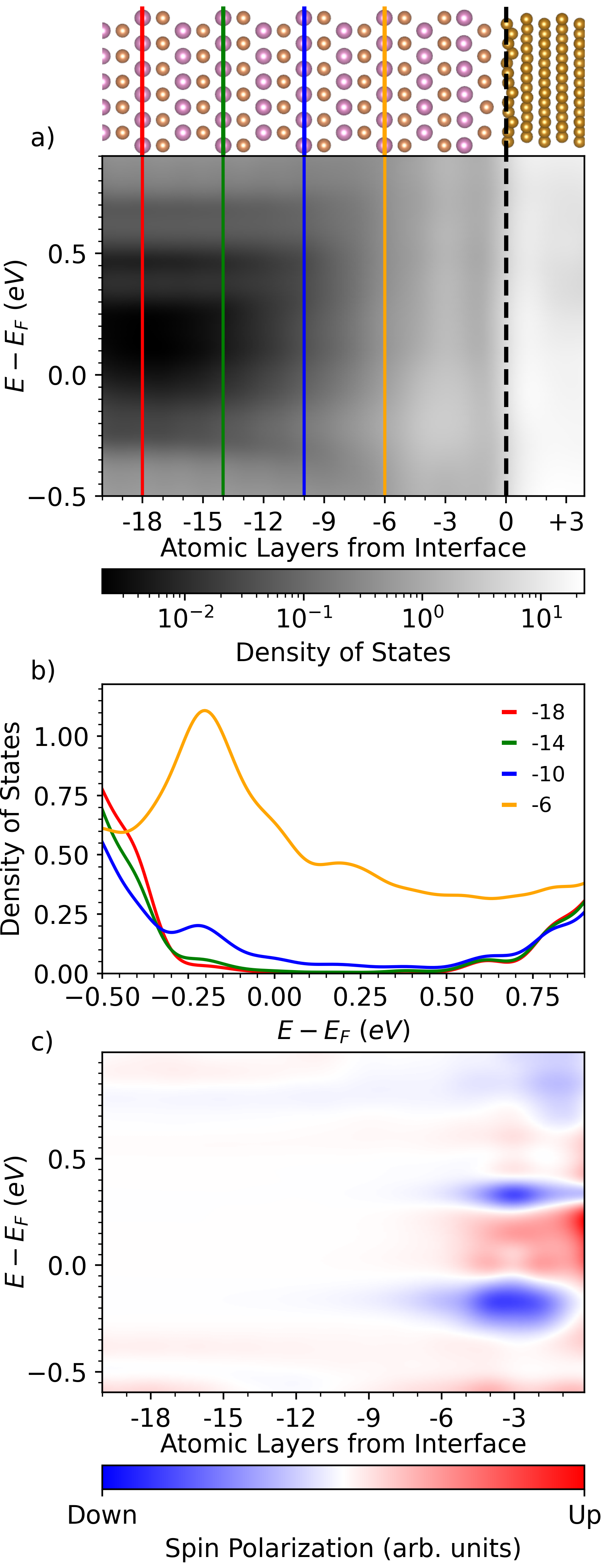}
\caption{Electronic structure of the most stable Fe(001)/InSb(001) interface structure: (a) the density of states as a function of distance from the interface with the interface structure illustrated on top; (b) the local density of states of the InSb at 6, 10, 14, and 18 layers from the interface, indicated in panel (a) by vertical lines in the same colors; and (c) the spin polarization (the difference between the majority DOS and the minority DOS) in the InSb as a function of the distance from the interface.
}
\label{fig:dos_InSbFe}
\end{figure}

Finally, electronic structure calculations were performed for the most stable interface configuration. To eliminate the effect of the surfaces, these calculations were performed for a periodic heterostructure with 11 atomic layers of Fe and 35 atomic layers of InSb.  Geometry relaxation was performed for the four atomic layers of InSb and the three atomic layers of Fe closest to the interface. The atoms were constrained in the $x$ and $y$ directions and allowed to relax along the $z$-direction. Figure \ref{fig:dos_InSbFe} shows the variation of the InSb local DOS as a function of the distance from the interface. The Fermi level position is at the edge of the InSb valence band. Close to the interface, (\textit{e.g.,} 6 atomic layers from the interface) a significant density of MIGS is observed in the band gap of the Insb. The MIGS decay gradually with the distance from the interface. About 18 layers away from the interface the bulk DOS of InSb is recovered. We note that due to the quantum size effect the band gap of InSb is 0.3 eV, which is 0.16 eV larger than the bulk PBE+U(BO) value of 0.14 eV.\cite{yang2021principles} 

A small magnetic moment of -0.052 $\mu$B is induced in the first InSb layer directly in contact with the Fe. The induced magnetic moment decays rapidly and completely vanishes beyond 5 atomic layers from the interface. Figure \ref{fig:dos_InSbFe}c shows the local spin polarization (the difference between the majority DOS and the minority DOS) as a function of the distance from the interface. Despite the small overall magnetization, the DOS of the InSb is spin polarized in the same region where significant presence of MIGS is observed. In this region, the DOS around the Fermi level is dominated by the majority spin channel, which may produce spin-polarized transport in agreement with Ref. \cite{yang2020spin}. At 0.1-0.3 eV below the Fermi level the DOS is dominated by the minority spin channel. This indicate that spin switching may be achieved by applying a bias.

\section{\label{sec:level1}Conclusion}
In summary, we have presented a new version of the Ogre Python package with the capability to predict the structure of epitaxial inorganic interfaces by lattice and surface matching. In the lattice matching step, all possible domain-matched interface structures are found within user-defined tolerances for the interface area and lattice mismatch. A Miller index scan is performed to determine the substrate and film orientations that would lead to the most favorable interface. In the surface matching step, Bayesian optimization (BO) is used to find the optimal configuration of each domain-matched interface, in terms of the interfacial distance in the $z$ direction and the registry in the $xy$ plane. For the BO objective function, we have formulated a geometric score function, based on the overlap and empty space between atomic spheres at the interface. We have demonstrated that the geometric score function reproduces the DFT results for the dependence of the energy on the interfacial distance, as well as the features of the DFT potential energy surface in the $xy$ plane at a fraction of the computational cost. For preliminary ranking of the optimized interfaces, we have formulated a ranking score based on the similarity between the overlap of atomic spheres at the interface to the overlap in the respective bulk crystal structures. We have demonstrated that the ranking score reproduces the DFT ranking and correctly predicts the order of stability of interface structures. After the lattice and surface matched structures are ranked based on the ranking score, DFT simulations can be performed for a small number of the most promising candidate structures. Ogre streamlines the evaluation of interface energies and calculation of electronic properties with DFT by automating the creation of interface models with a user-defined number of layers. For slab models, Ogre also automates the passivation of dangling bonds at the surface and adds a vacuum region. 

We have demonstrated the application of Ogre for two interfaces of interest in relation to quantum computing and spintronics: Al on As-terminated InAs and Fe on Sb-terminated InSb. Based on the results of a Miller index scan in the lattice matching step, the (111) orientation was selected for Al/InAs and the (001) orientation was selected for Fe/InSb. For Al/InAs (111) the top ranked structure produced by Ogre is in agreement with the experimentally observed structure. For Fe/InSb (001), whose structure has not been characterized experimentally, Ogre produces four structures that differ by the relative rotation of the Fe film on top of the InSb substrate. We have investigated the electronic structure of the most stable structures of both interfaces. In both cases, a significant density of metal-induced gap states is found in the semiconductor in the vicinity of the interface, which gradually decays within about 16 atomic layers. For Fe/InSb, although the induced magnetic moment is small and decays rapidly, the MIGS around the Fermi level are spin-polarized, which may produce spin-polarized transport.

Ogre may be used to interpret the results of experiments conducted on epitaxial inorganic interfaces by identifying the most likely interface configurations and correlating the structures with observed electronic properties and/or spectroscopic signatures. Moreover, Ogre may be used to predict the structure and properties of putative interfaces and guide synthesis efforts in promising directions. Ogre may be incorporated into an automated materials discovery workflow. Thus, Ogre can advance the understanding of the structure and properties of epitaxial inorganic interfaces, as well as the computational design and discovery of new interfaces for various applications, such as quantum computing and spintronic devices. 

\begin{acknowledgments}
 We thank Christian Ratsch from UCLA for helpful discussions of lattice matching. This research was funded by the Department of Energy through grant DE-SC0019274. This research used resources of the National Energy Research Scientific Computing Center (NERSC), a DOE Office of Science User Facility supported by the Office of Science of the U.S. Department of Energy under contract no. DE-AC02-05CH11231.
\end{acknowledgments}

\section*{Data availability}
The data that support the findings of this study are available from the corresponding author upon reasonable request. The Ogre code is available for download.

\renewcommand\refname{References}

\bibliography{paper}

\begin{thebibliography}{97}%
\makeatletter
\providecommand \@ifxundefined [1]{%
 \@ifx{#1\undefined}
}%
\providecommand \@ifnum [1]{%
 \ifnum #1\expandafter \@firstoftwo
 \else \expandafter \@secondoftwo
 \fi
}%
\providecommand \@ifx [1]{%
 \ifx #1\expandafter \@firstoftwo
 \else \expandafter \@secondoftwo
 \fi
}%
\providecommand \natexlab [1]{#1}%
\providecommand \enquote  [1]{``#1''}%
\providecommand \bibnamefont  [1]{#1}%
\providecommand \bibfnamefont [1]{#1}%
\providecommand \citenamefont [1]{#1}%
\providecommand \href@noop [0]{\@secondoftwo}%
\providecommand \href [0]{\begingroup \@sanitize@url \@href}%
\providecommand \@href[1]{\@@startlink{#1}\@@href}%
\providecommand \@@href[1]{\endgroup#1\@@endlink}%
\providecommand \@sanitize@url [0]{\catcode `\\12\catcode `\$12\catcode
  `\&12\catcode `\#12\catcode `\^12\catcode `\_12\catcode `\%12\relax}%
\providecommand \@@startlink[1]{}%
\providecommand \@@endlink[0]{}%
\providecommand \url  [0]{\begingroup\@sanitize@url \@url }%
\providecommand \@url [1]{\endgroup\@href {#1}{\urlprefix }}%
\providecommand \urlprefix  [0]{URL }%
\providecommand \Eprint [0]{\href }%
\providecommand \doibase [0]{http://dx.doi.org/}%
\providecommand \selectlanguage [0]{\@gobble}%
\providecommand \bibinfo  [0]{\@secondoftwo}%
\providecommand \bibfield  [0]{\@secondoftwo}%
\providecommand \translation [1]{[#1]}%
\providecommand \BibitemOpen [0]{}%
\providecommand \bibitemStop [0]{}%
\providecommand \bibitemNoStop [0]{.\EOS\space}%
\providecommand \EOS [0]{\spacefactor3000\relax}%
\providecommand \BibitemShut  [1]{\csname bibitem#1\endcsname}%
\let\auto@bib@innerbib\@empty
\bibitem [{\citenamefont {Antipov}\ \emph {et~al.}(2018)\citenamefont
  {Antipov}, \citenamefont {Bargerbos}, \citenamefont {Winkler}, \citenamefont
  {Bauer}, \citenamefont {Rossi},\ and\ \citenamefont
  {Lutchyn}}]{antipov2018effects}%
  \BibitemOpen
  \bibfield  {author} {\bibinfo {author} {\bibfnamefont {A.~E.}\ \bibnamefont
  {Antipov}}, \bibinfo {author} {\bibfnamefont {A.}~\bibnamefont {Bargerbos}},
  \bibinfo {author} {\bibfnamefont {G.~W.}\ \bibnamefont {Winkler}}, \bibinfo
  {author} {\bibfnamefont {B.}~\bibnamefont {Bauer}}, \bibinfo {author}
  {\bibfnamefont {E.}~\bibnamefont {Rossi}}, \ and\ \bibinfo {author}
  {\bibfnamefont {R.~M.}\ \bibnamefont {Lutchyn}},\ }\bibfield  {title}
  {\enquote {\bibinfo {title} {Effects of gate-induced electric fields on
  semiconductor majorana nanowires},}\ }\href@noop {} {\bibfield  {journal}
  {\bibinfo  {journal} {Physical Review X}\ }\textbf {\bibinfo {volume} {8}},\
  \bibinfo {pages} {031041} (\bibinfo {year} {2018})}\BibitemShut {NoStop}%
\bibitem [{\citenamefont {Chang}\ \emph {et~al.}(2015)\citenamefont {Chang},
  \citenamefont {Albrecht}, \citenamefont {Jespersen}, \citenamefont
  {Kuemmeth}, \citenamefont {Krogstrup}, \citenamefont {Nyg{\aa}rd},\ and\
  \citenamefont {Marcus}}]{chang2015hard}%
  \BibitemOpen
  \bibfield  {author} {\bibinfo {author} {\bibfnamefont {W.}~\bibnamefont
  {Chang}}, \bibinfo {author} {\bibfnamefont {S.}~\bibnamefont {Albrecht}},
  \bibinfo {author} {\bibfnamefont {T.}~\bibnamefont {Jespersen}}, \bibinfo
  {author} {\bibfnamefont {F.}~\bibnamefont {Kuemmeth}}, \bibinfo {author}
  {\bibfnamefont {P.}~\bibnamefont {Krogstrup}}, \bibinfo {author}
  {\bibfnamefont {J.}~\bibnamefont {Nyg{\aa}rd}}, \ and\ \bibinfo {author}
  {\bibfnamefont {C.~M.}\ \bibnamefont {Marcus}},\ }\bibfield  {title}
  {\enquote {\bibinfo {title} {Hard gap in epitaxial
  semiconductor--superconductor nanowires},}\ }\href@noop {} {\bibfield
  {journal} {\bibinfo  {journal} {Nature nanotechnology}\ }\textbf {\bibinfo
  {volume} {10}},\ \bibinfo {pages} {232--236} (\bibinfo {year}
  {2015})}\BibitemShut {NoStop}%
\bibitem [{\citenamefont {Das}\ \emph {et~al.}(2012)\citenamefont {Das},
  \citenamefont {Ronen}, \citenamefont {Most}, \citenamefont {Oreg},
  \citenamefont {Heiblum},\ and\ \citenamefont {Shtrikman}}]{das2012zero}%
  \BibitemOpen
  \bibfield  {author} {\bibinfo {author} {\bibfnamefont {A.}~\bibnamefont
  {Das}}, \bibinfo {author} {\bibfnamefont {Y.}~\bibnamefont {Ronen}}, \bibinfo
  {author} {\bibfnamefont {Y.}~\bibnamefont {Most}}, \bibinfo {author}
  {\bibfnamefont {Y.}~\bibnamefont {Oreg}}, \bibinfo {author} {\bibfnamefont
  {M.}~\bibnamefont {Heiblum}}, \ and\ \bibinfo {author} {\bibfnamefont
  {H.}~\bibnamefont {Shtrikman}},\ }\bibfield  {title} {\enquote {\bibinfo
  {title} {Zero-bias peaks and splitting in an al--inas nanowire topological
  superconductor as a signature of majorana fermions},}\ }\href@noop {}
  {\bibfield  {journal} {\bibinfo  {journal} {Nature Physics}\ }\textbf
  {\bibinfo {volume} {8}},\ \bibinfo {pages} {887--895} (\bibinfo {year}
  {2012})}\BibitemShut {NoStop}%
\bibitem [{\citenamefont {Mourik}\ \emph {et~al.}(2012)\citenamefont {Mourik},
  \citenamefont {Zuo}, \citenamefont {Frolov}, \citenamefont {Plissard},
  \citenamefont {Bakkers},\ and\ \citenamefont {Kouwenhoven}}]{Mourik2012}%
  \BibitemOpen
  \bibfield  {author} {\bibinfo {author} {\bibfnamefont {V.}~\bibnamefont
  {Mourik}}, \bibinfo {author} {\bibfnamefont {K.}~\bibnamefont {Zuo}},
  \bibinfo {author} {\bibfnamefont {S.}~\bibnamefont {Frolov}}, \bibinfo
  {author} {\bibfnamefont {S.}~\bibnamefont {Plissard}}, \bibinfo {author}
  {\bibfnamefont {E.}~\bibnamefont {Bakkers}}, \ and\ \bibinfo {author}
  {\bibfnamefont {L.~P.}\ \bibnamefont {Kouwenhoven}},\ }\bibfield  {title}
  {\enquote {\bibinfo {title} {Signatures of majorana fermions in hybrid
  superconductor-semiconductor nanowire devices},}\ }\href {\doibase
  10.1126/science.1222360} {\bibfield  {journal} {\bibinfo  {journal}
  {Science}\ }\textbf {\bibinfo {volume} {336}},\ \bibinfo {pages} {1003--1007}
  (\bibinfo {year} {2012})},\ \Eprint
  {http://arxiv.org/abs/https://science.sciencemag.org/content/336/6084/1003.full.pdf}
  {https://science.sciencemag.org/content/336/6084/1003.full.pdf} \BibitemShut
  {NoStop}%
\bibitem [{\citenamefont {G{\"u}l}\ \emph {et~al.}(2018)\citenamefont
  {G{\"u}l}, \citenamefont {Zhang}, \citenamefont {Bommer}, \citenamefont
  {de~Moor}, \citenamefont {Car}, \citenamefont {Plissard}, \citenamefont
  {Bakkers}, \citenamefont {Geresdi}, \citenamefont {Watanabe}, \citenamefont
  {Taniguchi},\ and\ \citenamefont {Kouwenhoven}}]{Zhang2016}%
  \BibitemOpen
  \bibfield  {author} {\bibinfo {author} {\bibfnamefont {{\"O}.}~\bibnamefont
  {G{\"u}l}}, \bibinfo {author} {\bibfnamefont {H.}~\bibnamefont {Zhang}},
  \bibinfo {author} {\bibfnamefont {J.~D.~S.}\ \bibnamefont {Bommer}}, \bibinfo
  {author} {\bibfnamefont {M.~W.~A.}\ \bibnamefont {de~Moor}}, \bibinfo
  {author} {\bibfnamefont {D.}~\bibnamefont {Car}}, \bibinfo {author}
  {\bibfnamefont {S.~R.}\ \bibnamefont {Plissard}}, \bibinfo {author}
  {\bibfnamefont {E.~P. A.~M.}\ \bibnamefont {Bakkers}}, \bibinfo {author}
  {\bibfnamefont {A.}~\bibnamefont {Geresdi}}, \bibinfo {author} {\bibfnamefont
  {K.}~\bibnamefont {Watanabe}}, \bibinfo {author} {\bibfnamefont
  {T.}~\bibnamefont {Taniguchi}}, \ and\ \bibinfo {author} {\bibfnamefont
  {L.~P.}\ \bibnamefont {Kouwenhoven}},\ }\bibfield  {title} {\enquote
  {\bibinfo {title} {Ballistic majorana nanowire devices},}\ }\href {\doibase
  10.1038/s41565-017-0032-8} {\bibfield  {journal} {\bibinfo  {journal} {Nature
  Nanotechnology}\ }\textbf {\bibinfo {volume} {13}},\ \bibinfo {pages} {192}
  (\bibinfo {year} {2018})}\BibitemShut {NoStop}%
\bibitem [{\citenamefont {Gul}\ \emph {et~al.}(2017)\citenamefont {Gul},
  \citenamefont {Zhang}, \citenamefont {de~Vries}, \citenamefont {van Veen},
  \citenamefont {Zuo}, \citenamefont {Mourik}, \citenamefont {Conesa-Boj},
  \citenamefont {Nowak}, \citenamefont {van Woerkom}, \citenamefont
  {Quintero-Pérez}, \citenamefont {Cassidy}, \citenamefont {Geresdi},
  \citenamefont {Koelling}, \citenamefont {Car}, \citenamefont {Plissard},
  \citenamefont {Bakkers},\ and\ \citenamefont {Kouwenhoven}}]{Guel2017}%
  \BibitemOpen
  \bibfield  {author} {\bibinfo {author} {\bibfnamefont {O.}~\bibnamefont
  {Gul}}, \bibinfo {author} {\bibfnamefont {H.}~\bibnamefont {Zhang}}, \bibinfo
  {author} {\bibfnamefont {F.~K.}\ \bibnamefont {de~Vries}}, \bibinfo {author}
  {\bibfnamefont {J.}~\bibnamefont {van Veen}}, \bibinfo {author}
  {\bibfnamefont {K.}~\bibnamefont {Zuo}}, \bibinfo {author} {\bibfnamefont
  {V.}~\bibnamefont {Mourik}}, \bibinfo {author} {\bibfnamefont
  {S.}~\bibnamefont {Conesa-Boj}}, \bibinfo {author} {\bibfnamefont {M.~P.}\
  \bibnamefont {Nowak}}, \bibinfo {author} {\bibfnamefont {D.~J.}\ \bibnamefont
  {van Woerkom}}, \bibinfo {author} {\bibfnamefont {M.}~\bibnamefont
  {Quintero-Pérez}}, \bibinfo {author} {\bibfnamefont {M.~C.}\ \bibnamefont
  {Cassidy}}, \bibinfo {author} {\bibfnamefont {A.}~\bibnamefont {Geresdi}},
  \bibinfo {author} {\bibfnamefont {S.}~\bibnamefont {Koelling}}, \bibinfo
  {author} {\bibfnamefont {D.}~\bibnamefont {Car}}, \bibinfo {author}
  {\bibfnamefont {S.~R.}\ \bibnamefont {Plissard}}, \bibinfo {author}
  {\bibfnamefont {E.~P. A.~M.}\ \bibnamefont {Bakkers}}, \ and\ \bibinfo
  {author} {\bibfnamefont {L.~P.}\ \bibnamefont {Kouwenhoven}},\ }\bibfield
  {title} {\enquote {\bibinfo {title} {Hard superconducting gap in insb
  nanowires},}\ }\href {\doibase 10.1021/acs.nanolett.7b00540} {\bibfield
  {journal} {\bibinfo  {journal} {Nano Letters}\ }\textbf {\bibinfo {volume}
  {17}},\ \bibinfo {pages} {2690--2696} (\bibinfo {year} {2017})},\ \bibinfo
  {note} {pMID: 28355877},\ \Eprint
  {http://arxiv.org/abs/https://doi.org/10.1021/acs.nanolett.7b00540}
  {https://doi.org/10.1021/acs.nanolett.7b00540} \BibitemShut {NoStop}%
\bibitem [{\citenamefont {Su}\ \emph {et~al.}(2020)\citenamefont {Su},
  \citenamefont {{\v{Z}}itko}, \citenamefont {Zhang}, \citenamefont {Wu},
  \citenamefont {Car}, \citenamefont {Plissard}, \citenamefont {Gazibegovic},
  \citenamefont {Badawy}, \citenamefont {Hocevar}, \citenamefont {Chen} \emph
  {et~al.}}]{su2020erasing}%
  \BibitemOpen
  \bibfield  {author} {\bibinfo {author} {\bibfnamefont {Z.}~\bibnamefont
  {Su}}, \bibinfo {author} {\bibfnamefont {R.}~\bibnamefont {{\v{Z}}itko}},
  \bibinfo {author} {\bibfnamefont {P.}~\bibnamefont {Zhang}}, \bibinfo
  {author} {\bibfnamefont {H.}~\bibnamefont {Wu}}, \bibinfo {author}
  {\bibfnamefont {D.}~\bibnamefont {Car}}, \bibinfo {author} {\bibfnamefont
  {S.}~\bibnamefont {Plissard}}, \bibinfo {author} {\bibfnamefont
  {S.}~\bibnamefont {Gazibegovic}}, \bibinfo {author} {\bibfnamefont
  {G.}~\bibnamefont {Badawy}}, \bibinfo {author} {\bibfnamefont
  {M.}~\bibnamefont {Hocevar}}, \bibinfo {author} {\bibfnamefont
  {J.}~\bibnamefont {Chen}},  \emph {et~al.},\ }\bibfield  {title} {\enquote
  {\bibinfo {title} {Erasing odd-parity states in semiconductor quantum dots
  coupled to superconductors},}\ }\href@noop {} {\bibfield  {journal} {\bibinfo
   {journal} {Physical Review B}\ }\textbf {\bibinfo {volume} {101}},\ \bibinfo
  {pages} {235315} (\bibinfo {year} {2020})}\BibitemShut {NoStop}%
\bibitem [{\citenamefont {Gazibegovic}\ \emph {et~al.}(2017)\citenamefont
  {Gazibegovic}, \citenamefont {Car}, \citenamefont {Zhang}, \citenamefont
  {Balk}, \citenamefont {Logan}, \citenamefont {de~Moor}, \citenamefont
  {Cassidy}, \citenamefont {Schmits}, \citenamefont {Xu}, \citenamefont {Wang}
  \emph {et~al.}}]{gazibegovic2017epitaxy}%
  \BibitemOpen
  \bibfield  {author} {\bibinfo {author} {\bibfnamefont {S.}~\bibnamefont
  {Gazibegovic}}, \bibinfo {author} {\bibfnamefont {D.}~\bibnamefont {Car}},
  \bibinfo {author} {\bibfnamefont {H.}~\bibnamefont {Zhang}}, \bibinfo
  {author} {\bibfnamefont {S.~C.}\ \bibnamefont {Balk}}, \bibinfo {author}
  {\bibfnamefont {J.~A.}\ \bibnamefont {Logan}}, \bibinfo {author}
  {\bibfnamefont {M.~W.}\ \bibnamefont {de~Moor}}, \bibinfo {author}
  {\bibfnamefont {M.~C.}\ \bibnamefont {Cassidy}}, \bibinfo {author}
  {\bibfnamefont {R.}~\bibnamefont {Schmits}}, \bibinfo {author} {\bibfnamefont
  {D.}~\bibnamefont {Xu}}, \bibinfo {author} {\bibfnamefont {G.}~\bibnamefont
  {Wang}},  \emph {et~al.},\ }\bibfield  {title} {\enquote {\bibinfo {title}
  {Epitaxy of advanced nanowire quantum devices},}\ }\href@noop {} {\bibfield
  {journal} {\bibinfo  {journal} {Nature}\ }\textbf {\bibinfo {volume} {548}},\
  \bibinfo {pages} {434--438} (\bibinfo {year} {2017})}\BibitemShut {NoStop}%
\bibitem [{\citenamefont {Anselmetti}\ \emph {et~al.}(2019)\citenamefont
  {Anselmetti}, \citenamefont {Martinez}, \citenamefont {M{\'e}nard},
  \citenamefont {Puglia}, \citenamefont {Malinowski}, \citenamefont {Lee},
  \citenamefont {Choi}, \citenamefont {Pendharkar}, \citenamefont
  {Palmstr{\o}m}, \citenamefont {Marcus} \emph {et~al.}}]{anselmetti2019end}%
  \BibitemOpen
  \bibfield  {author} {\bibinfo {author} {\bibfnamefont {G.}~\bibnamefont
  {Anselmetti}}, \bibinfo {author} {\bibfnamefont {E.}~\bibnamefont
  {Martinez}}, \bibinfo {author} {\bibfnamefont {G.}~\bibnamefont
  {M{\'e}nard}}, \bibinfo {author} {\bibfnamefont {D.}~\bibnamefont {Puglia}},
  \bibinfo {author} {\bibfnamefont {F.}~\bibnamefont {Malinowski}}, \bibinfo
  {author} {\bibfnamefont {J.}~\bibnamefont {Lee}}, \bibinfo {author}
  {\bibfnamefont {S.}~\bibnamefont {Choi}}, \bibinfo {author} {\bibfnamefont
  {M.}~\bibnamefont {Pendharkar}}, \bibinfo {author} {\bibfnamefont
  {C.}~\bibnamefont {Palmstr{\o}m}}, \bibinfo {author} {\bibfnamefont
  {C.}~\bibnamefont {Marcus}},  \emph {et~al.},\ }\bibfield  {title} {\enquote
  {\bibinfo {title} {End-to-end correlated subgap states in hybrid
  nanowires},}\ }\href@noop {} {\bibfield  {journal} {\bibinfo  {journal}
  {Physical Review B}\ }\textbf {\bibinfo {volume} {100}},\ \bibinfo {pages}
  {205412} (\bibinfo {year} {2019})}\BibitemShut {NoStop}%
\bibitem [{\citenamefont {de~Moor}\ \emph {et~al.}(2018)\citenamefont
  {de~Moor}, \citenamefont {Bommer}, \citenamefont {Xu}, \citenamefont
  {Winkler}, \citenamefont {Antipov}, \citenamefont {Bargerbos}, \citenamefont
  {Wang}, \citenamefont {Van~Loo}, \citenamefont {het Veld}, \citenamefont
  {Gazibegovic} \emph {et~al.}}]{de2018electric}%
  \BibitemOpen
  \bibfield  {author} {\bibinfo {author} {\bibfnamefont {M.~W.}\ \bibnamefont
  {de~Moor}}, \bibinfo {author} {\bibfnamefont {J.~D.}\ \bibnamefont {Bommer}},
  \bibinfo {author} {\bibfnamefont {D.}~\bibnamefont {Xu}}, \bibinfo {author}
  {\bibfnamefont {G.~W.}\ \bibnamefont {Winkler}}, \bibinfo {author}
  {\bibfnamefont {A.~E.}\ \bibnamefont {Antipov}}, \bibinfo {author}
  {\bibfnamefont {A.}~\bibnamefont {Bargerbos}}, \bibinfo {author}
  {\bibfnamefont {G.}~\bibnamefont {Wang}}, \bibinfo {author} {\bibfnamefont
  {N.}~\bibnamefont {Van~Loo}}, \bibinfo {author} {\bibfnamefont {R.~L.~O.}\
  \bibnamefont {het Veld}}, \bibinfo {author} {\bibfnamefont {S.}~\bibnamefont
  {Gazibegovic}},  \emph {et~al.},\ }\bibfield  {title} {\enquote {\bibinfo
  {title} {Electric field tunable superconductor-semiconductor coupling in
  majorana nanowires},}\ }\href@noop {} {\bibfield  {journal} {\bibinfo
  {journal} {New Journal of Physics}\ }\textbf {\bibinfo {volume} {20}},\
  \bibinfo {pages} {103049} (\bibinfo {year} {2018})}\BibitemShut {NoStop}%
\bibitem [{\citenamefont {Yang}\ \emph
  {et~al.}(2020{\natexlab{a}})\citenamefont {Yang}, \citenamefont {Heischmidt},
  \citenamefont {Gazibegovic}, \citenamefont {Badawy}, \citenamefont {Car},
  \citenamefont {Crowell}, \citenamefont {Bakkers},\ and\ \citenamefont
  {Pribiag}}]{yang2020spin}%
  \BibitemOpen
  \bibfield  {author} {\bibinfo {author} {\bibfnamefont {Z.}~\bibnamefont
  {Yang}}, \bibinfo {author} {\bibfnamefont {B.}~\bibnamefont {Heischmidt}},
  \bibinfo {author} {\bibfnamefont {S.}~\bibnamefont {Gazibegovic}}, \bibinfo
  {author} {\bibfnamefont {G.}~\bibnamefont {Badawy}}, \bibinfo {author}
  {\bibfnamefont {D.}~\bibnamefont {Car}}, \bibinfo {author} {\bibfnamefont
  {P.~A.}\ \bibnamefont {Crowell}}, \bibinfo {author} {\bibfnamefont {E.~P.}\
  \bibnamefont {Bakkers}}, \ and\ \bibinfo {author} {\bibfnamefont {V.~S.}\
  \bibnamefont {Pribiag}},\ }\bibfield  {title} {\enquote {\bibinfo {title}
  {Spin transport in ferromagnet-insb nanowire quantum devices},}\ }\href@noop
  {} {\bibfield  {journal} {\bibinfo  {journal} {Nano Letters}\ }\textbf
  {\bibinfo {volume} {20}},\ \bibinfo {pages} {3232--3239} (\bibinfo {year}
  {2020}{\natexlab{a}})}\BibitemShut {NoStop}%
\bibitem [{\citenamefont {Sands}\ \emph
  {et~al.}(1990{\natexlab{a}})\citenamefont {Sands}, \citenamefont
  {Palmstr{\o}m}, \citenamefont {Harbison}, \citenamefont {Keramidas},
  \citenamefont {Tabatabaie}, \citenamefont {Cheeks}, \citenamefont {Ramesh},\
  and\ \citenamefont {Silberberg}}]{sands1990stable}%
  \BibitemOpen
  \bibfield  {author} {\bibinfo {author} {\bibfnamefont {T.}~\bibnamefont
  {Sands}}, \bibinfo {author} {\bibfnamefont {C.}~\bibnamefont {Palmstr{\o}m}},
  \bibinfo {author} {\bibfnamefont {J.}~\bibnamefont {Harbison}}, \bibinfo
  {author} {\bibfnamefont {V.}~\bibnamefont {Keramidas}}, \bibinfo {author}
  {\bibfnamefont {N.}~\bibnamefont {Tabatabaie}}, \bibinfo {author}
  {\bibfnamefont {T.}~\bibnamefont {Cheeks}}, \bibinfo {author} {\bibfnamefont
  {R.}~\bibnamefont {Ramesh}}, \ and\ \bibinfo {author} {\bibfnamefont
  {Y.}~\bibnamefont {Silberberg}},\ }\bibfield  {title} {\enquote {\bibinfo
  {title} {Stable and epitaxial metal/iii-v semiconductor heterostructures},}\
  }\href@noop {} {\bibfield  {journal} {\bibinfo  {journal} {Materials Science
  Reports}\ }\textbf {\bibinfo {volume} {5}},\ \bibinfo {pages} {99--170}
  (\bibinfo {year} {1990}{\natexlab{a}})}\BibitemShut {NoStop}%
\bibitem [{\citenamefont {Zhu}\ \emph {et~al.}(2001)\citenamefont {Zhu},
  \citenamefont {Ramsteiner}, \citenamefont {Kostial}, \citenamefont
  {Wassermeier}, \citenamefont {Sch{\"o}nherr},\ and\ \citenamefont
  {Ploog}}]{zhu2001room}%
  \BibitemOpen
  \bibfield  {author} {\bibinfo {author} {\bibfnamefont {H.}~\bibnamefont
  {Zhu}}, \bibinfo {author} {\bibfnamefont {M.}~\bibnamefont {Ramsteiner}},
  \bibinfo {author} {\bibfnamefont {H.}~\bibnamefont {Kostial}}, \bibinfo
  {author} {\bibfnamefont {M.}~\bibnamefont {Wassermeier}}, \bibinfo {author}
  {\bibfnamefont {H.-P.}\ \bibnamefont {Sch{\"o}nherr}}, \ and\ \bibinfo
  {author} {\bibfnamefont {K.}~\bibnamefont {Ploog}},\ }\bibfield  {title}
  {\enquote {\bibinfo {title} {Room-temperature spin injection from fe into
  gaas},}\ }\href@noop {} {\bibfield  {journal} {\bibinfo  {journal} {Physical
  Review Letters}\ }\textbf {\bibinfo {volume} {87}},\ \bibinfo {pages}
  {016601} (\bibinfo {year} {2001})}\BibitemShut {NoStop}%
\bibitem [{\citenamefont {Lou}\ \emph {et~al.}(2007)\citenamefont {Lou},
  \citenamefont {Adelmann}, \citenamefont {Crooker}, \citenamefont {Garlid},
  \citenamefont {Zhang}, \citenamefont {Reddy}, \citenamefont {Flexner},
  \citenamefont {Palmstr{\o}m},\ and\ \citenamefont
  {Crowell}}]{lou2007electrical}%
  \BibitemOpen
  \bibfield  {author} {\bibinfo {author} {\bibfnamefont {X.}~\bibnamefont
  {Lou}}, \bibinfo {author} {\bibfnamefont {C.}~\bibnamefont {Adelmann}},
  \bibinfo {author} {\bibfnamefont {S.~A.}\ \bibnamefont {Crooker}}, \bibinfo
  {author} {\bibfnamefont {E.~S.}\ \bibnamefont {Garlid}}, \bibinfo {author}
  {\bibfnamefont {J.}~\bibnamefont {Zhang}}, \bibinfo {author} {\bibfnamefont
  {K.~M.}\ \bibnamefont {Reddy}}, \bibinfo {author} {\bibfnamefont {S.~D.}\
  \bibnamefont {Flexner}}, \bibinfo {author} {\bibfnamefont {C.~J.}\
  \bibnamefont {Palmstr{\o}m}}, \ and\ \bibinfo {author} {\bibfnamefont
  {P.~A.}\ \bibnamefont {Crowell}},\ }\bibfield  {title} {\enquote {\bibinfo
  {title} {Electrical detection of spin transport in lateral
  ferromagnet--semiconductor devices},}\ }\href@noop {} {\bibfield  {journal}
  {\bibinfo  {journal} {Nature Physics}\ }\textbf {\bibinfo {volume} {3}},\
  \bibinfo {pages} {197--202} (\bibinfo {year} {2007})}\BibitemShut {NoStop}%
\bibitem [{\citenamefont {Schultz}\ \emph {et~al.}(2009)\citenamefont
  {Schultz}, \citenamefont {Marom}, \citenamefont {Naveh}, \citenamefont {Lou},
  \citenamefont {Adelmann}, \citenamefont {Strand}, \citenamefont {Crowell},
  \citenamefont {Kronik},\ and\ \citenamefont
  {Palmstr{\o}m}}]{schultz2009spin}%
  \BibitemOpen
  \bibfield  {author} {\bibinfo {author} {\bibfnamefont {B.}~\bibnamefont
  {Schultz}}, \bibinfo {author} {\bibfnamefont {N.}~\bibnamefont {Marom}},
  \bibinfo {author} {\bibfnamefont {D.}~\bibnamefont {Naveh}}, \bibinfo
  {author} {\bibfnamefont {X.}~\bibnamefont {Lou}}, \bibinfo {author}
  {\bibfnamefont {C.}~\bibnamefont {Adelmann}}, \bibinfo {author}
  {\bibfnamefont {J.}~\bibnamefont {Strand}}, \bibinfo {author} {\bibfnamefont
  {P.}~\bibnamefont {Crowell}}, \bibinfo {author} {\bibfnamefont
  {L.}~\bibnamefont {Kronik}}, \ and\ \bibinfo {author} {\bibfnamefont
  {C.}~\bibnamefont {Palmstr{\o}m}},\ }\bibfield  {title} {\enquote {\bibinfo
  {title} {Spin injection across the fe/gaas interface: Role of interfacial
  ordering},}\ }\href@noop {} {\bibfield  {journal} {\bibinfo  {journal}
  {Physical Review B}\ }\textbf {\bibinfo {volume} {80}},\ \bibinfo {pages}
  {201309} (\bibinfo {year} {2009})}\BibitemShut {NoStop}%
\bibitem [{\citenamefont {Crooker}\ \emph {et~al.}(2005)\citenamefont
  {Crooker}, \citenamefont {Furis}, \citenamefont {Lou}, \citenamefont
  {Adelmann}, \citenamefont {Smith}, \citenamefont {Palmstr{\o}m},\ and\
  \citenamefont {Crowell}}]{crooker2005imaging}%
  \BibitemOpen
  \bibfield  {author} {\bibinfo {author} {\bibfnamefont {S.}~\bibnamefont
  {Crooker}}, \bibinfo {author} {\bibfnamefont {M.}~\bibnamefont {Furis}},
  \bibinfo {author} {\bibfnamefont {X.}~\bibnamefont {Lou}}, \bibinfo {author}
  {\bibfnamefont {C.}~\bibnamefont {Adelmann}}, \bibinfo {author}
  {\bibfnamefont {D.}~\bibnamefont {Smith}}, \bibinfo {author} {\bibfnamefont
  {C.}~\bibnamefont {Palmstr{\o}m}}, \ and\ \bibinfo {author} {\bibfnamefont
  {P.}~\bibnamefont {Crowell}},\ }\bibfield  {title} {\enquote {\bibinfo
  {title} {Imaging spin transport in lateral ferromagnet/semiconductor
  structures},}\ }\href@noop {} {\bibfield  {journal} {\bibinfo  {journal}
  {Science}\ }\textbf {\bibinfo {volume} {309}},\ \bibinfo {pages} {2191--2195}
  (\bibinfo {year} {2005})}\BibitemShut {NoStop}%
\bibitem [{\citenamefont {Rath}\ \emph {et~al.}(2018)\citenamefont {Rath},
  \citenamefont {Sivakumar}, \citenamefont {Sun}, \citenamefont {Patel},
  \citenamefont {Jeong}, \citenamefont {Feng}, \citenamefont {Stecklein},
  \citenamefont {Crowell}, \citenamefont {Palmstr\o{}m}, \citenamefont
  {Butler},\ and\ \citenamefont {Voyles}}]{rath2018reduced}%
  \BibitemOpen
  \bibfield  {author} {\bibinfo {author} {\bibfnamefont {A.}~\bibnamefont
  {Rath}}, \bibinfo {author} {\bibfnamefont {C.}~\bibnamefont {Sivakumar}},
  \bibinfo {author} {\bibfnamefont {C.}~\bibnamefont {Sun}}, \bibinfo {author}
  {\bibfnamefont {S.~J.}\ \bibnamefont {Patel}}, \bibinfo {author}
  {\bibfnamefont {J.~S.}\ \bibnamefont {Jeong}}, \bibinfo {author}
  {\bibfnamefont {J.}~\bibnamefont {Feng}}, \bibinfo {author} {\bibfnamefont
  {G.}~\bibnamefont {Stecklein}}, \bibinfo {author} {\bibfnamefont {P.~A.}\
  \bibnamefont {Crowell}}, \bibinfo {author} {\bibfnamefont {C.~J.}\
  \bibnamefont {Palmstr\o{}m}}, \bibinfo {author} {\bibfnamefont {W.~H.}\
  \bibnamefont {Butler}}, \ and\ \bibinfo {author} {\bibfnamefont {P.~M.}\
  \bibnamefont {Voyles}},\ }\bibfield  {title} {\enquote {\bibinfo {title}
  {Reduced interface spin polarization by antiferromagnetically coupled mn
  segregated to the $\mathrm{C}{\mathrm{o}}_{2}\mathrm{MnSi}$/gaas (001)
  interface},}\ }\href {\doibase 10.1103/PhysRevB.97.045304} {\bibfield
  {journal} {\bibinfo  {journal} {Phys. Rev. B}\ }\textbf {\bibinfo {volume}
  {97}},\ \bibinfo {pages} {045304} (\bibinfo {year} {2018})}\BibitemShut
  {NoStop}%
\bibitem [{\citenamefont {Tung}(2014)}]{tung2014the}%
  \BibitemOpen
  \bibfield  {author} {\bibinfo {author} {\bibfnamefont {R.~T.}\ \bibnamefont
  {Tung}},\ }\bibfield  {title} {\enquote {\bibinfo {title} {The physics and
  chemistry of the schottky barrier height},}\ }\href {\doibase
  10.1063/1.4858400} {\bibfield  {journal} {\bibinfo  {journal} {Applied
  Physics Reviews}\ }\textbf {\bibinfo {volume} {1}},\ \bibinfo {pages}
  {011304} (\bibinfo {year} {2014})},\ \Eprint
  {http://arxiv.org/abs/https://doi.org/10.1063/1.4858400}
  {https://doi.org/10.1063/1.4858400} \BibitemShut {NoStop}%
\bibitem [{\citenamefont {Yang}\ \emph {et~al.}(2021)\citenamefont {Yang},
  \citenamefont {Dardzinski}, \citenamefont {Hwang}, \citenamefont {Pikulin},
  \citenamefont {Winkler},\ and\ \citenamefont {Marom}}]{yang2021principles}%
  \BibitemOpen
  \bibfield  {author} {\bibinfo {author} {\bibfnamefont {S.}~\bibnamefont
  {Yang}}, \bibinfo {author} {\bibfnamefont {D.}~\bibnamefont {Dardzinski}},
  \bibinfo {author} {\bibfnamefont {A.}~\bibnamefont {Hwang}}, \bibinfo
  {author} {\bibfnamefont {D.~I.}\ \bibnamefont {Pikulin}}, \bibinfo {author}
  {\bibfnamefont {G.~W.}\ \bibnamefont {Winkler}}, \ and\ \bibinfo {author}
  {\bibfnamefont {N.}~\bibnamefont {Marom}},\ }\href@noop {} {\enquote
  {\bibinfo {title} {First principles feasibility assessment of a topological
  insulator at the inas/gasb interface},}\ } (\bibinfo {year} {2021}),\ \Eprint
  {http://arxiv.org/abs/2101.07873} {arXiv:2101.07873 [physics.comp-ph]}
  \BibitemShut {NoStop}%
\bibitem [{\citenamefont {Cho}(1983)}]{Cho1983}%
  \BibitemOpen
  \bibfield  {author} {\bibinfo {author} {\bibfnamefont {A.~Y.}\ \bibnamefont
  {Cho}},\ }\bibfield  {title} {\enquote {\bibinfo {title} {{Growth of III–V
  semiconductors by molecular beam epitaxy and their properties}},}\ }\href
  {\doibase https://doi.org/10.1016/0040-6090(83)90154-2} {\bibfield  {journal}
  {\bibinfo  {journal} {Thin Solid Films}\ }\textbf {\bibinfo {volume} {100}},\
  \bibinfo {pages} {291--317} (\bibinfo {year} {1983})}\BibitemShut {NoStop}%
\bibitem [{\citenamefont {Sands}\ \emph
  {et~al.}(1990{\natexlab{b}})\citenamefont {Sands}, \citenamefont
  {Palmstr{\o}m}, \citenamefont {Harbison}, \citenamefont {Keramidas},
  \citenamefont {Tabatabaie}, \citenamefont {Cheeks}, \citenamefont {Ramesh},\
  and\ \citenamefont {Silberberg}}]{Sands1990}%
  \BibitemOpen
  \bibfield  {author} {\bibinfo {author} {\bibfnamefont {T.}~\bibnamefont
  {Sands}}, \bibinfo {author} {\bibfnamefont {C.~J.}\ \bibnamefont
  {Palmstr{\o}m}}, \bibinfo {author} {\bibfnamefont {J.~P.}\ \bibnamefont
  {Harbison}}, \bibinfo {author} {\bibfnamefont {V.~G.}\ \bibnamefont
  {Keramidas}}, \bibinfo {author} {\bibfnamefont {N.}~\bibnamefont
  {Tabatabaie}}, \bibinfo {author} {\bibfnamefont {T.~L.}\ \bibnamefont
  {Cheeks}}, \bibinfo {author} {\bibfnamefont {R.}~\bibnamefont {Ramesh}}, \
  and\ \bibinfo {author} {\bibfnamefont {Y.}~\bibnamefont {Silberberg}},\
  }\bibfield  {title} {\enquote {\bibinfo {title} {{Stable and epitaxial
  metal/III-V semiconductor heterostructures}},}\ }\href {\doibase
  https://doi.org/10.1016/S0920-2307(05)80003-9} {\bibfield  {journal}
  {\bibinfo  {journal} {Materials Science Reports}\ }\textbf {\bibinfo {volume}
  {5}},\ \bibinfo {pages} {99--170} (\bibinfo {year}
  {1990}{\natexlab{b}})}\BibitemShut {NoStop}%
\bibitem [{\citenamefont {Narayan}\ and\ \citenamefont
  {Larson}(2002)}]{Narayan2002}%
  \BibitemOpen
  \bibfield  {author} {\bibinfo {author} {\bibfnamefont {J.}~\bibnamefont
  {Narayan}}\ and\ \bibinfo {author} {\bibfnamefont {B.~C.}\ \bibnamefont
  {Larson}},\ }\bibfield  {title} {\enquote {\bibinfo {title} {{Domain epitaxy:
  A unified paradigm for thin film growth}},}\ }\href {\doibase
  10.1063/1.1528301} {\bibfield  {journal} {\bibinfo  {journal} {Journal of
  Applied Physics}\ }\textbf {\bibinfo {volume} {93}},\ \bibinfo {pages}
  {278--285} (\bibinfo {year} {2002})}\BibitemShut {NoStop}%
\bibitem [{\citenamefont {Zheleva}, \citenamefont {Jagannadham},\ and\
  \citenamefont {Narayan}(1994)}]{Zheleva1994}%
  \BibitemOpen
  \bibfield  {author} {\bibinfo {author} {\bibfnamefont {T.}~\bibnamefont
  {Zheleva}}, \bibinfo {author} {\bibfnamefont {K.}~\bibnamefont
  {Jagannadham}}, \ and\ \bibinfo {author} {\bibfnamefont {J.}~\bibnamefont
  {Narayan}},\ }\bibfield  {title} {\enquote {\bibinfo {title} {{Epitaxial
  growth in large‐lattice‐mismatch systems}},}\ }\href {\doibase
  10.1063/1.356440} {\bibfield  {journal} {\bibinfo  {journal} {Journal of
  Applied Physics}\ }\textbf {\bibinfo {volume} {75}},\ \bibinfo {pages}
  {860--871} (\bibinfo {year} {1994})}\BibitemShut {NoStop}%
\bibitem [{\citenamefont {Bauer}\ and\ \citenamefont {van~der
  Merwe}(1986)}]{Bauer1986}%
  \BibitemOpen
  \bibfield  {author} {\bibinfo {author} {\bibfnamefont {E.}~\bibnamefont
  {Bauer}}\ and\ \bibinfo {author} {\bibfnamefont {J.~H.}\ \bibnamefont
  {van~der Merwe}},\ }\bibfield  {title} {\enquote {\bibinfo {title}
  {{Structure and growth of crystalline superlattices: From monolayer to
  superlattice}},}\ }\href {\doibase 10.1103/PhysRevB.33.3657} {\bibfield
  {journal} {\bibinfo  {journal} {Physical Review B}\ }\textbf {\bibinfo
  {volume} {33}},\ \bibinfo {pages} {3657--3671} (\bibinfo {year}
  {1986})}\BibitemShut {NoStop}%
\bibitem [{\citenamefont {Chow}\ and\ \citenamefont
  {Johnson}(1985)}]{Chow1985}%
  \BibitemOpen
  \bibfield  {author} {\bibinfo {author} {\bibfnamefont {P.~P.}\ \bibnamefont
  {Chow}}\ and\ \bibinfo {author} {\bibfnamefont {D.}~\bibnamefont {Johnson}},\
  }\bibfield  {title} {\enquote {\bibinfo {title} {{Growth and characterization
  of MBE grown HgTe–CdTe superlattices}},}\ }\href {\doibase
  10.1116/1.573247} {\bibfield  {journal} {\bibinfo  {journal} {Journal of
  Vacuum Science {\&} Technology A}\ }\textbf {\bibinfo {volume} {3}},\
  \bibinfo {pages} {67--70} (\bibinfo {year} {1985})}\BibitemShut {NoStop}%
\bibitem [{\citenamefont {Faurie}, \citenamefont {Million},\ and\ \citenamefont
  {Piaguet}(1982)}]{Faurie1982}%
  \BibitemOpen
  \bibfield  {author} {\bibinfo {author} {\bibfnamefont {J.~P.}\ \bibnamefont
  {Faurie}}, \bibinfo {author} {\bibfnamefont {A.}~\bibnamefont {Million}}, \
  and\ \bibinfo {author} {\bibfnamefont {J.}~\bibnamefont {Piaguet}},\
  }\bibfield  {title} {\enquote {\bibinfo {title} {{CdTe‐HgTe multilayers
  grown by molecular beam epitaxy}},}\ }\href {\doibase 10.1063/1.93644}
  {\bibfield  {journal} {\bibinfo  {journal} {Applied Physics Letters}\
  }\textbf {\bibinfo {volume} {41}},\ \bibinfo {pages} {713--715} (\bibinfo
  {year} {1982})}\BibitemShut {NoStop}%
\bibitem [{\citenamefont {West}(1999)}]{west1999basic}%
  \BibitemOpen
  \bibfield  {author} {\bibinfo {author} {\bibfnamefont {A.~R.}\ \bibnamefont
  {West}},\ }\href@noop {} {\emph {\bibinfo {title} {Basic solid state
  chemistry}}}\ (\bibinfo  {publisher} {John Wiley \& Sons Incorporated},\
  \bibinfo {year} {1999})\BibitemShut {NoStop}%
\bibitem [{\citenamefont {Xie}\ \emph {et~al.}(2016)\citenamefont {Xie},
  \citenamefont {Lucking}, \citenamefont {Chen}, \citenamefont {Bhat},
  \citenamefont {Wang}, \citenamefont {Lu},\ and\ \citenamefont
  {Zhang}}]{Xie2016}%
  \BibitemOpen
  \bibfield  {author} {\bibinfo {author} {\bibfnamefont {W.}~\bibnamefont
  {Xie}}, \bibinfo {author} {\bibfnamefont {M.}~\bibnamefont {Lucking}},
  \bibinfo {author} {\bibfnamefont {L.}~\bibnamefont {Chen}}, \bibinfo {author}
  {\bibfnamefont {I.}~\bibnamefont {Bhat}}, \bibinfo {author} {\bibfnamefont
  {G.-C.}\ \bibnamefont {Wang}}, \bibinfo {author} {\bibfnamefont {T.-M.}\
  \bibnamefont {Lu}}, \ and\ \bibinfo {author} {\bibfnamefont {S.}~\bibnamefont
  {Zhang}},\ }\bibfield  {title} {\enquote {\bibinfo {title} {{Modular Approach
  for Metal–Semiconductor Heterostructures with Very Large Interface Lattice
  Misfit: A First-Principles Perspective}},}\ }\href {\doibase
  10.1021/acs.cgd.6b00118} {\bibfield  {journal} {\bibinfo  {journal} {Crystal
  Growth {\&} Design}\ }\textbf {\bibinfo {volume} {16}},\ \bibinfo {pages}
  {2328--2334} (\bibinfo {year} {2016})}\BibitemShut {NoStop}%
\bibitem [{\citenamefont {Trampert}\ and\ \citenamefont
  {Ploog}(2000)}]{Trampert2000}%
  \BibitemOpen
  \bibfield  {author} {\bibinfo {author} {\bibfnamefont {A.}~\bibnamefont
  {Trampert}}\ and\ \bibinfo {author} {\bibfnamefont {K.~H.}\ \bibnamefont
  {Ploog}},\ }\bibfield  {title} {\enquote {\bibinfo {title} {{Heteroepitaxy of
  Large-Misfit Systems: Role of Coincidence Lattice}},}\ }\href {\doibase
  https://doi.org/10.1002/1521-4079(200007)35:6/7<793::AID-CRAT793>3.0.CO;2-3}
  {\bibfield  {journal} {\bibinfo  {journal} {Crystal Research and Technology}\
  }\textbf {\bibinfo {volume} {35}},\ \bibinfo {pages} {793--806} (\bibinfo
  {year} {2000})}\BibitemShut {NoStop}%
\bibitem [{\citenamefont {Erwin}\ \emph {et~al.}(2011)\citenamefont {Erwin},
  \citenamefont {Gao}, \citenamefont {Roder}, \citenamefont {L{\"{a}}hnemann},\
  and\ \citenamefont {Brandt}}]{Erwin2011}%
  \BibitemOpen
  \bibfield  {author} {\bibinfo {author} {\bibfnamefont {S.~C.}\ \bibnamefont
  {Erwin}}, \bibinfo {author} {\bibfnamefont {C.}~\bibnamefont {Gao}}, \bibinfo
  {author} {\bibfnamefont {C.}~\bibnamefont {Roder}}, \bibinfo {author}
  {\bibfnamefont {J.}~\bibnamefont {L{\"{a}}hnemann}}, \ and\ \bibinfo {author}
  {\bibfnamefont {O.}~\bibnamefont {Brandt}},\ }\bibfield  {title} {\enquote
  {\bibinfo {title} {{Epitaxial Interfaces between Crystallographically
  Mismatched Materials}},}\ }\href {\doibase 10.1103/PhysRevLett.107.026102}
  {\bibfield  {journal} {\bibinfo  {journal} {Physical Review Letters}\
  }\textbf {\bibinfo {volume} {107}},\ \bibinfo {pages} {26102} (\bibinfo
  {year} {2011})}\BibitemShut {NoStop}%
\bibitem [{\citenamefont {Krogstrup}\ \emph
  {et~al.}(2015{\natexlab{a}})\citenamefont {Krogstrup}, \citenamefont {Ziino},
  \citenamefont {Chang}, \citenamefont {Albrecht}, \citenamefont {Madsen},
  \citenamefont {Johnson}, \citenamefont {Nyg{\aa}rd}, \citenamefont {Marcus},\
  and\ \citenamefont {Jespersen}}]{Krogstrup2015}%
  \BibitemOpen
  \bibfield  {author} {\bibinfo {author} {\bibfnamefont {P.}~\bibnamefont
  {Krogstrup}}, \bibinfo {author} {\bibfnamefont {N.~L.~B.}\ \bibnamefont
  {Ziino}}, \bibinfo {author} {\bibfnamefont {W.}~\bibnamefont {Chang}},
  \bibinfo {author} {\bibfnamefont {S.~M.}\ \bibnamefont {Albrecht}}, \bibinfo
  {author} {\bibfnamefont {M.~H.}\ \bibnamefont {Madsen}}, \bibinfo {author}
  {\bibfnamefont {E.}~\bibnamefont {Johnson}}, \bibinfo {author} {\bibfnamefont
  {J.}~\bibnamefont {Nyg{\aa}rd}}, \bibinfo {author} {\bibfnamefont
  {C.}~\bibnamefont {Marcus}}, \ and\ \bibinfo {author} {\bibfnamefont {T.~S.}\
  \bibnamefont {Jespersen}},\ }\bibfield  {title} {\enquote {\bibinfo {title}
  {{Epitaxy of semiconductor–superconductor nanowires}},}\ }\href {\doibase
  10.1038/nmat4176} {\bibfield  {journal} {\bibinfo  {journal} {Nature
  Materials}\ }\textbf {\bibinfo {volume} {14}},\ \bibinfo {pages} {400--406}
  (\bibinfo {year} {2015}{\natexlab{a}})}\BibitemShut {NoStop}%
\bibitem [{\citenamefont {Zur}\ and\ \citenamefont {McGill}(1984)}]{Zur1984}%
  \BibitemOpen
  \bibfield  {author} {\bibinfo {author} {\bibfnamefont {A.}~\bibnamefont
  {Zur}}\ and\ \bibinfo {author} {\bibfnamefont {T.~C.}\ \bibnamefont
  {McGill}},\ }\bibfield  {title} {\enquote {\bibinfo {title} {{Lattice match:
  An application to heteroepitaxy}},}\ }\href {\doibase 10.1063/1.333084}
  {\bibfield  {journal} {\bibinfo  {journal} {Journal of Applied Physics}\
  }\textbf {\bibinfo {volume} {55}},\ \bibinfo {pages} {378--386} (\bibinfo
  {year} {1984})}\BibitemShut {NoStop}%
\bibitem [{\citenamefont {Kanne}\ \emph {et~al.}(2020)\citenamefont {Kanne},
  \citenamefont {Marnauza}, \citenamefont {Olsteins}, \citenamefont {Carrad},
  \citenamefont {Sestoft}, \citenamefont {de~Bruijckere}, \citenamefont {Zeng},
  \citenamefont {Johnson}, \citenamefont {Olsson}, \citenamefont
  {Grove-Rasmussen},\ and\ \citenamefont {Nygård}}]{kanne2020epitaxial}%
  \BibitemOpen
  \bibfield  {author} {\bibinfo {author} {\bibfnamefont {T.}~\bibnamefont
  {Kanne}}, \bibinfo {author} {\bibfnamefont {M.}~\bibnamefont {Marnauza}},
  \bibinfo {author} {\bibfnamefont {D.}~\bibnamefont {Olsteins}}, \bibinfo
  {author} {\bibfnamefont {D.~J.}\ \bibnamefont {Carrad}}, \bibinfo {author}
  {\bibfnamefont {J.~E.}\ \bibnamefont {Sestoft}}, \bibinfo {author}
  {\bibfnamefont {J.}~\bibnamefont {de~Bruijckere}}, \bibinfo {author}
  {\bibfnamefont {L.}~\bibnamefont {Zeng}}, \bibinfo {author} {\bibfnamefont
  {E.}~\bibnamefont {Johnson}}, \bibinfo {author} {\bibfnamefont
  {E.}~\bibnamefont {Olsson}}, \bibinfo {author} {\bibfnamefont
  {K.}~\bibnamefont {Grove-Rasmussen}}, \ and\ \bibinfo {author} {\bibfnamefont
  {J.}~\bibnamefont {Nygård}},\ }\href@noop {} {\enquote {\bibinfo {title}
  {Epitaxial pb on inas nanowires},}\ } (\bibinfo {year} {2020}),\ \Eprint
  {http://arxiv.org/abs/2002.11641} {arXiv:2002.11641 [cond-mat.mes-hall]}
  \BibitemShut {NoStop}%
\bibitem [{\citenamefont {Zega}\ \emph {et~al.}(2006)\citenamefont {Zega},
  \citenamefont {Hanbicki}, \citenamefont {Erwin}, \citenamefont {\ifmmode
  \check{Z}\else \v{Z}\fi{}uti\ifmmode~\acute{c}\else \'{c}\fi{}},
  \citenamefont {Kioseoglou}, \citenamefont {Li}, \citenamefont {Jonker},\ and\
  \citenamefont {Stroud}}]{zega2006determination}%
  \BibitemOpen
  \bibfield  {author} {\bibinfo {author} {\bibfnamefont {T.~J.}\ \bibnamefont
  {Zega}}, \bibinfo {author} {\bibfnamefont {A.~T.}\ \bibnamefont {Hanbicki}},
  \bibinfo {author} {\bibfnamefont {S.~C.}\ \bibnamefont {Erwin}}, \bibinfo
  {author} {\bibfnamefont {I.}~\bibnamefont {\ifmmode \check{Z}\else
  \v{Z}\fi{}uti\ifmmode~\acute{c}\else \'{c}\fi{}}}, \bibinfo {author}
  {\bibfnamefont {G.}~\bibnamefont {Kioseoglou}}, \bibinfo {author}
  {\bibfnamefont {C.~H.}\ \bibnamefont {Li}}, \bibinfo {author} {\bibfnamefont
  {B.~T.}\ \bibnamefont {Jonker}}, \ and\ \bibinfo {author} {\bibfnamefont
  {R.~M.}\ \bibnamefont {Stroud}},\ }\bibfield  {title} {\enquote {\bibinfo
  {title} {Determination of interface atomic structure and its impact on spin
  transport using $z$-contrast microscopy and density-functional theory},}\
  }\href {\doibase 10.1103/PhysRevLett.96.196101} {\bibfield  {journal}
  {\bibinfo  {journal} {Phys. Rev. Lett.}\ }\textbf {\bibinfo {volume} {96}},\
  \bibinfo {pages} {196101} (\bibinfo {year} {2006})}\BibitemShut {NoStop}%
\bibitem [{\citenamefont {Liu}\ \emph {et~al.}(2019)\citenamefont {Liu},
  \citenamefont {Luchini}, \citenamefont {Mart{\'\i}-S{\'a}nchez},
  \citenamefont {Koch}, \citenamefont {Schuwalow}, \citenamefont {Khan},
  \citenamefont {Stankevic}, \citenamefont {Francoual}, \citenamefont
  {Mardegan}, \citenamefont {Krieger} \emph {et~al.}}]{liu2019coherent}%
  \BibitemOpen
  \bibfield  {author} {\bibinfo {author} {\bibfnamefont {Y.}~\bibnamefont
  {Liu}}, \bibinfo {author} {\bibfnamefont {A.}~\bibnamefont {Luchini}},
  \bibinfo {author} {\bibfnamefont {S.}~\bibnamefont {Mart{\'\i}-S{\'a}nchez}},
  \bibinfo {author} {\bibfnamefont {C.}~\bibnamefont {Koch}}, \bibinfo {author}
  {\bibfnamefont {S.}~\bibnamefont {Schuwalow}}, \bibinfo {author}
  {\bibfnamefont {S.~A.}\ \bibnamefont {Khan}}, \bibinfo {author}
  {\bibfnamefont {T.}~\bibnamefont {Stankevic}}, \bibinfo {author}
  {\bibfnamefont {S.}~\bibnamefont {Francoual}}, \bibinfo {author}
  {\bibfnamefont {J.~R.}\ \bibnamefont {Mardegan}}, \bibinfo {author}
  {\bibfnamefont {J.~A.}\ \bibnamefont {Krieger}},  \emph {et~al.},\ }\bibfield
   {title} {\enquote {\bibinfo {title} {Coherent epitaxial
  semiconductor--ferromagnetic insulator inas/eus interfaces: Band alignment
  and magnetic structure},}\ }\href@noop {} {\bibfield  {journal} {\bibinfo
  {journal} {ACS Applied Materials \& Interfaces}\ }\textbf {\bibinfo {volume}
  {12}},\ \bibinfo {pages} {8780--8787} (\bibinfo {year} {2019})}\BibitemShut
  {NoStop}%
\bibitem [{\citenamefont {Wittkamper}\ \emph {et~al.}(2017)\citenamefont
  {Wittkamper}, \citenamefont {Xu}, \citenamefont {Kombaiah}, \citenamefont
  {Ram}, \citenamefont {De~Graef}, \citenamefont {Kitchin}, \citenamefont
  {Rohrer},\ and\ \citenamefont {Salvador}}]{Wittkamper2017}%
  \BibitemOpen
  \bibfield  {author} {\bibinfo {author} {\bibfnamefont {J.}~\bibnamefont
  {Wittkamper}}, \bibinfo {author} {\bibfnamefont {Z.}~\bibnamefont {Xu}},
  \bibinfo {author} {\bibfnamefont {B.}~\bibnamefont {Kombaiah}}, \bibinfo
  {author} {\bibfnamefont {F.}~\bibnamefont {Ram}}, \bibinfo {author}
  {\bibfnamefont {M.}~\bibnamefont {De~Graef}}, \bibinfo {author}
  {\bibfnamefont {J.~R.}\ \bibnamefont {Kitchin}}, \bibinfo {author}
  {\bibfnamefont {G.~S.}\ \bibnamefont {Rohrer}}, \ and\ \bibinfo {author}
  {\bibfnamefont {P.~A.}\ \bibnamefont {Salvador}},\ }\bibfield  {title}
  {\enquote {\bibinfo {title} {Competitive growth of scrutinyite
  ($\alpha$-pbo2) and rutile polymorphs of sno2 on all orientations of
  columbite conb2o6 substrates},}\ }\href {\doibase 10.1021/acs.cgd.7b00569}
  {\bibfield  {journal} {\bibinfo  {journal} {Crystal Growth {\&} Design}\
  }\textbf {\bibinfo {volume} {17}},\ \bibinfo {pages} {3929--3939} (\bibinfo
  {year} {2017})}\BibitemShut {NoStop}%
\bibitem [{\citenamefont {Xu}, \citenamefont {Salvador},\ and\ \citenamefont
  {Kitchin}(2017)}]{Xu2017}%
  \BibitemOpen
  \bibfield  {author} {\bibinfo {author} {\bibfnamefont {Z.}~\bibnamefont
  {Xu}}, \bibinfo {author} {\bibfnamefont {P.}~\bibnamefont {Salvador}}, \ and\
  \bibinfo {author} {\bibfnamefont {J.~R.}\ \bibnamefont {Kitchin}},\
  }\bibfield  {title} {\enquote {\bibinfo {title} {First-principles
  investigation of the epitaxial stabilization of oxide polymorphs: Tio2 on
  (sr,ba)tio3},}\ }\href {\doibase 10.1021/acsami.6b11791} {\bibfield
  {journal} {\bibinfo  {journal} {ACS Applied Materials {\&} Interfaces}\
  }\textbf {\bibinfo {volume} {9}},\ \bibinfo {pages} {4106--4118} (\bibinfo
  {year} {2017})}\BibitemShut {NoStop}%
\bibitem [{\citenamefont {Mehta}, \citenamefont {Salvador},\ and\ \citenamefont
  {Kitchin}(2014)}]{Mehta2014}%
  \BibitemOpen
  \bibfield  {author} {\bibinfo {author} {\bibfnamefont {P.}~\bibnamefont
  {Mehta}}, \bibinfo {author} {\bibfnamefont {P.~A.}\ \bibnamefont {Salvador}},
  \ and\ \bibinfo {author} {\bibfnamefont {J.~R.}\ \bibnamefont {Kitchin}},\
  }\bibfield  {title} {\enquote {\bibinfo {title} {Identifying potential bo2
  oxide polymorphs for epitaxial growth candidates},}\ }\href {\doibase
  10.1021/am4059149} {\bibfield  {journal} {\bibinfo  {journal} {ACS Applied
  Materials {\&} Interfaces}\ }\textbf {\bibinfo {volume} {6}},\ \bibinfo
  {pages} {3630--3639} (\bibinfo {year} {2014})}\BibitemShut {NoStop}%
\bibitem [{\citenamefont {Ding}\ \emph {et~al.}(2016)\citenamefont {Ding},
  \citenamefont {Dwaraknath}, \citenamefont {Garten}, \citenamefont {Ndione},
  \citenamefont {Ginley},\ and\ \citenamefont {Persson}}]{Ding2016}%
  \BibitemOpen
  \bibfield  {author} {\bibinfo {author} {\bibfnamefont {H.}~\bibnamefont
  {Ding}}, \bibinfo {author} {\bibfnamefont {S.~S.}\ \bibnamefont
  {Dwaraknath}}, \bibinfo {author} {\bibfnamefont {L.}~\bibnamefont {Garten}},
  \bibinfo {author} {\bibfnamefont {P.}~\bibnamefont {Ndione}}, \bibinfo
  {author} {\bibfnamefont {D.}~\bibnamefont {Ginley}}, \ and\ \bibinfo {author}
  {\bibfnamefont {K.~A.}\ \bibnamefont {Persson}},\ }\bibfield  {title}
  {\enquote {\bibinfo {title} {{Computational Approach for Epitaxial Polymorph
  Stabilization through Substrate Selection}},}\ }\href {\doibase
  10.1021/acsami.6b01630} {\bibfield  {journal} {\bibinfo  {journal} {ACS
  Applied Materials and Interfaces}\ }\textbf {\bibinfo {volume} {8}},\
  \bibinfo {pages} {13086--13093} (\bibinfo {year} {2016})}\BibitemShut
  {NoStop}%
\bibitem [{\citenamefont {Oganov}\ \emph {et~al.}(2019)\citenamefont {Oganov},
  \citenamefont {Pickard}, \citenamefont {Zhu},\ and\ \citenamefont
  {Needs}}]{Oganov2019}%
  \BibitemOpen
  \bibfield  {author} {\bibinfo {author} {\bibfnamefont {A.~R.}\ \bibnamefont
  {Oganov}}, \bibinfo {author} {\bibfnamefont {C.~J.}\ \bibnamefont {Pickard}},
  \bibinfo {author} {\bibfnamefont {Q.}~\bibnamefont {Zhu}}, \ and\ \bibinfo
  {author} {\bibfnamefont {R.~J.}\ \bibnamefont {Needs}},\ }\href {\doibase
  10.1038/s41578-019-0101-8} {\enquote {\bibinfo {title} {{Structure prediction
  drives materials discovery}},}\ } (\bibinfo {year} {2019})\BibitemShut
  {NoStop}%
\bibitem [{\citenamefont {Glass}, \citenamefont {Oganov},\ and\ \citenamefont
  {Hansen}(2006)}]{glass2006uspex}%
  \BibitemOpen
  \bibfield  {author} {\bibinfo {author} {\bibfnamefont {C.~W.}\ \bibnamefont
  {Glass}}, \bibinfo {author} {\bibfnamefont {A.~R.}\ \bibnamefont {Oganov}}, \
  and\ \bibinfo {author} {\bibfnamefont {N.}~\bibnamefont {Hansen}},\
  }\bibfield  {title} {\enquote {\bibinfo {title} {Uspex—evolutionary crystal
  structure prediction},}\ }\href@noop {} {\bibfield  {journal} {\bibinfo
  {journal} {Computer physics communications}\ }\textbf {\bibinfo {volume}
  {175}},\ \bibinfo {pages} {713--720} (\bibinfo {year} {2006})}\BibitemShut
  {NoStop}%
\bibitem [{\citenamefont {Lonie}\ and\ \citenamefont
  {Zurek}(2011)}]{lonie2011xtalopt}%
  \BibitemOpen
  \bibfield  {author} {\bibinfo {author} {\bibfnamefont {D.~C.}\ \bibnamefont
  {Lonie}}\ and\ \bibinfo {author} {\bibfnamefont {E.}~\bibnamefont {Zurek}},\
  }\bibfield  {title} {\enquote {\bibinfo {title} {Xtalopt: An open-source
  evolutionary algorithm for crystal structure prediction},}\ }\href@noop {}
  {\bibfield  {journal} {\bibinfo  {journal} {Computer Physics Communications}\
  }\textbf {\bibinfo {volume} {182}},\ \bibinfo {pages} {372--387} (\bibinfo
  {year} {2011})}\BibitemShut {NoStop}%
\bibitem [{\citenamefont {Trimarchi}\ and\ \citenamefont
  {Zunger}(2007)}]{trimarchi2007global}%
  \BibitemOpen
  \bibfield  {author} {\bibinfo {author} {\bibfnamefont {G.}~\bibnamefont
  {Trimarchi}}\ and\ \bibinfo {author} {\bibfnamefont {A.}~\bibnamefont
  {Zunger}},\ }\bibfield  {title} {\enquote {\bibinfo {title} {Global
  space-group optimization problem: Finding the stablest crystal structure
  without constraints},}\ }\href@noop {} {\bibfield  {journal} {\bibinfo
  {journal} {Physical Review B}\ }\textbf {\bibinfo {volume} {75}},\ \bibinfo
  {pages} {104113} (\bibinfo {year} {2007})}\BibitemShut {NoStop}%
\bibitem [{\citenamefont {Chua}\ \emph {et~al.}(2010)\citenamefont {Chua},
  \citenamefont {Benedek}, \citenamefont {Chen}, \citenamefont {Finnis},\ and\
  \citenamefont {Sutton}}]{Chua2010}%
  \BibitemOpen
  \bibfield  {author} {\bibinfo {author} {\bibfnamefont {A.~L.}\ \bibnamefont
  {Chua}}, \bibinfo {author} {\bibfnamefont {N.~A.}\ \bibnamefont {Benedek}},
  \bibinfo {author} {\bibfnamefont {L.}~\bibnamefont {Chen}}, \bibinfo {author}
  {\bibfnamefont {M.~W.}\ \bibnamefont {Finnis}}, \ and\ \bibinfo {author}
  {\bibfnamefont {A.~P.}\ \bibnamefont {Sutton}},\ }\bibfield  {title}
  {\enquote {\bibinfo {title} {{A genetic algorithm for predicting the
  structures of interfaces in multicomponent systems}},}\ }\href {\doibase
  10.1038/nmat2712} {\bibfield  {journal} {\bibinfo  {journal} {Nature
  Materials}\ }\textbf {\bibinfo {volume} {9}},\ \bibinfo {pages} {418--422}
  (\bibinfo {year} {2010})}\BibitemShut {NoStop}%
\bibitem [{\citenamefont {Zhu}\ \emph {et~al.}(2018)\citenamefont {Zhu},
  \citenamefont {Samanta}, \citenamefont {Li}, \citenamefont {Rudd},\ and\
  \citenamefont {Frolov}}]{Zhu2018}%
  \BibitemOpen
  \bibfield  {author} {\bibinfo {author} {\bibfnamefont {Q.}~\bibnamefont
  {Zhu}}, \bibinfo {author} {\bibfnamefont {A.}~\bibnamefont {Samanta}},
  \bibinfo {author} {\bibfnamefont {B.}~\bibnamefont {Li}}, \bibinfo {author}
  {\bibfnamefont {R.~E.}\ \bibnamefont {Rudd}}, \ and\ \bibinfo {author}
  {\bibfnamefont {T.}~\bibnamefont {Frolov}},\ }\bibfield  {title} {\enquote
  {\bibinfo {title} {{Predicting phase behavior of grain boundaries with
  evolutionary search and machine learning}},}\ }\href {\doibase
  10.1038/s41467-018-02937-2} {\bibfield  {journal} {\bibinfo  {journal}
  {Nature Communications}\ }\textbf {\bibinfo {volume} {9}},\ \bibinfo {pages}
  {1--9} (\bibinfo {year} {2018})},\ \Eprint {http://arxiv.org/abs/1707.09699}
  {arXiv:1707.09699} \BibitemShut {NoStop}%
\bibitem [{\citenamefont {Raclariu}\ \emph {et~al.}(2015)\citenamefont
  {Raclariu}, \citenamefont {Deshpande}, \citenamefont {Bruggemann},
  \citenamefont {Zhuge}, \citenamefont {Yu}, \citenamefont {Ratsch},\ and\
  \citenamefont {Shankar}}]{Raclariu2015}%
  \BibitemOpen
  \bibfield  {author} {\bibinfo {author} {\bibfnamefont {A.~M.}\ \bibnamefont
  {Raclariu}}, \bibinfo {author} {\bibfnamefont {S.}~\bibnamefont {Deshpande}},
  \bibinfo {author} {\bibfnamefont {J.}~\bibnamefont {Bruggemann}}, \bibinfo
  {author} {\bibfnamefont {W.}~\bibnamefont {Zhuge}}, \bibinfo {author}
  {\bibfnamefont {T.~H.}\ \bibnamefont {Yu}}, \bibinfo {author} {\bibfnamefont
  {C.}~\bibnamefont {Ratsch}}, \ and\ \bibinfo {author} {\bibfnamefont
  {S.}~\bibnamefont {Shankar}},\ }\bibfield  {title} {\enquote {\bibinfo
  {title} {{A fast method for predicting the formation of crystal interfaces
  and heterocrystals}},}\ }\href {\doibase 10.1016/j.commatsci.2015.05.023}
  {\bibfield  {journal} {\bibinfo  {journal} {Computational Materials Science}\
  }\textbf {\bibinfo {volume} {108}},\ \bibinfo {pages} {88--93} (\bibinfo
  {year} {2015})}\BibitemShut {NoStop}%
\bibitem [{\citenamefont {Mathew}\ \emph {et~al.}(2016)\citenamefont {Mathew},
  \citenamefont {Singh}, \citenamefont {Gabriel}, \citenamefont {Choudhary},
  \citenamefont {Sinnott}, \citenamefont {Davydov}, \citenamefont {Tavazza},\
  and\ \citenamefont {Hennig}}]{Mathew2016}%
  \BibitemOpen
  \bibfield  {author} {\bibinfo {author} {\bibfnamefont {K.}~\bibnamefont
  {Mathew}}, \bibinfo {author} {\bibfnamefont {A.~K.}\ \bibnamefont {Singh}},
  \bibinfo {author} {\bibfnamefont {J.~J.}\ \bibnamefont {Gabriel}}, \bibinfo
  {author} {\bibfnamefont {K.}~\bibnamefont {Choudhary}}, \bibinfo {author}
  {\bibfnamefont {S.~B.}\ \bibnamefont {Sinnott}}, \bibinfo {author}
  {\bibfnamefont {A.~V.}\ \bibnamefont {Davydov}}, \bibinfo {author}
  {\bibfnamefont {F.}~\bibnamefont {Tavazza}}, \ and\ \bibinfo {author}
  {\bibfnamefont {R.~G.}\ \bibnamefont {Hennig}},\ }\bibfield  {title}
  {\enquote {\bibinfo {title} {{MPInterfaces: A Materials Project based Python
  tool for high-throughput computational screening of interfacial systems}},}\
  }\href {\doibase 10.1016/j.commatsci.2016.05.020} {\bibfield  {journal}
  {\bibinfo  {journal} {Computational Materials Science}\ }\textbf {\bibinfo
  {volume} {122}},\ \bibinfo {pages} {183--190} (\bibinfo {year} {2016})},\
  \Eprint {http://arxiv.org/abs/1602.07784} {arXiv:1602.07784} \BibitemShut
  {NoStop}%
\bibitem [{\citenamefont {Gao}\ \emph {et~al.}(2019)\citenamefont {Gao},
  \citenamefont {Gao}, \citenamefont {Lu}, \citenamefont {Lv}, \citenamefont
  {Wang},\ and\ \citenamefont {Ma}}]{Gao2019}%
  \BibitemOpen
  \bibfield  {author} {\bibinfo {author} {\bibfnamefont {B.}~\bibnamefont
  {Gao}}, \bibinfo {author} {\bibfnamefont {P.}~\bibnamefont {Gao}}, \bibinfo
  {author} {\bibfnamefont {S.}~\bibnamefont {Lu}}, \bibinfo {author}
  {\bibfnamefont {J.}~\bibnamefont {Lv}}, \bibinfo {author} {\bibfnamefont
  {Y.}~\bibnamefont {Wang}}, \ and\ \bibinfo {author} {\bibfnamefont
  {Y.}~\bibnamefont {Ma}},\ }\bibfield  {title} {\enquote {\bibinfo {title}
  {{Interface structure prediction via CALYPSO method}},}\ }\href {\doibase
  10.1016/j.scib.2019.02.009} {\bibfield  {journal} {\bibinfo  {journal}
  {Science Bulletin}\ }\textbf {\bibinfo {volume} {64}},\ \bibinfo {pages}
  {301--309} (\bibinfo {year} {2019})}\BibitemShut {NoStop}%
\bibitem [{\citenamefont {Ong}\ \emph {et~al.}(2013)\citenamefont {Ong},
  \citenamefont {Richards}, \citenamefont {Jain}, \citenamefont {Hautier},
  \citenamefont {Kocher}, \citenamefont {Cholia}, \citenamefont {Gunter},
  \citenamefont {Chevrier}, \citenamefont {Persson},\ and\ \citenamefont
  {Ceder}}]{ong2013python}%
  \BibitemOpen
  \bibfield  {author} {\bibinfo {author} {\bibfnamefont {S.~P.}\ \bibnamefont
  {Ong}}, \bibinfo {author} {\bibfnamefont {W.~D.}\ \bibnamefont {Richards}},
  \bibinfo {author} {\bibfnamefont {A.}~\bibnamefont {Jain}}, \bibinfo {author}
  {\bibfnamefont {G.}~\bibnamefont {Hautier}}, \bibinfo {author} {\bibfnamefont
  {M.}~\bibnamefont {Kocher}}, \bibinfo {author} {\bibfnamefont
  {S.}~\bibnamefont {Cholia}}, \bibinfo {author} {\bibfnamefont
  {D.}~\bibnamefont {Gunter}}, \bibinfo {author} {\bibfnamefont {V.~L.}\
  \bibnamefont {Chevrier}}, \bibinfo {author} {\bibfnamefont {K.~A.}\
  \bibnamefont {Persson}}, \ and\ \bibinfo {author} {\bibfnamefont
  {G.}~\bibnamefont {Ceder}},\ }\bibfield  {title} {\enquote {\bibinfo {title}
  {Python materials genomics (pymatgen): A robust, open-source python library
  for materials analysis},}\ }\href@noop {} {\bibfield  {journal} {\bibinfo
  {journal} {Computational Materials Science}\ }\textbf {\bibinfo {volume}
  {68}},\ \bibinfo {pages} {314--319} (\bibinfo {year} {2013})}\BibitemShut
  {NoStop}%
\bibitem [{\citenamefont {Larsen}\ \emph {et~al.}(2017)\citenamefont {Larsen},
  \citenamefont {Mortensen}, \citenamefont {Blomqvist}, \citenamefont
  {Castelli}, \citenamefont {Christensen}, \citenamefont {Du{\l}ak},
  \citenamefont {Friis}, \citenamefont {Groves}, \citenamefont {Hammer},
  \citenamefont {Hargus} \emph {et~al.}}]{larsen2017atomic}%
  \BibitemOpen
  \bibfield  {author} {\bibinfo {author} {\bibfnamefont {A.~H.}\ \bibnamefont
  {Larsen}}, \bibinfo {author} {\bibfnamefont {J.~J.}\ \bibnamefont
  {Mortensen}}, \bibinfo {author} {\bibfnamefont {J.}~\bibnamefont
  {Blomqvist}}, \bibinfo {author} {\bibfnamefont {I.~E.}\ \bibnamefont
  {Castelli}}, \bibinfo {author} {\bibfnamefont {R.}~\bibnamefont
  {Christensen}}, \bibinfo {author} {\bibfnamefont {M.}~\bibnamefont
  {Du{\l}ak}}, \bibinfo {author} {\bibfnamefont {J.}~\bibnamefont {Friis}},
  \bibinfo {author} {\bibfnamefont {M.~N.}\ \bibnamefont {Groves}}, \bibinfo
  {author} {\bibfnamefont {B.}~\bibnamefont {Hammer}}, \bibinfo {author}
  {\bibfnamefont {C.}~\bibnamefont {Hargus}},  \emph {et~al.},\ }\bibfield
  {title} {\enquote {\bibinfo {title} {The atomic simulation environment—a
  python library for working with atoms},}\ }\href@noop {} {\bibfield
  {journal} {\bibinfo  {journal} {Journal of Physics: Condensed Matter}\
  }\textbf {\bibinfo {volume} {29}},\ \bibinfo {pages} {273002} (\bibinfo
  {year} {2017})}\BibitemShut {NoStop}%
\bibitem [{\citenamefont {Frederiksen}\ \emph {et~al.}(2004)\citenamefont
  {Frederiksen}, \citenamefont {Jacobsen}, \citenamefont {Brown},\ and\
  \citenamefont {Sethna}}]{frederiksen2004bayesian}%
  \BibitemOpen
  \bibfield  {author} {\bibinfo {author} {\bibfnamefont {S.~L.}\ \bibnamefont
  {Frederiksen}}, \bibinfo {author} {\bibfnamefont {K.~W.}\ \bibnamefont
  {Jacobsen}}, \bibinfo {author} {\bibfnamefont {K.~S.}\ \bibnamefont {Brown}},
  \ and\ \bibinfo {author} {\bibfnamefont {J.~P.}\ \bibnamefont {Sethna}},\
  }\bibfield  {title} {\enquote {\bibinfo {title} {Bayesian ensemble approach
  to error estimation of interatomic potentials},}\ }\href@noop {} {\bibfield
  {journal} {\bibinfo  {journal} {Physical review letters}\ }\textbf {\bibinfo
  {volume} {93}},\ \bibinfo {pages} {165501} (\bibinfo {year}
  {2004})}\BibitemShut {NoStop}%
\bibitem [{\citenamefont {Chen}\ \emph {et~al.}(2017)\citenamefont {Chen},
  \citenamefont {Deng}, \citenamefont {Tran}, \citenamefont {Tang},
  \citenamefont {Chu},\ and\ \citenamefont {Ong}}]{chen2017accurate}%
  \BibitemOpen
  \bibfield  {author} {\bibinfo {author} {\bibfnamefont {C.}~\bibnamefont
  {Chen}}, \bibinfo {author} {\bibfnamefont {Z.}~\bibnamefont {Deng}}, \bibinfo
  {author} {\bibfnamefont {R.}~\bibnamefont {Tran}}, \bibinfo {author}
  {\bibfnamefont {H.}~\bibnamefont {Tang}}, \bibinfo {author} {\bibfnamefont
  {I.-H.}\ \bibnamefont {Chu}}, \ and\ \bibinfo {author} {\bibfnamefont
  {S.~P.}\ \bibnamefont {Ong}},\ }\bibfield  {title} {\enquote {\bibinfo
  {title} {Accurate force field for molybdenum by machine learning large
  materials data},}\ }\href {\doibase 10.1103/PhysRevMaterials.1.043603}
  {\bibfield  {journal} {\bibinfo  {journal} {Phys. Rev. Materials}\ }\textbf
  {\bibinfo {volume} {1}},\ \bibinfo {pages} {043603} (\bibinfo {year}
  {2017})}\BibitemShut {NoStop}%
\bibitem [{\citenamefont {Li}\ \emph {et~al.}(2016)\citenamefont {Li},
  \citenamefont {Hui}, \citenamefont {Shao}, \citenamefont {Chen},
  \citenamefont {Wang}, \citenamefont {Jia}, \citenamefont {Li}, \citenamefont
  {Chen},\ and\ \citenamefont {Cheng}}]{li2016first}%
  \BibitemOpen
  \bibfield  {author} {\bibinfo {author} {\bibfnamefont {X.}~\bibnamefont
  {Li}}, \bibinfo {author} {\bibfnamefont {Q.}~\bibnamefont {Hui}}, \bibinfo
  {author} {\bibfnamefont {D.}~\bibnamefont {Shao}}, \bibinfo {author}
  {\bibfnamefont {J.}~\bibnamefont {Chen}}, \bibinfo {author} {\bibfnamefont
  {P.}~\bibnamefont {Wang}}, \bibinfo {author} {\bibfnamefont {Z.}~\bibnamefont
  {Jia}}, \bibinfo {author} {\bibfnamefont {C.}~\bibnamefont {Li}}, \bibinfo
  {author} {\bibfnamefont {Z.}~\bibnamefont {Chen}}, \ and\ \bibinfo {author}
  {\bibfnamefont {N.}~\bibnamefont {Cheng}},\ }\bibfield  {title} {\enquote
  {\bibinfo {title} {First-principles study on the stability and electronic
  structure of mg/zrb 2 interfaces},}\ }\href@noop {} {\bibfield  {journal}
  {\bibinfo  {journal} {Science China Materials}\ }\textbf {\bibinfo {volume}
  {59}},\ \bibinfo {pages} {28--37} (\bibinfo {year} {2016})}\BibitemShut
  {NoStop}%
\bibitem [{\citenamefont {Liu}, \citenamefont {Wang},\ and\ \citenamefont
  {Ye}(2004)}]{liu2004first}%
  \BibitemOpen
  \bibfield  {author} {\bibinfo {author} {\bibfnamefont {L.}~\bibnamefont
  {Liu}}, \bibinfo {author} {\bibfnamefont {S.}~\bibnamefont {Wang}}, \ and\
  \bibinfo {author} {\bibfnamefont {H.}~\bibnamefont {Ye}},\ }\bibfield
  {title} {\enquote {\bibinfo {title} {First-principles study of polar al/tin
  (1 1 1) interfaces},}\ }\href@noop {} {\bibfield  {journal} {\bibinfo
  {journal} {Acta materialia}\ }\textbf {\bibinfo {volume} {52}},\ \bibinfo
  {pages} {3681--3688} (\bibinfo {year} {2004})}\BibitemShut {NoStop}%
\bibitem [{\citenamefont {Zhuo}\ \emph {et~al.}(2018)\citenamefont {Zhuo},
  \citenamefont {Mao}, \citenamefont {Xu},\ and\ \citenamefont
  {Fu}}]{zhuo2018density}%
  \BibitemOpen
  \bibfield  {author} {\bibinfo {author} {\bibfnamefont {Z.}~\bibnamefont
  {Zhuo}}, \bibinfo {author} {\bibfnamefont {H.}~\bibnamefont {Mao}}, \bibinfo
  {author} {\bibfnamefont {H.}~\bibnamefont {Xu}}, \ and\ \bibinfo {author}
  {\bibfnamefont {Y.}~\bibnamefont {Fu}},\ }\bibfield  {title} {\enquote
  {\bibinfo {title} {Density functional theory study of al/nbb2 heterogeneous
  nucleation interface},}\ }\href@noop {} {\bibfield  {journal} {\bibinfo
  {journal} {Applied Surface Science}\ }\textbf {\bibinfo {volume} {456}},\
  \bibinfo {pages} {37--42} (\bibinfo {year} {2018})}\BibitemShut {NoStop}%
\bibitem [{\citenamefont {Wang}, \citenamefont {Li},\ and\ \citenamefont
  {Xu}(2020)}]{wang2020first}%
  \BibitemOpen
  \bibfield  {author} {\bibinfo {author} {\bibfnamefont {J.}~\bibnamefont
  {Wang}}, \bibinfo {author} {\bibfnamefont {Y.}~\bibnamefont {Li}}, \ and\
  \bibinfo {author} {\bibfnamefont {R.}~\bibnamefont {Xu}},\ }\bibfield
  {title} {\enquote {\bibinfo {title} {First-principles calculations on
  electronic structure and interfacial stability of mg/nbb2 heterogeneous
  nucleation interface},}\ }\href@noop {} {\bibfield  {journal} {\bibinfo
  {journal} {Surface Science}\ }\textbf {\bibinfo {volume} {691}},\ \bibinfo
  {pages} {121487} (\bibinfo {year} {2020})}\BibitemShut {NoStop}%
\bibitem [{\citenamefont {Yang}\ \emph
  {et~al.}(2020{\natexlab{b}})\citenamefont {Yang}, \citenamefont {Schröter},
  \citenamefont {Schuwalow}, \citenamefont {Rajpalk}, \citenamefont {Ohtani},
  \citenamefont {KrogstrupGeorg}, \citenamefont {Winkler}, \citenamefont
  {Gukelberger}, \citenamefont {Gresch}, \citenamefont {Aeppli}, \citenamefont
  {Lutchyn}, \citenamefont {Strocov},\ and\ \citenamefont
  {Marom}}]{yang2020electronic}%
  \BibitemOpen
  \bibfield  {author} {\bibinfo {author} {\bibfnamefont {S.}~\bibnamefont
  {Yang}}, \bibinfo {author} {\bibfnamefont {N.~B.~M.}\ \bibnamefont
  {Schröter}}, \bibinfo {author} {\bibfnamefont {S.}~\bibnamefont
  {Schuwalow}}, \bibinfo {author} {\bibfnamefont {M.}~\bibnamefont {Rajpalk}},
  \bibinfo {author} {\bibfnamefont {K.}~\bibnamefont {Ohtani}}, \bibinfo
  {author} {\bibfnamefont {P.}~\bibnamefont {KrogstrupGeorg}}, \bibinfo
  {author} {\bibfnamefont {W.}~\bibnamefont {Winkler}}, \bibinfo {author}
  {\bibfnamefont {J.}~\bibnamefont {Gukelberger}}, \bibinfo {author}
  {\bibfnamefont {D.}~\bibnamefont {Gresch}}, \bibinfo {author} {\bibfnamefont
  {G.}~\bibnamefont {Aeppli}}, \bibinfo {author} {\bibfnamefont {R.~M.}\
  \bibnamefont {Lutchyn}}, \bibinfo {author} {\bibfnamefont {V.~N.}\
  \bibnamefont {Strocov}}, \ and\ \bibinfo {author} {\bibfnamefont
  {N.}~\bibnamefont {Marom}},\ }\href@noop {} {\enquote {\bibinfo {title}
  {Electronic structure of inas and insb surfaces: density functional theory
  and angle-resolved photoemission spectroscopy},}\ } (\bibinfo {year}
  {2020}{\natexlab{b}}),\ \Eprint {http://arxiv.org/abs/2012.14935}
  {arXiv:2012.14935 [cond-mat.mtrl-sci]} \BibitemShut {NoStop}%
\bibitem [{\citenamefont {Yang}, \citenamefont {Wu},\ and\ \citenamefont
  {Marom}(2020)}]{yang2020topological}%
  \BibitemOpen
  \bibfield  {author} {\bibinfo {author} {\bibfnamefont {S.}~\bibnamefont
  {Yang}}, \bibinfo {author} {\bibfnamefont {C.}~\bibnamefont {Wu}}, \ and\
  \bibinfo {author} {\bibfnamefont {N.}~\bibnamefont {Marom}},\ }\bibfield
  {title} {\enquote {\bibinfo {title} {Topological properties of snse/eus and
  snte/cate interfaces},}\ }\href {\doibase 10.1103/PhysRevMaterials.4.034203}
  {\bibfield  {journal} {\bibinfo  {journal} {Phys. Rev. Materials}\ }\textbf
  {\bibinfo {volume} {4}},\ \bibinfo {pages} {034203} (\bibinfo {year}
  {2020})}\BibitemShut {NoStop}%
\bibitem [{\citenamefont {Yang}\ \emph
  {et~al.}(2020{\natexlab{c}})\citenamefont {Yang}, \citenamefont {Bier},
  \citenamefont {Wen}, \citenamefont {Zhan}, \citenamefont {Moayedpour},\ and\
  \citenamefont {Marom}}]{yang2020ogre}%
  \BibitemOpen
  \bibfield  {author} {\bibinfo {author} {\bibfnamefont {S.}~\bibnamefont
  {Yang}}, \bibinfo {author} {\bibfnamefont {I.}~\bibnamefont {Bier}}, \bibinfo
  {author} {\bibfnamefont {W.}~\bibnamefont {Wen}}, \bibinfo {author}
  {\bibfnamefont {J.}~\bibnamefont {Zhan}}, \bibinfo {author} {\bibfnamefont
  {S.}~\bibnamefont {Moayedpour}}, \ and\ \bibinfo {author} {\bibfnamefont
  {N.}~\bibnamefont {Marom}},\ }\bibfield  {title} {\enquote {\bibinfo {title}
  {Ogre: A python package for molecular crystal surface generation with
  applications to surface energy and crystal habit prediction},}\ }\href@noop
  {} {\bibfield  {journal} {\bibinfo  {journal} {The Journal of Chemical
  Physics}\ }\textbf {\bibinfo {volume} {152}},\ \bibinfo {pages} {244122}
  (\bibinfo {year} {2020}{\natexlab{c}})}\BibitemShut {NoStop}%
\bibitem [{\citenamefont {Blum}\ \emph {et~al.}(2009)\citenamefont {Blum},
  \citenamefont {Gehrke}, \citenamefont {Hanke}, \citenamefont {Havu},
  \citenamefont {Havu}, \citenamefont {Ren}, \citenamefont {Reuter},\ and\
  \citenamefont {Scheffler}}]{Blum2009}%
  \BibitemOpen
  \bibfield  {author} {\bibinfo {author} {\bibfnamefont {V.}~\bibnamefont
  {Blum}}, \bibinfo {author} {\bibfnamefont {R.}~\bibnamefont {Gehrke}},
  \bibinfo {author} {\bibfnamefont {F.}~\bibnamefont {Hanke}}, \bibinfo
  {author} {\bibfnamefont {P.}~\bibnamefont {Havu}}, \bibinfo {author}
  {\bibfnamefont {V.}~\bibnamefont {Havu}}, \bibinfo {author} {\bibfnamefont
  {X.}~\bibnamefont {Ren}}, \bibinfo {author} {\bibfnamefont {K.}~\bibnamefont
  {Reuter}}, \ and\ \bibinfo {author} {\bibfnamefont {M.}~\bibnamefont
  {Scheffler}},\ }\bibfield  {title} {\enquote {\bibinfo {title} {{Ab initio
  molecular simulations with numeric atom-centered orbitals}},}\ }\href
  {\doibase 10.1016/j.cpc.2009.06.022} {\bibfield  {journal} {\bibinfo
  {journal} {Computer Physics Communications}\ }\textbf {\bibinfo {volume}
  {180}},\ \bibinfo {pages} {2175--2196} (\bibinfo {year} {2009})}\BibitemShut
  {NoStop}%
\bibitem [{\citenamefont {Joubert}(1999)}]{Joubert1999}%
  \BibitemOpen
  \bibfield  {author} {\bibinfo {author} {\bibfnamefont {D.}~\bibnamefont
  {Joubert}},\ }\bibfield  {title} {\enquote {\bibinfo {title} {{From ultrasoft
  pseudopotentials to the projector augmented-wave method}},}\ }\href {\doibase
  10.1103/PhysRevB.59.1758} {\bibfield  {journal} {\bibinfo  {journal}
  {Physical Review B - Condensed Matter and Materials Physics}\ }\textbf
  {\bibinfo {volume} {59}},\ \bibinfo {pages} {1758--1775} (\bibinfo {year}
  {1999})}\BibitemShut {NoStop}%
\bibitem [{\citenamefont {Kresse}\ and\ \citenamefont
  {Furthm{\"{u}}ller}(1996{\natexlab{a}})}]{Kresse1996}%
  \BibitemOpen
  \bibfield  {author} {\bibinfo {author} {\bibfnamefont {G.}~\bibnamefont
  {Kresse}}\ and\ \bibinfo {author} {\bibfnamefont {J.}~\bibnamefont
  {Furthm{\"{u}}ller}},\ }\bibfield  {title} {\enquote {\bibinfo {title}
  {{Efficient iterative schemes for ab initio total-energy calculations using a
  plane-wave basis set}},}\ }\href {\doibase 10.1103/PhysRevB.54.11169}
  {\bibfield  {journal} {\bibinfo  {journal} {Physical Review B - Condensed
  Matter and Materials Physics}\ }\textbf {\bibinfo {volume} {54}},\ \bibinfo
  {pages} {11169--11186} (\bibinfo {year} {1996}{\natexlab{a}})}\BibitemShut
  {NoStop}%
\bibitem [{\citenamefont {Kresse}\ and\ \citenamefont
  {Furthm{\"{u}}ller}(1996{\natexlab{b}})}]{Kresse1996a}%
  \BibitemOpen
  \bibfield  {author} {\bibinfo {author} {\bibfnamefont {G.}~\bibnamefont
  {Kresse}}\ and\ \bibinfo {author} {\bibfnamefont {J.}~\bibnamefont
  {Furthm{\"{u}}ller}},\ }\bibfield  {title} {\enquote {\bibinfo {title}
  {{Efficiency of ab-initio total energy calculations for metals and
  semiconductors using a plane-wave basis set}},}\ }\href {\doibase
  10.1016/0927-0256(96)00008-0} {\bibfield  {journal} {\bibinfo  {journal}
  {Computational Materials Science}\ }\textbf {\bibinfo {volume} {6}},\
  \bibinfo {pages} {15--50} (\bibinfo {year} {1996}{\natexlab{b}})}\BibitemShut
  {NoStop}%
\bibitem [{\citenamefont {Kresse}\ and\ \citenamefont
  {Hafner}(1993)}]{Kresse1993}%
  \BibitemOpen
  \bibfield  {author} {\bibinfo {author} {\bibfnamefont {G.}~\bibnamefont
  {Kresse}}\ and\ \bibinfo {author} {\bibfnamefont {J.}~\bibnamefont
  {Hafner}},\ }\bibfield  {title} {\enquote {\bibinfo {title} {{Ab initio
  molecular dynamics for open-shell transition metals}},}\ }\href {\doibase
  10.1103/PhysRevB.48.13115} {\bibfield  {journal} {\bibinfo  {journal}
  {Physical Review B}\ }\textbf {\bibinfo {volume} {48}},\ \bibinfo {pages}
  {13115--13118} (\bibinfo {year} {1993})}\BibitemShut {NoStop}%
\bibitem [{\citenamefont {Kresse}\ and\ \citenamefont
  {Hafner}(1994)}]{Kresse1994}%
  \BibitemOpen
  \bibfield  {author} {\bibinfo {author} {\bibfnamefont {G.}~\bibnamefont
  {Kresse}}\ and\ \bibinfo {author} {\bibfnamefont {J.}~\bibnamefont
  {Hafner}},\ }\bibfield  {title} {\enquote {\bibinfo {title} {{Ab initio
  molecular-dynamics simulation of the liquid-metalamorphous- semiconductor
  transition in germanium}},}\ }\href {\doibase 10.1103/PhysRevB.49.14251}
  {\bibfield  {journal} {\bibinfo  {journal} {Physical Review B}\ }\textbf
  {\bibinfo {volume} {49}},\ \bibinfo {pages} {14251--14269} (\bibinfo {year}
  {1994})}\BibitemShut {NoStop}%
\bibitem [{\citenamefont {Chang}\ and\ \citenamefont
  {Ploog}(2012)}]{chang2012molecular}%
  \BibitemOpen
  \bibfield  {author} {\bibinfo {author} {\bibfnamefont {L.~L.}\ \bibnamefont
  {Chang}}\ and\ \bibinfo {author} {\bibfnamefont {K.}~\bibnamefont {Ploog}},\
  }\href@noop {} {\emph {\bibinfo {title} {Molecular beam epitaxy and
  heterostructures}}},\ Vol.~\bibinfo {volume} {87}\ (\bibinfo  {publisher}
  {Springer Science \& Business Media},\ \bibinfo {year} {2012})\BibitemShut
  {NoStop}%
\bibitem [{\citenamefont {Spackman}\ and\ \citenamefont
  {Jayatilaka}(2009)}]{Spackman2009}%
  \BibitemOpen
  \bibfield  {author} {\bibinfo {author} {\bibfnamefont {M.~A.}\ \bibnamefont
  {Spackman}}\ and\ \bibinfo {author} {\bibfnamefont {D.}~\bibnamefont
  {Jayatilaka}},\ }\bibfield  {title} {\enquote {\bibinfo {title} {{Hirshfeld
  surface analysis}},}\ }\href {\doibase 10.1039/b818330a} {\bibfield
  {journal} {\bibinfo  {journal} {CrystEngComm}\ }\textbf {\bibinfo {volume}
  {11}},\ \bibinfo {pages} {19--32} (\bibinfo {year} {2009})}\BibitemShut
  {NoStop}%
\bibitem [{\citenamefont {Scholz}\ and\ \citenamefont
  {Stirner}(2019)}]{Scholz2019}%
  \BibitemOpen
  \bibfield  {author} {\bibinfo {author} {\bibfnamefont {D.}~\bibnamefont
  {Scholz}}\ and\ \bibinfo {author} {\bibfnamefont {T.}~\bibnamefont
  {Stirner}},\ }\bibfield  {title} {\enquote {\bibinfo {title} {{Convergence of
  surface energy calculations for various methods: (0 0 1) hematite as
  benchmark}},}\ }\href {\doibase 10.1088/1361-648X/ab069d} {\bibfield
  {journal} {\bibinfo  {journal} {Journal of Physics Condensed Matter}\
  }\textbf {\bibinfo {volume} {31}},\ \bibinfo {pages} {195901} (\bibinfo
  {year} {2019})}\BibitemShut {NoStop}%
\bibitem [{\citenamefont {Sun}\ and\ \citenamefont {Ceder}(2013)}]{Sun2013}%
  \BibitemOpen
  \bibfield  {author} {\bibinfo {author} {\bibfnamefont {W.}~\bibnamefont
  {Sun}}\ and\ \bibinfo {author} {\bibfnamefont {G.}~\bibnamefont {Ceder}},\
  }\bibfield  {title} {\enquote {\bibinfo {title} {{Efficient creation and
  convergence of surface slabs}},}\ }\href {\doibase
  10.1016/j.susc.2013.05.016} {\bibfield  {journal} {\bibinfo  {journal}
  {Surface Science}\ }\textbf {\bibinfo {volume} {617}},\ \bibinfo {pages}
  {53--59} (\bibinfo {year} {2013})}\BibitemShut {NoStop}%
\bibitem [{\citenamefont {Krogstrup}\ \emph
  {et~al.}(2015{\natexlab{b}})\citenamefont {Krogstrup}, \citenamefont {Ziino},
  \citenamefont {Chang}, \citenamefont {Albrecht}, \citenamefont {Madsen},
  \citenamefont {Johnson}, \citenamefont {Nyg{\aa}rd}, \citenamefont {Marcus},\
  and\ \citenamefont {Jespersen}}]{krogstrup2015epitaxy}%
  \BibitemOpen
  \bibfield  {author} {\bibinfo {author} {\bibfnamefont {P.}~\bibnamefont
  {Krogstrup}}, \bibinfo {author} {\bibfnamefont {N.}~\bibnamefont {Ziino}},
  \bibinfo {author} {\bibfnamefont {W.}~\bibnamefont {Chang}}, \bibinfo
  {author} {\bibfnamefont {S.}~\bibnamefont {Albrecht}}, \bibinfo {author}
  {\bibfnamefont {M.}~\bibnamefont {Madsen}}, \bibinfo {author} {\bibfnamefont
  {E.}~\bibnamefont {Johnson}}, \bibinfo {author} {\bibfnamefont
  {J.}~\bibnamefont {Nyg{\aa}rd}}, \bibinfo {author} {\bibfnamefont {C.~M.}\
  \bibnamefont {Marcus}}, \ and\ \bibinfo {author} {\bibfnamefont
  {T.}~\bibnamefont {Jespersen}},\ }\bibfield  {title} {\enquote {\bibinfo
  {title} {Epitaxy of semiconductor--superconductor nanowires},}\ }\href@noop
  {} {\bibfield  {journal} {\bibinfo  {journal} {Nature materials}\ }\textbf
  {\bibinfo {volume} {14}},\ \bibinfo {pages} {400--406} (\bibinfo {year}
  {2015}{\natexlab{b}})}\BibitemShut {NoStop}%
\bibitem [{\citenamefont {G{\"u}sken}\ \emph {et~al.}(2017)\citenamefont
  {G{\"u}sken}, \citenamefont {Rieger}, \citenamefont {Zellekens},
  \citenamefont {Bennemann}, \citenamefont {Neumann}, \citenamefont {Lepsa},
  \citenamefont {Sch{\"a}pers},\ and\ \citenamefont
  {Gr{\"u}tzmacher}}]{gusken2017mbe}%
  \BibitemOpen
  \bibfield  {author} {\bibinfo {author} {\bibfnamefont {N.~A.}\ \bibnamefont
  {G{\"u}sken}}, \bibinfo {author} {\bibfnamefont {T.}~\bibnamefont {Rieger}},
  \bibinfo {author} {\bibfnamefont {P.}~\bibnamefont {Zellekens}}, \bibinfo
  {author} {\bibfnamefont {B.}~\bibnamefont {Bennemann}}, \bibinfo {author}
  {\bibfnamefont {E.}~\bibnamefont {Neumann}}, \bibinfo {author} {\bibfnamefont
  {M.~I.}\ \bibnamefont {Lepsa}}, \bibinfo {author} {\bibfnamefont
  {T.}~\bibnamefont {Sch{\"a}pers}}, \ and\ \bibinfo {author} {\bibfnamefont
  {D.}~\bibnamefont {Gr{\"u}tzmacher}},\ }\bibfield  {title} {\enquote
  {\bibinfo {title} {Mbe growth of al/inas and nb/inas superconducting hybrid
  nanowire structures},}\ }\href@noop {} {\bibfield  {journal} {\bibinfo
  {journal} {Nanoscale}\ }\textbf {\bibinfo {volume} {9}},\ \bibinfo {pages}
  {16735--16741} (\bibinfo {year} {2017})}\BibitemShut {NoStop}%
\bibitem [{\citenamefont {Marom}\ \emph {et~al.}(2010)\citenamefont {Marom},
  \citenamefont {Bernstein}, \citenamefont {Garel}, \citenamefont {Tkatchenko},
  \citenamefont {Joselevich}, \citenamefont {Kronik},\ and\ \citenamefont
  {Hod}}]{Marom2010}%
  \BibitemOpen
  \bibfield  {author} {\bibinfo {author} {\bibfnamefont {N.}~\bibnamefont
  {Marom}}, \bibinfo {author} {\bibfnamefont {J.}~\bibnamefont {Bernstein}},
  \bibinfo {author} {\bibfnamefont {J.}~\bibnamefont {Garel}}, \bibinfo
  {author} {\bibfnamefont {A.}~\bibnamefont {Tkatchenko}}, \bibinfo {author}
  {\bibfnamefont {E.}~\bibnamefont {Joselevich}}, \bibinfo {author}
  {\bibfnamefont {L.}~\bibnamefont {Kronik}}, \ and\ \bibinfo {author}
  {\bibfnamefont {O.}~\bibnamefont {Hod}},\ }\bibfield  {title} {\enquote
  {\bibinfo {title} {{Stacking and registry effects in layered materials: The
  case of hexagonal boron nitride}},}\ }\href {\doibase
  10.1103/PhysRevLett.105.046801} {\bibfield  {journal} {\bibinfo  {journal}
  {Physical Review Letters}\ }\textbf {\bibinfo {volume} {105}},\ \bibinfo
  {pages} {046801} (\bibinfo {year} {2010})},\ \Eprint
  {http://arxiv.org/abs/1002.1728} {arXiv:1002.1728} \BibitemShut {NoStop}%
\bibitem [{\citenamefont {Frazier}(2018)}]{Frazier2018}%
  \BibitemOpen
  \bibfield  {author} {\bibinfo {author} {\bibfnamefont {P.~I.}\ \bibnamefont
  {Frazier}},\ }\bibfield  {title} {\enquote {\bibinfo {title} {{A Tutorial on
  Bayesian Optimization}},}\ }\href {http://arxiv.org/abs/1807.02811} {\
  (\bibinfo {year} {2018})},\ \Eprint {http://arxiv.org/abs/1807.02811}
  {arXiv:1807.02811} \BibitemShut {NoStop}%
\bibitem [{\citenamefont {Brochu}, \citenamefont {Cora},\ and\ \citenamefont
  {De~Freitas}(2010)}]{brochu2010tutorial}%
  \BibitemOpen
  \bibfield  {author} {\bibinfo {author} {\bibfnamefont {E.}~\bibnamefont
  {Brochu}}, \bibinfo {author} {\bibfnamefont {V.~M.}\ \bibnamefont {Cora}}, \
  and\ \bibinfo {author} {\bibfnamefont {N.}~\bibnamefont {De~Freitas}},\
  }\bibfield  {title} {\enquote {\bibinfo {title} {A tutorial on bayesian
  optimization of expensive cost functions, with application to active user
  modeling and hierarchical reinforcement learning},}\ }\href@noop {}
  {\bibfield  {journal} {\bibinfo  {journal} {arXiv preprint arXiv:1012.2599}\
  } (\bibinfo {year} {2010})}\BibitemShut {NoStop}%
\bibitem [{\citenamefont {Nogueira}(2014)}]{Nogueira2014}%
  \BibitemOpen
  \bibfield  {author} {\bibinfo {author} {\bibfnamefont {F.}~\bibnamefont
  {Nogueira}},\ }\href {https://github.com/fmfn/BayesianOptimization} {\enquote
  {\bibinfo {title} {Bayesian optimization: Open source constrained global
  optimization tool for python},}\ } (\bibinfo {year} {2014})\BibitemShut
  {NoStop}%
\bibitem [{\citenamefont {Williams}\ and\ \citenamefont
  {Rasmussen}(2006)}]{williams2006gaussian}%
  \BibitemOpen
  \bibfield  {author} {\bibinfo {author} {\bibfnamefont {C.}~\bibnamefont
  {Williams}}\ and\ \bibinfo {author} {\bibfnamefont {C.~E.}\ \bibnamefont
  {Rasmussen}},\ }\bibfield  {title} {\enquote {\bibinfo {title} {Gaussian
  processes for machine learning, vol. 2},}\ }\href@noop {} {\bibfield
  {journal} {\bibinfo  {journal} {MIT press Cambridge, MA}\ }\textbf {\bibinfo
  {volume} {302}},\ \bibinfo {pages} {303} (\bibinfo {year}
  {2006})}\BibitemShut {NoStop}%
\bibitem [{\citenamefont {Huang}, \citenamefont {Lindgren},\ and\ \citenamefont
  {Chelikowsky}(2005)}]{huang2005surface}%
  \BibitemOpen
  \bibfield  {author} {\bibinfo {author} {\bibfnamefont {X.}~\bibnamefont
  {Huang}}, \bibinfo {author} {\bibfnamefont {E.}~\bibnamefont {Lindgren}}, \
  and\ \bibinfo {author} {\bibfnamefont {J.~R.}\ \bibnamefont {Chelikowsky}},\
  }\bibfield  {title} {\enquote {\bibinfo {title} {Surface passivation method
  for semiconductor nanostructures},}\ }\href@noop {} {\bibfield  {journal}
  {\bibinfo  {journal} {Physical Review B}\ }\textbf {\bibinfo {volume} {71}},\
  \bibinfo {pages} {165328} (\bibinfo {year} {2005})}\BibitemShut {NoStop}%
\bibitem [{\citenamefont {Deng}\ \emph {et~al.}(2012)\citenamefont {Deng},
  \citenamefont {Li}, \citenamefont {Li},\ and\ \citenamefont
  {Wei}}]{deng2012effect}%
  \BibitemOpen
  \bibfield  {author} {\bibinfo {author} {\bibfnamefont {H.-X.}\ \bibnamefont
  {Deng}}, \bibinfo {author} {\bibfnamefont {S.-S.}\ \bibnamefont {Li}},
  \bibinfo {author} {\bibfnamefont {J.}~\bibnamefont {Li}}, \ and\ \bibinfo
  {author} {\bibfnamefont {S.-H.}\ \bibnamefont {Wei}},\ }\bibfield  {title}
  {\enquote {\bibinfo {title} {Effect of hydrogen passivation on the electronic
  structure of ionic semiconductor nanostructures},}\ }\href@noop {} {\bibfield
   {journal} {\bibinfo  {journal} {Physical Review B}\ }\textbf {\bibinfo
  {volume} {85}},\ \bibinfo {pages} {195328} (\bibinfo {year}
  {2012})}\BibitemShut {NoStop}%
\bibitem [{\citenamefont {Zhang}\ \emph {et~al.}(2016)\citenamefont {Zhang},
  \citenamefont {Zhang}, \citenamefont {Tse}, \citenamefont {Wong},
  \citenamefont {Chan}, \citenamefont {Deng},\ and\ \citenamefont
  {Zhu}}]{zhang2016pseudo}%
  \BibitemOpen
  \bibfield  {author} {\bibinfo {author} {\bibfnamefont {Y.}~\bibnamefont
  {Zhang}}, \bibinfo {author} {\bibfnamefont {J.}~\bibnamefont {Zhang}},
  \bibinfo {author} {\bibfnamefont {K.}~\bibnamefont {Tse}}, \bibinfo {author}
  {\bibfnamefont {L.}~\bibnamefont {Wong}}, \bibinfo {author} {\bibfnamefont
  {C.}~\bibnamefont {Chan}}, \bibinfo {author} {\bibfnamefont {B.}~\bibnamefont
  {Deng}}, \ and\ \bibinfo {author} {\bibfnamefont {J.}~\bibnamefont {Zhu}},\
  }\bibfield  {title} {\enquote {\bibinfo {title} {Pseudo-hydrogen passivation:
  A novel way to calculate absolute surface energy of zinc blende (111)/(111)
  surface},}\ }\href@noop {} {\bibfield  {journal} {\bibinfo  {journal}
  {Scientific reports}\ }\textbf {\bibinfo {volume} {6}},\ \bibinfo {pages}
  {1--7} (\bibinfo {year} {2016})}\BibitemShut {NoStop}%
\bibitem [{\citenamefont {Xiong}\ \emph {et~al.}(2017)\citenamefont {Xiong},
  \citenamefont {Liu}, \citenamefont {Zhang}, \citenamefont {Du},\ and\
  \citenamefont {Chen}}]{xiong2017first}%
  \BibitemOpen
  \bibfield  {author} {\bibinfo {author} {\bibfnamefont {H.}~\bibnamefont
  {Xiong}}, \bibinfo {author} {\bibfnamefont {Z.}~\bibnamefont {Liu}}, \bibinfo
  {author} {\bibfnamefont {H.}~\bibnamefont {Zhang}}, \bibinfo {author}
  {\bibfnamefont {Z.}~\bibnamefont {Du}}, \ and\ \bibinfo {author}
  {\bibfnamefont {C.}~\bibnamefont {Chen}},\ }\bibfield  {title} {\enquote
  {\bibinfo {title} {First principles calculation of interfacial stability,
  energy and electronic properties of sic/zrb2 interface},}\ }\href@noop {}
  {\bibfield  {journal} {\bibinfo  {journal} {Journal of Physics and Chemistry
  of Solids}\ }\textbf {\bibinfo {volume} {107}},\ \bibinfo {pages} {162--169}
  (\bibinfo {year} {2017})}\BibitemShut {NoStop}%
\bibitem [{\citenamefont {Christensen}, \citenamefont {Dudiy},\ and\
  \citenamefont {Wahnstr{\"o}m}(2002)}]{christensen2002first}%
  \BibitemOpen
  \bibfield  {author} {\bibinfo {author} {\bibfnamefont {M.}~\bibnamefont
  {Christensen}}, \bibinfo {author} {\bibfnamefont {S.}~\bibnamefont {Dudiy}},
  \ and\ \bibinfo {author} {\bibfnamefont {G.}~\bibnamefont {Wahnstr{\"o}m}},\
  }\bibfield  {title} {\enquote {\bibinfo {title} {First-principles simulations
  of metal-ceramic interface adhesion: Co/wc versus co/tic},}\ }\href@noop {}
  {\bibfield  {journal} {\bibinfo  {journal} {Physical Review B}\ }\textbf
  {\bibinfo {volume} {65}},\ \bibinfo {pages} {045408} (\bibinfo {year}
  {2002})}\BibitemShut {NoStop}%
\bibitem [{\citenamefont {Arya}\ and\ \citenamefont {Carter}(2003)}]{Arya2003}%
  \BibitemOpen
  \bibfield  {author} {\bibinfo {author} {\bibfnamefont {A.}~\bibnamefont
  {Arya}}\ and\ \bibinfo {author} {\bibfnamefont {E.~A.}\ \bibnamefont
  {Carter}},\ }\bibfield  {title} {\enquote {\bibinfo {title} {Structure,
  bonding, and adhesion at the tic(100)/fe(110) interface from first
  principles},}\ }\href {\doibase 10.1063/1.1565323} {\bibfield  {journal}
  {\bibinfo  {journal} {The Journal of Chemical Physics}\ }\textbf {\bibinfo
  {volume} {118}},\ \bibinfo {pages} {8982--8996} (\bibinfo {year} {2003})},\
  \Eprint {http://arxiv.org/abs/https://doi.org/10.1063/1.1565323}
  {https://doi.org/10.1063/1.1565323} \BibitemShut {NoStop}%
\bibitem [{\citenamefont {Arya}\ and\ \citenamefont {Carter}(2004)}]{Arya2004}%
  \BibitemOpen
  \bibfield  {author} {\bibinfo {author} {\bibfnamefont {A.}~\bibnamefont
  {Arya}}\ and\ \bibinfo {author} {\bibfnamefont {E.~A.}\ \bibnamefont
  {Carter}},\ }\bibfield  {title} {\enquote {\bibinfo {title} {Structure,
  bonding, and adhesion at the zrc(100)/fe(110) interface from first
  principles},}\ }\href {\doibase https://doi.org/10.1016/j.susc.2004.04.022}
  {\bibfield  {journal} {\bibinfo  {journal} {Surface Science}\ }\textbf
  {\bibinfo {volume} {560}},\ \bibinfo {pages} {103 -- 120} (\bibinfo {year}
  {2004})}\BibitemShut {NoStop}%
\bibitem [{\citenamefont {Cosandey}, \citenamefont {Zhang},\ and\ \citenamefont
  {Madey}(2001)}]{COSANDEY20011}%
  \BibitemOpen
  \bibfield  {author} {\bibinfo {author} {\bibfnamefont {F.}~\bibnamefont
  {Cosandey}}, \bibinfo {author} {\bibfnamefont {L.}~\bibnamefont {Zhang}}, \
  and\ \bibinfo {author} {\bibfnamefont {T.}~\bibnamefont {Madey}},\ }\bibfield
   {title} {\enquote {\bibinfo {title} {Effect of substrate temperature on the
  epitaxial growth of au on tio2(110)},}\ }\href {\doibase
  https://doi.org/10.1016/S0039-6028(00)01063-3} {\bibfield  {journal}
  {\bibinfo  {journal} {Surface Science}\ }\textbf {\bibinfo {volume} {474}},\
  \bibinfo {pages} {1--13} (\bibinfo {year} {2001})}\BibitemShut {NoStop}%
\bibitem [{\citenamefont {Bansal}\ \emph {et~al.}(2011)\citenamefont {Bansal},
  \citenamefont {Kim}, \citenamefont {Edrey}, \citenamefont {Brahlek},
  \citenamefont {Horibe}, \citenamefont {Iida}, \citenamefont {Tanimura},
  \citenamefont {Li}, \citenamefont {Feng}, \citenamefont {Lee}, \citenamefont
  {Gustafsson}, \citenamefont {Andrei},\ and\ \citenamefont
  {Oh}}]{BANSAL2011224}%
  \BibitemOpen
  \bibfield  {author} {\bibinfo {author} {\bibfnamefont {N.}~\bibnamefont
  {Bansal}}, \bibinfo {author} {\bibfnamefont {Y.~S.}\ \bibnamefont {Kim}},
  \bibinfo {author} {\bibfnamefont {E.}~\bibnamefont {Edrey}}, \bibinfo
  {author} {\bibfnamefont {M.}~\bibnamefont {Brahlek}}, \bibinfo {author}
  {\bibfnamefont {Y.}~\bibnamefont {Horibe}}, \bibinfo {author} {\bibfnamefont
  {K.}~\bibnamefont {Iida}}, \bibinfo {author} {\bibfnamefont {M.}~\bibnamefont
  {Tanimura}}, \bibinfo {author} {\bibfnamefont {G.-H.}\ \bibnamefont {Li}},
  \bibinfo {author} {\bibfnamefont {T.}~\bibnamefont {Feng}}, \bibinfo {author}
  {\bibfnamefont {H.-D.}\ \bibnamefont {Lee}}, \bibinfo {author} {\bibfnamefont
  {T.}~\bibnamefont {Gustafsson}}, \bibinfo {author} {\bibfnamefont
  {E.}~\bibnamefont {Andrei}}, \ and\ \bibinfo {author} {\bibfnamefont
  {S.}~\bibnamefont {Oh}},\ }\bibfield  {title} {\enquote {\bibinfo {title}
  {Epitaxial growth of topological insulator bi2se3 film on si(111) with
  atomically sharp interface},}\ }\href {\doibase
  https://doi.org/10.1016/j.tsf.2011.07.033} {\bibfield  {journal} {\bibinfo
  {journal} {Thin Solid Films}\ }\textbf {\bibinfo {volume} {520}},\ \bibinfo
  {pages} {224--229} (\bibinfo {year} {2011})}\BibitemShut {NoStop}%
\bibitem [{\citenamefont {Bl{\"o}chl}(1994)}]{blochl1994projector}%
  \BibitemOpen
  \bibfield  {author} {\bibinfo {author} {\bibfnamefont {P.~E.}\ \bibnamefont
  {Bl{\"o}chl}},\ }\bibfield  {title} {\enquote {\bibinfo {title} {Projector
  augmented-wave method},}\ }\href@noop {} {\bibfield  {journal} {\bibinfo
  {journal} {Physical review B}\ }\textbf {\bibinfo {volume} {50}},\ \bibinfo
  {pages} {17953} (\bibinfo {year} {1994})}\BibitemShut {NoStop}%
\bibitem [{\citenamefont {Kresse}\ and\ \citenamefont
  {Joubert}(1999)}]{Kresse1999}%
  \BibitemOpen
  \bibfield  {author} {\bibinfo {author} {\bibfnamefont {G.}~\bibnamefont
  {Kresse}}\ and\ \bibinfo {author} {\bibfnamefont {D.}~\bibnamefont
  {Joubert}},\ }\bibfield  {title} {\enquote {\bibinfo {title} {From ultrasoft
  pseudopotentials to the projector augmented-wave method},}\ }\href {\doibase
  10.1103/PhysRevB.59.1758} {\bibfield  {journal} {\bibinfo  {journal} {Phys.
  Rev. B}\ }\textbf {\bibinfo {volume} {59}},\ \bibinfo {pages} {1758--1775}
  (\bibinfo {year} {1999})}\BibitemShut {NoStop}%
\bibitem [{\citenamefont {Perdew}, \citenamefont {Burke},\ and\ \citenamefont
  {Ernzerhof}(1996)}]{perdew1996generalized}%
  \BibitemOpen
  \bibfield  {author} {\bibinfo {author} {\bibfnamefont {J.~P.}\ \bibnamefont
  {Perdew}}, \bibinfo {author} {\bibfnamefont {K.}~\bibnamefont {Burke}}, \
  and\ \bibinfo {author} {\bibfnamefont {M.}~\bibnamefont {Ernzerhof}},\
  }\bibfield  {title} {\enquote {\bibinfo {title} {Generalized gradient
  approximation made simple},}\ }\href@noop {} {\bibfield  {journal} {\bibinfo
  {journal} {Physical review letters}\ }\textbf {\bibinfo {volume} {77}},\
  \bibinfo {pages} {3865} (\bibinfo {year} {1996})}\BibitemShut {NoStop}%
\bibitem [{\citenamefont {Dudarev}\ \emph {et~al.}(1998)\citenamefont
  {Dudarev}, \citenamefont {Botton}, \citenamefont {Savrasov}, \citenamefont
  {Humphreys},\ and\ \citenamefont {Sutton}}]{dudarev1998electron}%
  \BibitemOpen
  \bibfield  {author} {\bibinfo {author} {\bibfnamefont {S.~L.}\ \bibnamefont
  {Dudarev}}, \bibinfo {author} {\bibfnamefont {G.~A.}\ \bibnamefont {Botton}},
  \bibinfo {author} {\bibfnamefont {S.~Y.}\ \bibnamefont {Savrasov}}, \bibinfo
  {author} {\bibfnamefont {C.~J.}\ \bibnamefont {Humphreys}}, \ and\ \bibinfo
  {author} {\bibfnamefont {A.~P.}\ \bibnamefont {Sutton}},\ }\bibfield  {title}
  {\enquote {\bibinfo {title} {Electron-energy-loss spectra and the structural
  stability of nickel oxide: An lsda+u study},}\ }\href {\doibase
  10.1103/PhysRevB.57.1505} {\bibfield  {journal} {\bibinfo  {journal} {Phys.
  Rev. B}\ }\textbf {\bibinfo {volume} {57}},\ \bibinfo {pages} {1505--1509}
  (\bibinfo {year} {1998})}\BibitemShut {NoStop}%
\bibitem [{\citenamefont {Yu}\ \emph {et~al.}(2020)\citenamefont {Yu},
  \citenamefont {Yang}, \citenamefont {Wu},\ and\ \citenamefont
  {Marom}}]{Yu2020}%
  \BibitemOpen
  \bibfield  {author} {\bibinfo {author} {\bibfnamefont {M.}~\bibnamefont
  {Yu}}, \bibinfo {author} {\bibfnamefont {S.}~\bibnamefont {Yang}}, \bibinfo
  {author} {\bibfnamefont {C.}~\bibnamefont {Wu}}, \ and\ \bibinfo {author}
  {\bibfnamefont {N.}~\bibnamefont {Marom}},\ }\bibfield  {title} {\enquote
  {\bibinfo {title} {Machine learning the hubbard u parameter in dft+u using
  bayesian optimization},}\ }\href {\doibase 10.1038/s41524-020-00446-9}
  {\bibfield  {journal} {\bibinfo  {journal} {npj Computational Materials}\
  }\textbf {\bibinfo {volume} {6}},\ \bibinfo {pages} {180} (\bibinfo {year}
  {2020})}\BibitemShut {NoStop}%
\bibitem [{\citenamefont {Tkatchenko}\ and\ \citenamefont
  {Scheffler}(2009{\natexlab{a}})}]{tkatchenko2009accurate}%
  \BibitemOpen
  \bibfield  {author} {\bibinfo {author} {\bibfnamefont {A.}~\bibnamefont
  {Tkatchenko}}\ and\ \bibinfo {author} {\bibfnamefont {M.}~\bibnamefont
  {Scheffler}},\ }\bibfield  {title} {\enquote {\bibinfo {title} {Accurate
  molecular van der waals interactions from ground-state electron density and
  free-atom reference data},}\ }\href@noop {} {\bibfield  {journal} {\bibinfo
  {journal} {Physical review letters}\ }\textbf {\bibinfo {volume} {102}},\
  \bibinfo {pages} {073005} (\bibinfo {year} {2009}{\natexlab{a}})}\BibitemShut
  {NoStop}%
\bibitem [{\citenamefont {Neugebauer}\ and\ \citenamefont
  {Scheffler}(1992)}]{neugebauer1992adsorbate}%
  \BibitemOpen
  \bibfield  {author} {\bibinfo {author} {\bibfnamefont {J.}~\bibnamefont
  {Neugebauer}}\ and\ \bibinfo {author} {\bibfnamefont {M.}~\bibnamefont
  {Scheffler}},\ }\bibfield  {title} {\enquote {\bibinfo {title}
  {Adsorbate-substrate and adsorbate-adsorbate interactions of na and k
  adlayers on al (111)},}\ }\href@noop {} {\bibfield  {journal} {\bibinfo
  {journal} {Physical Review B}\ }\textbf {\bibinfo {volume} {46}},\ \bibinfo
  {pages} {16067} (\bibinfo {year} {1992})}\BibitemShut {NoStop}%
\bibitem [{\citenamefont {Steiner}\ \emph {et~al.}(2016)\citenamefont
  {Steiner}, \citenamefont {Khmelevskyi}, \citenamefont {Marsmann},\ and\
  \citenamefont {Kresse}}]{steiner2016calculation}%
  \BibitemOpen
  \bibfield  {author} {\bibinfo {author} {\bibfnamefont {S.}~\bibnamefont
  {Steiner}}, \bibinfo {author} {\bibfnamefont {S.}~\bibnamefont
  {Khmelevskyi}}, \bibinfo {author} {\bibfnamefont {M.}~\bibnamefont
  {Marsmann}}, \ and\ \bibinfo {author} {\bibfnamefont {G.}~\bibnamefont
  {Kresse}},\ }\bibfield  {title} {\enquote {\bibinfo {title} {Calculation of
  the magnetic anisotropy with projected-augmented-wave methodology and the
  case study of disordered ${\mathrm{fe}}_{1\ensuremath{-}x}{\mathrm{co}}_{x}$
  alloys},}\ }\href {\doibase 10.1103/PhysRevB.93.224425} {\bibfield  {journal}
  {\bibinfo  {journal} {Phys. Rev. B}\ }\textbf {\bibinfo {volume} {93}},\
  \bibinfo {pages} {224425} (\bibinfo {year} {2016})}\BibitemShut {NoStop}%
\bibitem [{\citenamefont {Tkatchenko}\ and\ \citenamefont
  {Scheffler}(2009{\natexlab{b}})}]{Tkatchenko2009}%
  \BibitemOpen
  \bibfield  {author} {\bibinfo {author} {\bibfnamefont {A.}~\bibnamefont
  {Tkatchenko}}\ and\ \bibinfo {author} {\bibfnamefont {M.}~\bibnamefont
  {Scheffler}},\ }\bibfield  {title} {\enquote {\bibinfo {title} {{Accurate
  molecular van der Waals interactions from ground-state electron density and
  free-atom reference data}},}\ }\href {\doibase
  10.1103/PhysRevLett.102.073005} {\bibfield  {journal} {\bibinfo  {journal}
  {Physical Review Letters}\ }\textbf {\bibinfo {volume} {102}},\ \bibinfo
  {pages} {073005} (\bibinfo {year} {2009}{\natexlab{b}})}\BibitemShut
  {NoStop}%
\bibitem [{\citenamefont {Heine}(1965)}]{volker1965theory}%
  \BibitemOpen
  \bibfield  {author} {\bibinfo {author} {\bibfnamefont {V.}~\bibnamefont
  {Heine}},\ }\bibfield  {title} {\enquote {\bibinfo {title} {Theory of surface
  states},}\ }\href {\doibase 10.1103/PhysRev.138.A1689} {\bibfield  {journal}
  {\bibinfo  {journal} {Phys. Rev.}\ }\textbf {\bibinfo {volume} {138}},\
  \bibinfo {pages} {A1689--A1696} (\bibinfo {year} {1965})}\BibitemShut
  {NoStop}%
\bibitem [{\citenamefont {Mönch}(1999)}]{monch1999winfried}%
  \BibitemOpen
  \bibfield  {author} {\bibinfo {author} {\bibfnamefont {W.}~\bibnamefont
  {Mönch}},\ }\bibfield  {title} {\enquote {\bibinfo {title} {Barrier heights
  of real schottky contacts explained by metal-induced gap states and lateral
  inhomogeneities},}\ }\href {\doibase 10.1116/1.590839} {\bibfield  {journal}
  {\bibinfo  {journal} {Journal of Vacuum Science \& Technology B:
  Microelectronics and Nanometer Structures Processing, Measurement, and
  Phenomena}\ }\textbf {\bibinfo {volume} {17}},\ \bibinfo {pages} {1867--1876}
  (\bibinfo {year} {1999})},\ \Eprint
  {http://arxiv.org/abs/https://avs.scitation.org/doi/pdf/10.1116/1.590839}
  {https://avs.scitation.org/doi/pdf/10.1116/1.590839} \BibitemShut {NoStop}%
\bibitem [{\citenamefont {Nishimura}, \citenamefont {Kita},\ and\ \citenamefont
  {Toriumi}(2007)}]{nishimura2007evidence}%
  \BibitemOpen
  \bibfield  {author} {\bibinfo {author} {\bibfnamefont {T.}~\bibnamefont
  {Nishimura}}, \bibinfo {author} {\bibfnamefont {K.}~\bibnamefont {Kita}}, \
  and\ \bibinfo {author} {\bibfnamefont {A.}~\bibnamefont {Toriumi}},\
  }\bibfield  {title} {\enquote {\bibinfo {title} {Evidence for strong
  fermi-level pinning due to metal-induced gap states at metal/germanium
  interface},}\ }\href {\doibase 10.1063/1.2789701} {\bibfield  {journal}
  {\bibinfo  {journal} {Applied Physics Letters}\ }\textbf {\bibinfo {volume}
  {91}},\ \bibinfo {pages} {123123} (\bibinfo {year} {2007})},\ \Eprint
  {http://arxiv.org/abs/https://doi.org/10.1063/1.2789701}
  {https://doi.org/10.1063/1.2789701} \BibitemShut {NoStop}%
\end{thebibliography}%

\end{document}